\documentclass[aps,prb,twocolumn,showpacs,eqsecnum]{revtex4}
\usepackage{bm}

\usepackage{amsmath}
\usepackage{amssymb}
\usepackage{graphicx}
\usepackage{psfrag}

\newcommand{\beq}{\begin{equation}}
\newcommand{\eeq}{\end{equation}}
\newcommand{\beqar}{\begin{eqnarray*}}
\newcommand{\eeqar}{\end{eqnarray*}}

\newcommand{\sgn}{{\rm sgn}\,}
\newcommand{\ua}{\uparrow}
\newcommand{\da}{\downarrow}

\newcommand{\dm}{\diamond}

\newcommand{\dt}{{\text d}}
\newcommand{\etxt}{{\text e}}

\newcommand{\itxt}{{\text i}}

\newcommand{\Stxt}{{\text S}}
\newcommand{\Ttxt}{{\text T}}
\newcommand{\Utxt}{{\text U}}

\newcommand{\Hh}{\hat{H}}

\newcommand{\psih}{\hat{\psi}}

\newcommand{\Abr}{\bar{A}}
\newcommand{\Bbr}{\bar{B}}

\newcommand{\Tc}{\mathcal{T}}

\newcommand{\Fc}{\mathcal{F}}

\newcommand{\taub}{{\mbox{\boldmath{$\tau$}}}}

\newcommand{\pd}{\partial}
\newcommand{\rtarr}{\rightarrow}

\newcommand{\gb}{{\bar{g}}}

\newcommand{\Kbr}{{\bar{K}}}

\newcommand{\psit}{\tilde{\psi}}

\newcommand{\ch}{\hat{c}}

\newcommand{\ph}{\hat{p}}

\newcommand{\Kb}{{\bf K}}

\newcommand{\nb}{{\bf n}}

\newcommand{\ab}{{\bf a}}

\newcommand{\s}{{\bf s}}
\newcommand{\qb}{{\bf q}}

\newcommand{\Ab}{{\bf A}}

\newcommand{\rb}{{\bf r}}

\newcommand{\fb}{\bar{f}}

\newcommand{\dg}{\dagger}

\newcommand{\lan}{\langle}
\newcommand{\ran}{\rangle}
\newcommand{\om}{\omega}

\newcommand{\al}{\alpha}
\newcommand{\be}{\beta}

\newcommand{\Ga}{\Gamma}
\newcommand{\de}{\delta}

\newcommand{\la}{\lambda}

\newcommand{\sig}{\sigma}

\newcommand{\vphi}{\varphi}

\newcommand{\e}{\epsilon}
\newcommand{\Hc}{{\cal H}}
\newcommand{\lt}{\left}
\newcommand{\rt}{\right}
\newcommand{\tr}{\text{tr}}

\newcommand{\rv}{{\vec{r}}}
\newcommand{\Psih}{\hat{\Psi}}
\newcommand{\Ec}{{\cal{E}}}

\begin{document}
\title{
Phase diagram for the $\nu=0$ quantum Hall state
in monolayer graphene}

\author{Maxim Kharitonov}

\address{
\centerline{Center for Materials Theory, Department of Physics and Astronomy, Rutgers University, Piscataway, NJ 08854, USA}
}
\date{\today}
\begin{abstract}

The $\nu=0$ quantum Hall state in a defect-free graphene sample is studied within the framework of quantum Hall ferromagnetism.
We perform a systematic analysis of the ``isospin''  anisotropies, which arise from the
valley and sublattice asymmetric short-range electron-electron  (e-e) and electron-phonon (e-ph)
interactions. The phase diagram,
obtained in the presence of generic isospin anisotropy and the Zeeman effect,
consists of four phases characterized by the following orders:
spin-polarized ferromagnetic, canted antiferromagnetic,
charge density wave, and Kekul\'{e} distortion.
We take into account
the Landau level mixing effects
and show that they result in the key renormalizations of parameters.
First, the absolute values of the anisotropy energies become greatly
enhanced and can significantly exceed the Zeeman energy.
Second, the signs of the anisotropy energies due to e-e interactions can change upon renormalization.
A crucial consequence of the latter is  that the short-range e-e interactions alone could favor any state on the phase diagram,
depending on the details of interactions at the lattice scale.
On the other hand, the leading e-ph interactions always favor the Kekul\'{e} distortion order.
The possibility of inducing phase transitions by tilting the magnetic field is discussed.

\end{abstract}
\maketitle

\section{Introduction}

\begin{figure}
\includegraphics[width=.28\textwidth]{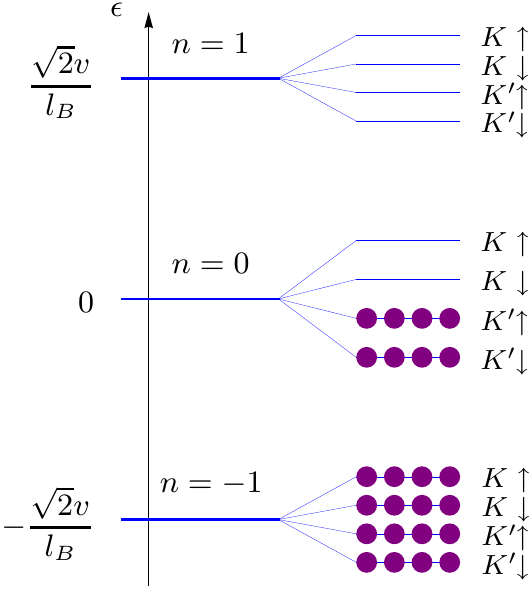}
\caption{Landau levels (LLs) [Eq.~(\ref{eq:en})] in monolayer graphene. Neglecting the Zeeman effect,
the orbitals  (i.e., eigenstates of the orbital part of the single-particle Hamiltonian)
of each LL are four-fold degenerate due to two projections of spin and two valleys (splitting shown for illustration purpose and not implied).
At $\nu=0$ filling factor, each orbital of the $n=0$ LL is occupied on average by two electrons, while LLs with $n<0$ ($n>0$) are filled (empty).
The many-body states with identical occupation of the $KK'\otimes s$ isospin-spin subspaces of all orbitals
(only one possible occupation is shown) exactly minimize the
energy of the Coulomb interactions, SU(4)-symmetric in the $KK'\otimes s$ space.
Such states form the family of degenerate ground states of the $\nu=0$ quantum Hall ferromagnet.}
\label{fig:LLs}
\end{figure}

\begin{figure}
\centerline{
\includegraphics[width=.22\textwidth]{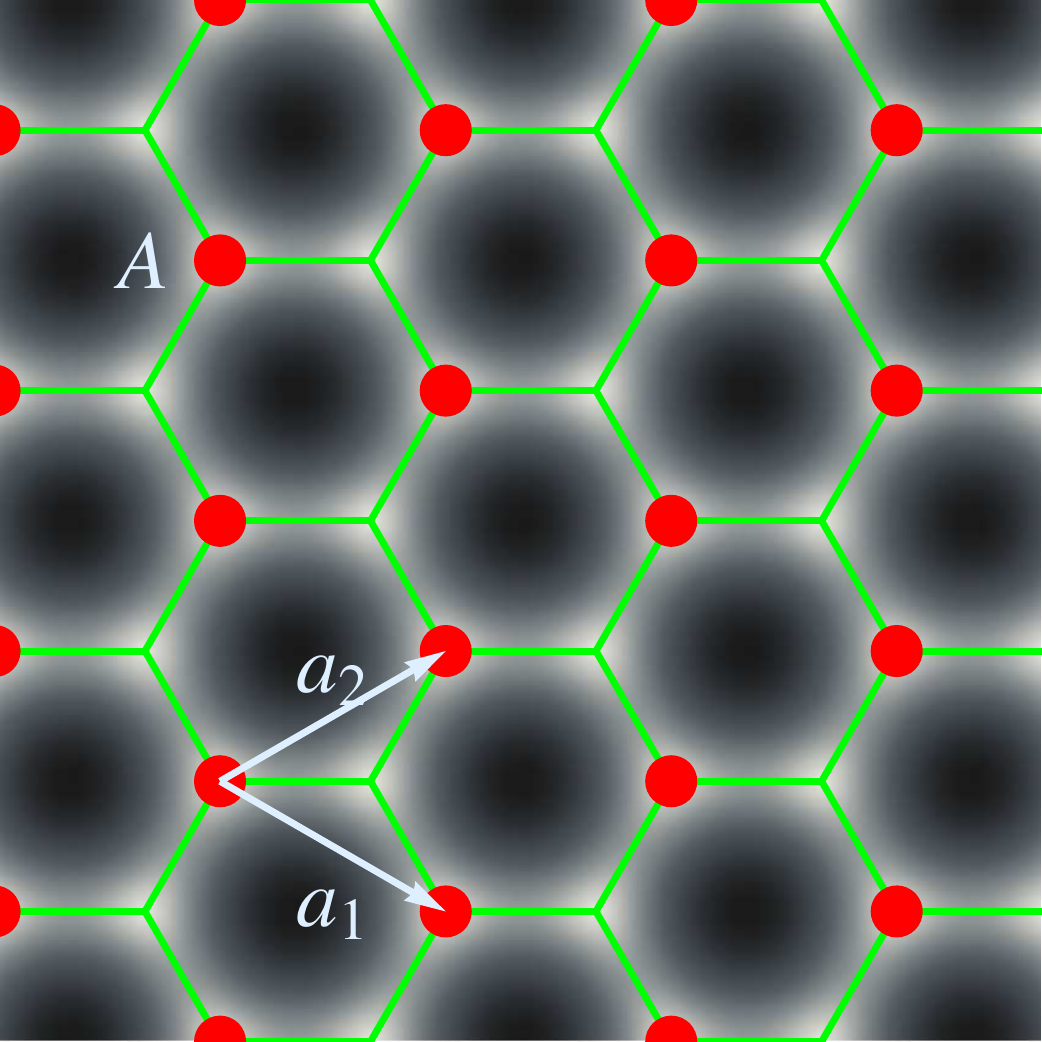}
\hspace{.3cm}
\includegraphics[width=.22\textwidth]{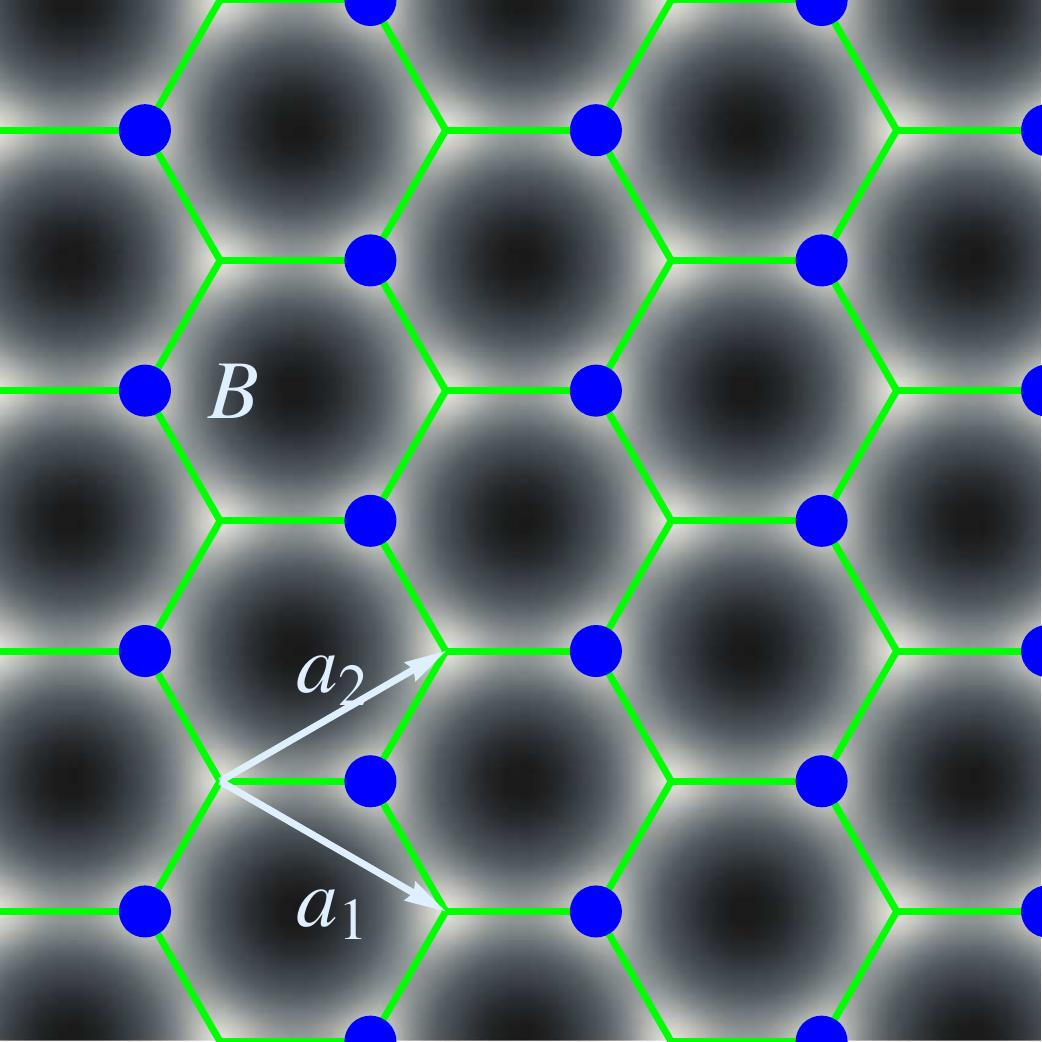}
}
\caption{Graphene honeycomb lattice and the structure of the wave-functions of the $n=0$ LL at the lattice scale.
In each valley, $K$ or $K'$,
the wave-functions reside on only one sublattice, $A$ (left) or $B$ (right).
}
\label{fig:n0wfs}
\end{figure}

Quantum Hall effects in graphene~\cite{Novoselov_etal,Zhang_etal,Berger_etal}
were observed shortly after~\cite{QHE1,QHE2} the discovery of the material and have attracted a lot of attention ever since
(for general reviews on graphene, see Refs.~\onlinecite{GeimNovoselov,CastroNetoetalRMP,DasSarmaetalRMP}).
Besides the ``anomalous'' sequence of the orbital Landau levels~\cite{ZhengAndo,GusyninSharapov,PGCN} (LLs), Fig.~\ref{fig:LLs},
\beq
    \e_n = \frac{v}{l_B} \sqrt{2 |n|} \sgn n, \mbox{ } l_B = \sqrt{\frac{c}{e B_\perp}},
\label{eq:en}
\eeq
characteristic of the Dirac nature of electron spectrum in graphene,
an additional to spin, two-fold valley degeneracy brings in extra richness to physical phenomena
[In Eq.~(\ref{eq:en}), $n$ takes  integer values, $v$ is the Dirac velocity, $l_B$ is the magnetic length dependent on the perpendicular component $B_\perp$ of the magnetic field, and we put $\hbar=1$ throughout the paper].
Among all, the $n=0$ LL really stands out:
it is located exactly at the Dirac point $\e_0 =0$ and, in each valley, $K$ or $K'$, its wave-functions reside solely on one of the sublattices,
$A$ or $B$, Fig.~\ref{fig:n0wfs}.
For $n=0$ LL, the valley $KK'$ ``isospin'' is, therefore, equivalent to the sublattice $AB$ ``pseudospin'',
and is referred to as just  the ``isospin'' below.

As the quality of graphene devices progressed,
the strongly correlated quantum Hall physics
clearly emerged~\cite{Columbia0,Kim2,Rutgers2,nu0Ong,nu0Brookhaven,nu0Andrei,nu0Kim,ColumbiaBN} at integer and fractional filling factors $\nu$.
To date, one of the most intriguing questions concerns the nature of the $\nu=0$ quantum Hall
state, in which the orbitals of the $n=0$ LL,
four-fold degenerate in the $KK' \otimes s$ isospin-spin space in the absence of the Zeeman effect,
are occupied on average by two electrons, Fig.~\ref{fig:LLs}.
The interest is largely motivated  by the strongly insulating behavior
of the state, initially observed in samples on SiO$_2$ substrate~\cite{nu0Ong,nu0Brookhaven},
later -- in suspended samples~\cite{nu0Andrei,nu0Kim}, and quite recently -- in samples on boron nitride substrate~\cite{ColumbiaBN}.
The tendency of the higher quality samples to be more resistive
signifies that the insulating behavior is an intrinsic property
of an ideal defect-free system, rather than it is due to disorder-induced
localization effects.
Although the interaction-driven character of the insulating $\nu=0$ state is apparent, its precise nature remains an open challenge in the graphene field.

A lot of theoretical activity has been devoted~\cite{NM,YDM,AF,Abanin,FB,Goerbig,DY,JM,Nomura,Chamon,FuchsLederer,MC01,MC02,MC1,Herbut,MC2,Karlsruhe}
to the properties of the $\nu=0$ quantum Hall state in graphene (see also Refs.~\onlinecite{Yangreview,GoerbigHT}
for reviews).
Due to the quenched kinetic energy, in sufficiently clean samples and/or high enough magnetic fields, electrons in a partially filled LL
present a strongly interacting system.
While at arbitrary (e.g., fractional) filling factors $\nu$ such systems pose a formidable theoretical challenge,
an appealing property of the integer fillings is that, in the leading approximation,
the family of the many-body ground states can be found exactly.
For all integer $\nu$ in graphene, including $\nu=0$, this
solution is  provided
by the general theory of quantum Hall ferromagnetism~\cite{Arovas_etal,QHB,Ezawa_etal,NM,YDM} (QHFM) for multicomponent systems,
in which spin and valley degrees of freedom are united into one SU(4) ``spin''.
(Alternative views on the $\nu=0$ state based on the idea of ``magnetic catalysis'' can be found in Refs.~\onlinecite{MC01,MC02,MC1,Herbut,MC2}.)

The idea of constructing the exact solution can be traced back to the Hund's rule in atomic physics:
specifically for the $\nu=0$ state, the energy of the repulsive Coulomb interactions is minimized by the many-body states,
in which the four-dimensional (4D) $KK' \otimes s$ isospin-spin subspace of each orbital
of the $n=0$ LL is occupied by two electrons in exactly the same way, Fig.~\ref{fig:LLs}.
Two key provisos that make this result exact are
(i) absence of the single-particle or many-body perturbations that break the SU(4) symmetry in the $KK' \otimes s$ space and
(ii) absence of the interaction-induced electron transitions between different LLs (LL ``mixing'').
For $\Stxt \Utxt (4)$-symmetric Coulomb interactions,
the ground state is
degenerate with respect to the choice of occupation of the isospin-spin space and
there is no preference between  ordering of the spin vs. isospin degrees of freedom.

This exact result is the cornerstone of the theory of the $\nu=0$ state.
In a real graphene system, neither of the conditions (i) or (ii) is satisfied precisely,
and the result is generally not exact.
However, it is possible to develop the low-energy quantum field theory for the $\nu=0$ quantum Hall ferromagnet (QHFM),
in which the deviations from these conditions are consistently taken into account.
As we show here, for Coulomb interactions of moderate strength $e^2/v\sim 1$, as is the case in graphene,
the effects of LL mixing can be systematically taken into account in the large-$N$ approximation.
Further, the  factors
that break the symmetry in the $AB \otimes s$
space typically have smaller energy scales than the Coulomb interactions and can be taken into account perturbatively.
They are, nonetheless, extremely important, as they lift the rich degeneracy of the ground state favoring certain orders,
the physical properties of which may differ substantially.

The simplest single-particle mechanism is the
Zeeman effect, which breaks the spin symmetry and naturally favors the fully spin-polarized ferromagnetic (F) state~\cite{Abanin,FB,YDM}.
Since spin-orbit interaction  is weak in graphene,
the Zeeman effect is practically the only relevant
factor that affects the spin symmetry,
while other key perturbations break the isospin symmetry.
An analogous regular ``Zeeman'' field for the $KK'$ isospin is virtually impossible to implement in a controlled way in real graphene,
although a random one can arise from the short-range disorder.

However, even in an ideal defect-free sample electron-electron (e-e) and electron-phonon (e-ph) interactions
necessarily break the valley and sublattice symmetry at the lattice scales~\cite{AF,Goerbig,JM,Nomura,Chamon}.
Having many-body origin, these mechanisms give rise to the isospin anisotropy
rather than the Zeeman-type fields.
Existing studies~\cite{AF,JM} of the lattice effects of e-e interactions on the $\nu=0$ state in the framework of the QHFM theory
were carried out using the tight-binding extended Hubbard model,
with adjustable interactions at the lattice scale and Coulomb asymptotic at larger distances.
Depending on the interactions at the lattice scale,
the competition between the F and charge-density-wave (CDW) ground states was predicted in Ref.~\onlinecite{AF},
while the numerical mean-field analysis of Ref.~\onlinecite{JM}
predicted either the CDW or the antiferromagnetic (AF) ground state.
Electron-phonon interactions, on the other hand, were predicted to favor the fully isospin-polarized states
with either the Kekul\'{e} distortion~\cite{Nomura,Chamon} (KD) or CDW~\cite{FuchsLederer} orders.
We also mention that similar phases were predicted in Ref.~\onlinecite{Herbut} within a different framework of ``magnetic catalysis''.

The variety of the proposed ground states poses a question
whether the above list is exhaustive, i.e., whether, for a given source of the isospin anisotropy,
one can find all possible orders that could be realized in a general case scenario.
Attempting to answer this question, in this paper, we perform a systematic analysis of the isospin anisotropies of the $\nu=0$ QHFM
arising from the short-range e-e and e-ph interactions, without appealing to any specific lattice model.
Starting from the most general form of e-e interactions allowed by symmetry in the Dirac Hamiltonian
and taking the leading e-ph interactions into account,
we derive the low-energy QHFM theory in the presence of the isospin anisotropy
and obtain a phase diagram for the $\nu=0$ state.
The diagram obtained by separately minimizing the energy
of the generic isospin anisotropy
(i.e., considered without any assumptions about the nature or properties of the underlying
interactions, whether e-e or e-ph, repulsive or attractive) consists of four phases:
$\Stxt\Utxt(2)$-spin-degenerate F and  AF phases, and spin-singlet CDW and KD phases with $\text{Z}_2$ and $\Utxt(1)$ isospin
degeneracies, respectively.
Including the Zeeman effect  does not alter the
CDW or KD phases, but removes the degeneracy of the F phase
and transforms the AF phase into a canted antiferromagnetic (CAF) phase with noncollinear spin polarizations of sublattices.

We also consider the critical renormalizations of the isospin anisotropy by the long-range Coulomb interactions
and address the question whether one can rule out certain states from the phase diagram
based on the repulsive nature of the Coulomb interactions
and attractive nature of the phonon-mediated interactions.
We arrive at an important conclusion that
the short-range e-e interactions could favor essentially any state on the generic phase diagram: F, CAF, CDW, or KD.
The reason for this are peculiar properties
of the renormalizations, which allow for sign changes of the e-e coupling constants, switching the interactions
from repulsive to attractive in certain valley-sublattice channels.
As a result, unless a reliable numerical estimate for the bare short-range e-e coupling constants is provided,
one cannot theoretically rule out any possibility.
In contrast, the phonon-mediated interactions remain attractive under renormalizations
and the leading e-ph interactions always favor the KD order.

Among potential practical applications of the present work,
we also study the transitions that could be induced between the obtained phases by the tilting magnetic field.
Once combined with a more details analysis of the charge excitations of these phases, to be presented elsewhere~\cite{MKunpub},
our findings could be used to identify the particular phase realized
in the experimentally observed~\cite{nu0Ong,nu0Brookhaven,nu0Andrei,nu0Kim,ColumbiaBN} insulating $\nu=0$ state.

The rest of the paper is organized as follows. In Sec.~\ref{sec:model},
the low-energy Hamiltonian, which describes electron dynamics in the vicinity of  the Dirac point,
is presented, the basic properties of Landau levels in graphene are discussed,
and the projected Hamiltonian for the $n=0$ Landau level is derived.
In Sec.~\ref{sec:QHFM},
the low-energy quantum field theory for the $\nu=0$ quantum Hall ferromagnet, which includes the effects of the isospin anisotropy, is derived.
In Sec.~\ref{sec:renorm}, the Landau level mixing effects are considered in the large-$N$ approximation
and the critical renormalizations of the isospin anisotropies are studied.
In Sec.~\ref{sec:ground},
the phase diagram of the $\nu=0$ quantum Hall ferromagnet in the presence of the isospin anisotropy and Zeeman effect is obtained.
The possibility of inducing phase transitions by tilting the magnetic field is discussed.
Concluding remarks and connection to the experiment are presented in Sec.~\ref{sec:conclusions}.

\section{Model and Hamiltonian \label{sec:model}}

We start the analysis by writing down the low-energy Hamiltonian
\beq
    \Hh = \Hh_0 +\Hh_\text{e-e} + \Hh_\text{e-ph},
\label{eq:H}
\eeq
which describes electron dynamics in graphene in the vicinity of  the Dirac point.
The terms $\Hh_0$, $\Hh_\text{e-e}$, and $\Hh_\text{e-ph}$, describing noninteracting electrons, e-e and e-ph interactions,
respectively, are discussed in the next three subsections.
Our Hamiltonian and the choice of basis are identical to those of Refs.~\onlinecite{AKT,BA}, with some differences in notation.

\subsection{Basis and single-particle Hamiltonian \label{sec:H0}}

At atomic scales, the single-particle electron Hamiltonian can be written as
\beq
    \hat{\Hc}_0 = \int \dt^3 \rv\, \Psih^\dg_\sig (\rv)   \lt[ - \frac{\pd_\rv^2}{2 m}  +  U(\rv) \rt]    \Psih_\sig(\rv).
\label{eq:H0a}
\eeq
Here, $\Psih_\sig(\rv)$ is the electron field operator,
$\rv=(x,y,z)$ is a continuous three-dimensional (3D) radius-vector,
and $\sig=\ua,\da$ is the spin projection (summation over $\sig$ is implied).
The self-consistent periodic potential $U(\rv)$ of the graphene honeycomb lattice constrains electrons around the  $z=0$ plane
and has a $C_{6v}$ point group symmetry within the plane.
This symmetry dictates the following properties of the graphene band structure.
Exactly at the Dirac point, taken to be at zero energy $\e=0$,
there are four orthogonal Bloch-wave solutions $u_{KA}(\rv)$, $ u_{KB}(\rv)$, $u_{K'A}(\rv)$, $u_{K'B}(\rv)$
of the Schr\"{o}dinger equation associated with Eq.~(\ref{eq:H0a}).
The indices $K$ and $K'$ refer to different valleys in the Brillouin with wave vectors $\Kb= \frac{4 \pi}{3 a_0^2} (\ab_1-\ab_2)$
and $\Kb' =-\Kb$, respectively ($\ab_{1,2}$ are the primitive translations of the honeycomb lattice, shown in Fig.~\ref{fig:n0wfs}
and $a_0= |\ab_{1,2}| \approx 2.46 \AA$ is the lattice constant), and the indices $A$ and $B$
indicate that the wave-functions are predominantly localized at the positions of the $A$ and $B$ sites.
The solutions corresponding to different valleys
are related as $u_{K'A}(\rv)=u_{KA}^*(\rv)$ and  $u_{K'B}(\rv)=u_{KB}^*(\rv)$.

For the excitation energies $\e$ much smaller than the bandwidth, one may expand the electron field in terms of
the $\e=0$ solutions,
\begin{eqnarray}
    \Psih_\sig(\rv) & = &  \psih_{KA\sig}(\rb) u_{KA}(\rv) + \psih_{KB\sig}(\rb) u_{KB}(\rv) \nonumber \\
    & + &  \psih_{K'A\sig}(\rb) u_{K'A}(\rv) + \psih_{K'B\sig}(\rb) u_{K'B}(\rv).
\label{eq:Psiexp}
\end{eqnarray}
The Dirac field operators $\psi_{\la\sig}(\rb)$, $\la= KA$, $KB$, $K'A$, $K'B$,  are functions of a two-dimensional (2D) continuous radius vector $\rb=(x,y)$
and vary at scales much larger than the atomic one $a_0$.
To ensure the standard normalization of the fields reflected in anticommutation relation
\[
    \{ \psi_{\la\sig}(\rb), \psi_{\sig'\la'}^\dg (\rb') \} = \de_{\la\la'}\de_{\sig\sig'}\de(\rb-\rb'),
\]
where $\de(\rb-\rb')$ is a 2D delta function at large scales
and $\{  ,  \}$ is the anticommutator,
the Bloch wave-functions $u_\la(\rv)$ must be normalized as
\beq
    \int_\text{3uc} \dt^3 \rv\, u^*_\la(\rv) u_{\la'}(\rv) = 3 |[\ab_1 \times \ab_2]| \de_{\la\la'},
\label{eq:uorth}
\eeq
where the integration is performed over the tripled unit cell (3uc), which contains six atoms.

The Dirac fields can be joined in a vector as
\beq
    \psih_\sig(\rb) =
    \lt(\begin{array}{c} \psih_{KA\sig}(\rb) \\ \psih_{KB\sig}(\rb)
            \\ \psih_{K'B\sig}(\rb) \\ -\psih_{K'A\sig}(\rb) \end{array}\rt)_{KK'\otimes\Abr\Bbr}.
\label{eq:psisigdef}
\eeq
The advantage of this ordering is that it gives the most symmetric representation of the Dirac Hamiltonian.
Since this way the sublattice indices in the $K'$ valley are interchanged,
to avoid confusion,
we denote the sublattice space of the basis (\ref{eq:psisigdef}) as $\Abr \Bbr$,
and the whole 4D space -- as the direct product $KK' \otimes \Abr\Bbr$.
With spin included, the low-energy electron degrees of freedom are described by
the eight-component field operator
\beq
    \psih(\rb) = \lt(\begin{array}{c} \psih_\ua(\rb) \\ \psih_\da(\rb) \end{array}\rt)_s
\label{eq:psidef}
\eeq
in the direct product  $KK' \otimes \Abr \Bbr \otimes s$ of the valley ($KK'$), sublattice ($\Abr \Bbr$), and spin ($s$) spaces.

The symmetry properties of the Bloch wave-functions $u_\la(\rv)$ at the Dirac point are sufficient
to derive the many-body low-energy Hamiltonian in the basis of $\psih(\rb)$.
The single-particle Hamiltonian, obtained from Eq.~(\ref{eq:H0a}), has the form
\beq
    \Hh_0 =  \int \dt^2 \rb \,\psih^\dg (\rb) \lt[ v \sum_{\al=x,y} \Tc_{0\al} \lt( \ph_\al-\frac{e}{c} A_\al \rt)   - \e_Z S_z  \rt] \psih(\rb),
\label{eq:H0}
\eeq
where $\ph_\al = -\itxt \nabla_\al$, $\nabla = (\pd_x,\pd_y)$, and $v \approx 10^8 \text{cm/s}$ is the velocity of the Dirac spectrum.
Here and below, for $\al,\be=0,x,y,z$,
\[
    \Tc_{\al\be} = \tau^{KK'}_\al \otimes \tau^{\Abr\Bbr}_\be \otimes \hat{1}^s,
\]
with the unity ($\tau_0 = \hat{1}$) and Pauli ($\tau_x$, $\tau_y$, $\tau_z$) matrices in the corresponding 2D subspaces.
In Eq.~(\ref{eq:H0}), we introduced the orbital and spin effects of the magnetic field [not written in Eq.~(\ref{eq:H0a})], described by the vector potential $A_\al(\rb)$,
$\text{rot}\, \Ab = (0,0,B_\perp)$, and the Zeeman term with $\e_Z =\mu_B B$, $B=\sqrt{B_\perp^2+B_\parallel^2}$, and
\[
    S_z = \hat{1}^{KK'}_\al \otimes 1^{\Abr\Bbr}_\be \otimes \tau_z^s.
\]
We assume arbitrary orientation of the total magnetic field ${\bf B}= (B_\parallel, 0, B_\perp)$ relative to the plane $z=0$ of graphene sample;
the $z$ direction in the spin space points along ${\bf B}$ and
is not necessarily perpendicular to the sample.

\subsection{Electron-electron interactions\label{sec:Hee}}

The most general form of the spin-symmetric e-e interactions in the low-energy Hamiltonian (\ref{eq:H})
can be written down~\cite{AKT} solely based on the symmetry considerations as
\beq
    \Hh_\text{e-e} = \Hh_\text{e-e,0} + \Hh_\text{e-e,1}.
\label{eq:Hee}
\eeq
Here,
\beq
    \Hh_\text{e-e,0} = \frac{1}{2} \int \dt^2 \rb \dt^2 \rb' \, [\psih^\dg(\rb) \psih(\rb)] V_0(\rb-\rb') [\psih^\dg(\rb') \psih(\rb')]
\label{eq:Hee0}
\eeq
describes the long-range Coulomb interactions, $V_0(\rb) = e^2/|\rb|$,
symmetric in valley-sublattice space $KK'\otimes \Abr\Bbr$ and
\beq
    \Hh_\text{e-e,1} = \frac{1}{2} \int \dt^2 \rb \, \sideset{}{'}\sum_{\al,\be} g_{\al\be} [\psih^\dg(\rb) \Tc_{\al\be} \psih(\rb)]^2
\label{eq:Hee1}
\eeq
describes the short-range e-e interactions that break the valley and/or sublattice symmetry.
The summation $\sideset{}{'}\sum_{\al,\be}$ includes all combinations $\al,\be=0,x,y,z$,
of the valley $\al$ and sublattice $\be$ channels, except for the symmetric one $\al=\be=0$, which is given by Eq.~(\ref{eq:Hee0}).
In Eqs.~(\ref{eq:Hee0}) and  (\ref{eq:Hee1}), and below, normal ordering of the operators is understood.

The symmetry of the honeycomb lattice yields the following relations between the couplings~\cite{AKT},
\[
    g_{\perp\perp} \equiv g_{xx} = g_{xy} = g_{yx} = g_{yy},
\]
\[
    g_{\perp z } \equiv g_{xz} = g_{yz}, \mbox{ }
    g_{z \perp } \equiv g_{zx} = g_{zy},
\]
\[
    g_{\perp 0 } \equiv g_{x0} = g_{y0}, \mbox{ }
    g_{0 \perp } \equiv  g_{0x} = g_{0y}.
\]
Thus, the asymmetry of the interactions in $KK'\otimes \Abr\Bbr$ space is described by eight independent coupling constants
$g_{\perp\perp}$, $g_{\perp z}$, $g_{z \perp}$, $g_{zz}$, $g_{\perp 0}$, $g_{z 0 }$, $g_{ 0 \perp }$, $g_{ 0 z}$.

Although, of course, the origin of both symmetric~[Eq.~(\ref{eq:Hee0})] and asymmetric~[Eq.~(\ref{eq:Hee1})] e-e interactions
are the actual Coulomb interactions, for brevity,
we will refer to them as the ``Coulomb'' and ``short-range/asymmetric  e-e'' interactions, respectively.

To lowest orders,
the expressions for the coupling constants $g_{\al\be}$ can be obtained by considering
the Coulomb interactions in the atomic-scale model,
\beq
    \hat{\Hc}_\text{e-e}= \frac{1}{2} \int \dt^3 \rv \dt^3 \rv'\,
        \Psih^\dg_\sig(\rv) \Psih^\dg_{\sig'}(\rv') \frac{e^2}{|\rv-\rv'|} \Psih_{\sig'}(\rv') \Psih_\sig(\rv).
\label{eq:Heea}
\eeq
Substituting the expansion (\ref{eq:Psiexp}) for $\Psi_\sig(\rv)$ into Eq.~(\ref{eq:Heea})
and using the slow variation of $\psih(\rb)$ at atomic scales, one obtains Eqs.~(\ref{eq:Hee}), (\ref{eq:Hee0}),  and (\ref{eq:Hee1})
with the first-order expressions~\cite{AKT}
\beq
    g_{\al\be}^{(1)} = \int_\text{3uc} \frac{ \dt^3 \rv }{3 |[\ab_1 \times \ab_2 ]|}
    \int \dt^3 \rv' \rho_{\al\be}(\rv) \frac{e^2}{|\rv-\rv'|} \rho_{\al\be}(\rv')
\label{eq:g1expr}
\eeq
for the couplings $g_{\al\be}$.
In Eq.~(\ref{eq:g1expr}),
\[
    \rho_{\al\be}(\rv) = \frac{1}{2} u^\dg(\rv) \Tc_{\al\be} u(\rv)
\]
are the {\em real} densities  in a given valley-sublattice channel $\al\be$ and
\[
     u(\rv) = \lt(\begin{array}{c} u_{KA}(\rv) \\ u_{KB}(\rv) \\ u_{K'B}(\rv) \\ -u_{K'A}(\rv) \end{array}\rt)_{KK'\otimes \Abr\Bbr}.
\]
From the orthogonality properties (\ref{eq:uorth}) of the Bloch wave-functions, it is clear that the integrands in Eq.~(\ref{eq:g1expr}),
as functions of $\rv-\rv'$, decay over several unit cells and the  asymmetric interactions are indeed short-ranged.
This emphasizes the fact that breaking of the valley-sublattice symmetry arises from atomic scales.

Using the relations $u_{K'A}^*(\rv)=u_{KA}(\rv)$ and $u_{K'B}^*(\rv)=u_{KB}(\rv)$, we see that
the densities $\rho_{0\al}(\rv) = \rho_{\al 0}(\rv) \equiv  0 $, $\al =x,y,z,$ vanish identically.
Therefore, $g_{0 \perp}^{(1)}=g_{0 z}^{(1)}=g_{\perp 0}^{(1)}=g_{z 0}^{(1)}=0$, and
the couplings $g_{0 \perp}$, $g_{0 z}$, $g_{\perp 0}$, $g_{z 0}$, although not prohibited by symmetry, vanish in the first order.
The nonvanishing   expressions $g_{\al \be}^{(2)}$ for $g_{\al\be}$, with $\al=0$ or $\be=0$,
arise in the second order in the Coulomb interactions
and involve virtual transitions to other bands; we do not present these expressions here.

For future discussion in Secs.~\ref{sec:renorm} and \ref{sec:ground}, we note the following properties.
The first-order microscopic expressions (\ref{eq:g1expr}) for the coupling constants have the form of the electrostatic
Coulomb energy for the density distributions $\rho_{\al\be}(\rv)$.
Since it is well known~\cite{LL8} that the electrostatic energy is positive-definite,
i.e., positive for any nonvanishing charge distribution,
we conclude that all nonvanishing first-order expressions  (\ref{eq:g1expr}) must be positive,
\beq
    g_{\perp\perp}^{(1)}>0, \mbox{ } g_{\perp z}^{(1)}>0,  \mbox{ } g_{z\perp}^{(1)}>0, \mbox{ } g_{zz}^{(1)}>0.
\label{eq:g1sign}
\eeq
On the other hand, the second-order expressions have to be negative,
\beq
    g_{0\perp}^{(2)}<0, \mbox{ } g_{0z}^{(2)}<0, \mbox{ } g_{\perp 0}^{(2)}<0, \mbox{ } g_{z 0}^{(2)} <0.
\label{eq:g2sign}
\eeq

One can expect the lowest-order expressions $g_{\al\be}^{(1)}$ [Eq.~(\ref{eq:g1expr})] and $g_{\al\be}^{(2)}$ to provide
accurate estimates for the couplings $g_{\al\be}$ in the limit of weak Coulomb interactions, $e^2/v \ll 1$.
For stronger interactions, $e^2/v \sim 1$, as in real graphene, this is not necessarily the case.
The reason is that the short-range interactions renormalize themselves at energies on the order of bandwidth $v/a_0$.
As an illustration of this fact,
the diagrams of the  low-energy Dirac theory involving just the short-range interactions contain ultraviolet divergencies:
schematically, each extra order produces a relative factor $ g_{\al\be} \int^{1/a_0} \dt q /v \sim g_{\al\be} /(v a_0) \sim e^2/v $.
These renormalizations change the magnitude and, possibly, the signs of the couplings $g_{\al\be}$ in certain channels.

Thus, it is more reasonable to treat the couplings $g_{\al\be}$ in Eq.~(\ref{eq:Hee1}) as the bare inputs of the low-energy theory,
without any assumptions about their signs and relative values, and consider all possibilities.
This is the approach we choose in the of paper.
An order-of-magnitude estimate for the bare couplings, valid for both weak ($e^2/v \ll 1$) and moderate ($e^2/v \sim 1$) Coulomb interactions, is
\beq
    g_{\al\be} \sim e^2 a_0.
\label{eq:gest}
\eeq

Anticipating the results of the next sections,
the asymmetric short-range e-e interactions (\ref{eq:Hee1}), although weaker than the symmetric Coulomb ones (\ref{eq:Hee0}),
appear to be play a crucial role in the  physics of the $\nu=0$ QHFM.
Possible relations between and the signs of the couplings $g_{\al\be}$ become especially important,
as they determine the properties of the isospin anisotropy and, as a result, the favored ground state order.
The implications of the potential sign restrictions on $g_{\al\be}$, suggested by
Eqs.~(\ref{eq:g1sign}) and (\ref{eq:g2sign}),
will be discussed in Sec.~\ref{sec:renorm} and \ref{sec:ground}.

\subsection{Electron-phonon interactions\label{sec:Heph}}

\begin{figure}
\centerline{
\includegraphics[width=.22\textwidth]{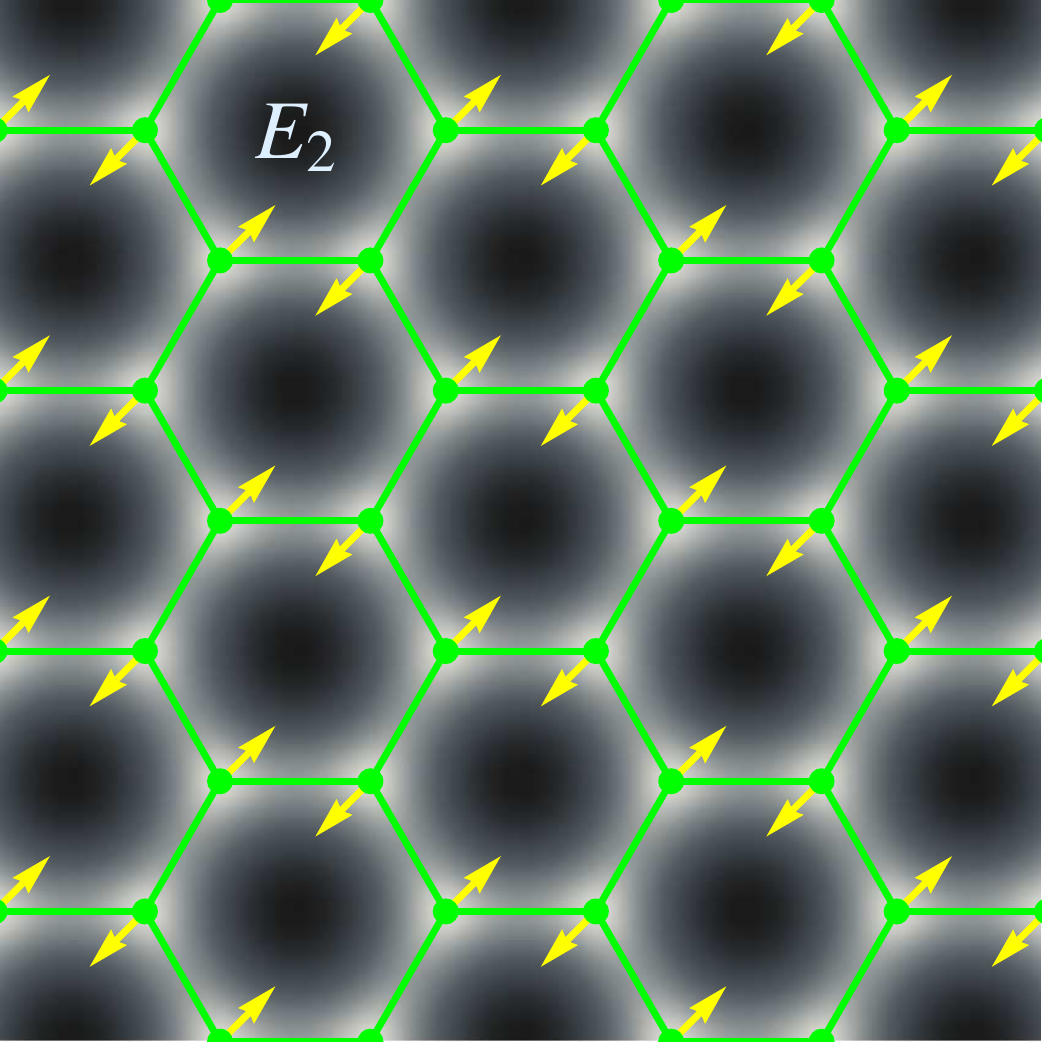}
\hspace{.3cm}
\includegraphics[width=.22\textwidth]{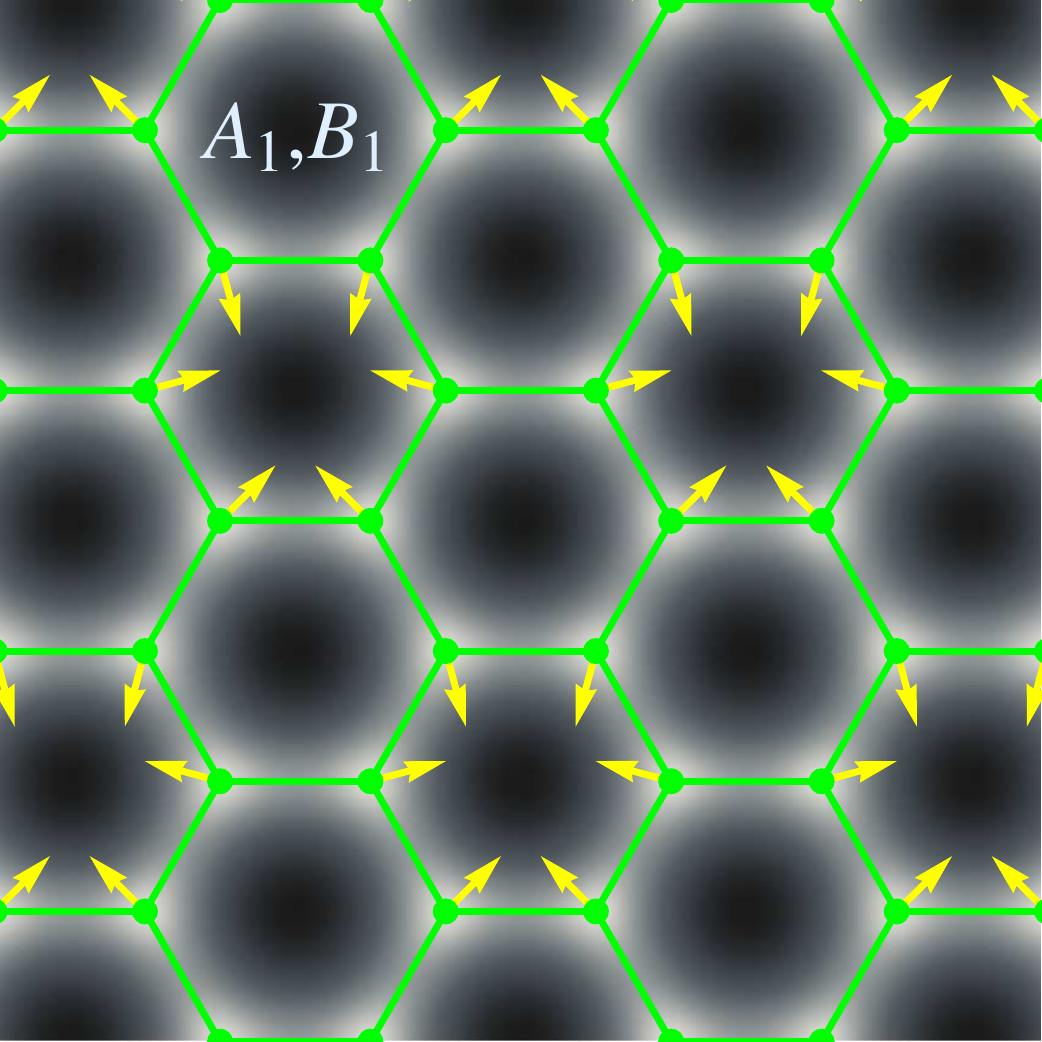}
}
\caption{In-plane optical phonon modes with the strongest e-ph coupling. (left)
Linear combination of the two degenerate $E_2$ modes $(\hat{u}_x(\rb), \hat{u}_y(\rb))$ with the phonon wavevector at $\Ga$ point. (right)
Linear combination of degenerate $A_1$, $B_1$ modes $(\hat{u}_a(\rb), \hat{u}_b(\rb))$ with the wavevector at $K,K'$ points.
}
\label{fig:phmodes}
\end{figure}

Besides the short-range e-e interactions (\ref{eq:Hee1}), another source of the isospin anisotropy in the $\nu=0$ QHFM
comes from e-ph interactions.
Electrons in graphene couple most efficiently to the following in-plane optical phonons:
two $E_2$ modes with the phonon wave-vector at the  $\Ga$ point and $A_1,B_1$ modes with wave-vector at $K,K'$ point  (following the classification of Ref.~\onlinecite{BA}), shown in Fig.~\ref{fig:phmodes}.
The corresponding e-ph interactions can be described by the Hamiltonian
\begin{eqnarray}
    \Hh_\text{e-ph} &=& \int \dt^2 \rb \,  \psih^\dg(\rb) \{ F_{E_2} [ \Tc_{zy} \hat{u}_x(\rb) - \Tc_{zx} \hat{u}_y(\rb)  ]  \nonumber \\
                     & + &     F_{A_1} [ \Tc_{xz} \hat{u}_a(\rb) + \Tc_{yz} \hat{u}_b(\rb)  ] \}\psih(\rb).
\label{eq:Heph}
\end{eqnarray}
The two degenerate $E_2$ modes have the frequency $\om_{E_2} \approx 0.196 \text{eV}$.
The $A_1$ and $B_1$ modes are also related by symmetry and have the same frequency $\om_{A_1} \approx 0.170 \text{eV} $
and same coupling constant $F_{A_1}=F_{B_1}$.
The lattice deformation due to $A_1$, $B_1$ modes is that of the Kekul\'{e} distortion.
The two-fold degeneracy of the modes allows for arbitrary in-plane displacement of a given atom in Fig.~\ref{fig:phmodes}, (left) and (right),
while the displacements of the remaining atoms in a tripled unit cell are related by the corresponding symmetry.

The phonon dynamics is described by the correlation functions of the displacement operators $\hat{u}_\mu(\rb)$,
\beq
    - \lan \text{T}_\tau \hat{u}_\mu(\tau,\rb) \hat{u}_\mu(0,0) \ran =
    \frac{s_0}{2 M \om_\mu} \de(\rb)
    \int_{-\infty}^{\infty} \frac{\dt \om }{2 \pi}
    \etxt^{- \itxt \om \tau}
     D_\mu(\om),
\label{eq:Dph}
\eeq
\[
    D_\mu(\om) =  - \frac{2 \om_\mu}{\om^2+\om_\mu^2},
\]
in the Matsubara representation.
Here, $M$ is the mass of the carbon atom
and $s_0 =\frac{\sqrt{3}}{4} a_0^2 $ is the area per carbon atom.
We will perform calculations at zero temperature only, in which case $\om$ is a continuous frequency.

The order-of-magnitude estimate
\beq
    F_\mu \sim e^2 /a_0^2
\label{eq:Fest}
\eeq
for the coupling constants follows from the dimensional analysis of Eq.~(\ref{eq:Heph})
and the electrostatic origin of e-ph interactions.

\subsection{Landau levels in graphene\label{sec:LLs}}
In this section, we briefly present the single-particle basis of the problem and emphasize the key properties
of the $n=0$ LL.

Solving the Dirac equation associated with Eq.~(\ref{eq:H0}) in the gauge $\Ab(\rb)=(0,B_\perp x,0)$,
in the $\Abr \Bbr$ sublattice space of each valley $K$ and $K'$, one obtains~\cite{ZhengAndo,GusyninSharapov,PGCN} the wave-functions
\beq
    \lan \rb | n p\ran = \frac{1}{\sqrt{2}} \lt( \begin{array}{c}  \phi_{|n|}(x-x_p) \\ \sgn n\, \phi_{|n|-1}(x-x_p) \end{array} \rt)_{\Abr \Bbr}
            \frac{\etxt^{\itxt p y}}{\sqrt{L_y}}
\label{eq:np}
\eeq
for all integer $n \neq 0$
and
\beq
     \lan \rb | 0 p\ran  = \lt( \begin{array}{c}  \phi_0(x-x_p) \\ 0\end{array} \rt)_{\Abr\Bbr}
            \frac{\etxt^{\itxt p y}}{\sqrt{L_y}},
\label{eq:0p}
\eeq
with the orbital energies $\e_n$ given by Eq.~(\ref{eq:en}).
Here, $\phi_{|n|}(x)$ are the harmonic oscillator wavefunctions,
$x_p = p l_B^2$ is the ``guiding center'',
and $L_y$ is the size of the sample in the $y$ direction, introduced to discretize the momentum quantum number $p$.

The complete set of the single-particle eigenstates in the $KK' \otimes \Abr\Bbr\otimes s$ space is
given by
\beq
    |np\mu\sig\ran = |\mu\ran_{KK'} \otimes |np\ran_{\Abr\Bbr} \otimes |\sig\ran_s,
\label{eq:psibasis}
\eeq
with  $\mu =K,K'$, $\sig=\ua,\da$, and
\[
    |K\ran = \lt(\begin{array}{c} 1\\  0\end{array}\rt)_{KK'},
    |K'\ran = \lt(\begin{array}{c} 0\\  1\end{array}\rt)_{KK'},
\]
\[
    |\ua\ran = \lt(\begin{array}{c} 1\\  0\end{array}\rt)_s,
    |\da\ran = \lt(\begin{array}{c} 0\\  1\end{array}\rt)_s.
\]

The field operator (\ref{eq:psidef}) can be expanded in the basis~(\ref{eq:psibasis})
as
\beq
    \psih(\rb)
    = \sum_{n=-\infty}^{\infty} \psih_n(\rb), \mbox{ } \psih_n (\rb)= \sum_{p\mu\sig} \lan \rb | np\mu\sig\ran \ch_{np\mu \sig},
\label{eq:psiexpr}
\eeq
where $\ch_{np\mu\sig}$ are the annihilation operators.

The $n=0$ LL with $\e_0=0$ is located exactly at the Dirac point and
possesses arguably the most peculiar properties:
in each valley, $K$ or $K'$, its wave-functions (\ref{eq:psibasis}) resides solely on one (actual) sublattice,  $A$ or $B$,
respectively [see Fig.~\ref{fig:n0wfs}, Eq.~(\ref{eq:0p}), and recall the accepted ordering (\ref{eq:psidef})].
Therefore, for each spin projection $\sig$, the part
\beq
    \psih_{0\sig}(\rb) =
    \sum_{p\mu} \lan\rb \sig |0 p\mu\sig\ran \ch_{0p\mu\sig}=
    \lt( \begin{array}{c} \psih_{0KA\sig} (\rb) \\ 0 \\ \psih_{0K'B\sig} (\rb)\\ 0 \end{array} \rt)_{KK' \otimes \Abr \Bbr}
\label{eq:psi0}
\eeq
of the field operator (\ref{eq:psiexpr}) pertaining to $n=0$ LL
has only two nonvanishing components, $\psih_{0KA\sig}(\rb)$  and $\psih_{0K'B\sig}(\rb)$,
whereas
\beq
    \psih_{0KB\sig}(\rb) = \psih_{0K'A\sig}(\rb) \equiv 0.
\eeq
Thus, for $n=0$ LL the valley ($KK'$) and sublattice ($AB$) degrees of freedom are essentially equivalent,
$K\leftrightarrow A$, $K' \leftrightarrow B$.
Further on, when discussing $n=0$ LL below, we refer to this 2D degree of freedom $(\psih_{0KA\sig}, \psih_{0K'B\sig})$
as just the ``$KK'$ valley isospin''.
Accordingly, we join  the nonvanishing components of the $n=0$ LL operator (\ref{eq:psi0})  in a 4D vector
\beq
    \psit_0(\rb) = \lt( \begin{array}{c}
        \psih_{0KA\ua}(\rb) \\ \psih_{0KA\da}(\rb) \\
        \psih_{0K'B\ua}(\rb) \\ \psih_{0K'B\da}(\rb) \end{array} \rt)_{KK'\otimes s}
\label{eq:psit0}
\eeq
in the isospin-spin space $KK'\otimes s$.

\subsection{Projected Hamiltonian for $n=0$ LL}

When addressing the many-body aspects of the $\nu=0$ state,
as a starting point, one may neglect the contributions from $n\neq 0$ LLs and restrict oneself to the dynamics within
$n=0$ LL, described in terms of the field (\ref{eq:psit0}).
From the form (\ref{eq:psi0}) of $\psih_0(\rb)$, we obtain
\beq
    \psih_0^\dg \Tc_{\al \be} \psih_0 = \lt\{ \begin{array}{ll} \psit_0^\dg \Tc_\al \psit_0, & \be =0,z, \\
                                            0, & \be=x,y, \end{array}\rt.
\label{eq:Talbe0}
\eeq
for $\al =0,x,y,z$, where
\beq
    \Tc_\al =\tau^{KK'}_\al \otimes 1^s, \mbox{ } \al=0,x,y,z.
\label{eq:Tal}
\eeq
are the $KK'$-isospin matrices.
I.e., electrons in the $n=0$ LL couple directly only to the source fields
with $\Tc_{\al 0}$ and $\Tc_{\al z}$ vertex structure in $KK' \otimes \Abr \Bbr$ space.
This is a consequence of the properties of the $n=0$ LL wave-functions.

Substituting $\psih(\rb)$ in the form (\ref{eq:psiexpr}) into Eqs.~(\ref{eq:Hee0}), (\ref{eq:Hee1}), and (\ref{eq:Heph})
and retaining only the $n=0$ LL component (\ref{eq:psi0}),
we obtain the bare projected Hamiltonian in terms of the field (\ref{eq:psit0}),
\beq
    \Hh^{(0)} = \Hh^{(0)}_0 + \Hh^{(0)}_\text{e-e,0} + \Hh^{(0)}_\text{e-e,1}+  \Hh^{(0)}_\text{e-ph},
\label{eq:Hn0}
\eeq
\beq
    \Hh^{(0)}_0 = -\e_Z \int \dt^2 \rb \, \psit^\dg_0(\rb) S_z \psit_0(\rb), \mbox{ }    S_z = \hat{1}^{AB} \otimes \tau_z^s ,
\label{eq:H00}
\eeq
\beq
    \Hh_\text{e-e,0}^{(0)} = \frac{1}{2} \int \dt^2 \rb \dt^2 \rb' \, [\psit^\dg_0(\rb) \psit_0(\rb)] V_0(\rb-\rb') [\psit^\dg_0(\rb') \psit_0(\rb')],
\label{eq:Hee00}
\eeq
\beq
    \Hh_\text{e-e,1}^{(0)} = \frac{1}{2} \int \dt^2 \rb \sum_{\al =x,y,z} g_{\al} [\psit^\dg_0(\rb) \Tc_\al  \psit_0(\rb)]^2,
\label{eq:Hee10}
\eeq
with $g_{\al} = g_{\al 0} + g_{\al z}$, and
\beq
    \Hh_\text{e-ph}^{(0)} = \int \dt^2 \rb \, F_{A_2} \psit^\dg_0(\rb) [\Tc_x u_a(\rb) + \Tc_y u_b(\rb)]  \psit_0(\rb).
\label{eq:Heph0}
\eeq

In the single-particle Hamiltonian $\Hh_0^{(0)}$, since the kinetic energy $\e_0=0$, only the Zeeman term is present.
In $\Hh_\text{e-e,1}^{(0)}$ and $\Hh_\text{e-ph}^{(0)}$,
due to the property (\ref{eq:Talbe0}), only the short-range e-e interactions
with $g_{\al 0}$ and $g_{\al z}$,  $\al=x,y,z$, couplings and e-ph interactions with
$F_{A_1}$ and $F_{B_1}$ couplings remain.
The coupling $g_{0z}$, although also does not vanish,
produces a symmetric term $\propto g_{0z} [\psit_0^\dg(\rb) \psit_0(\rb)]^2$, which may be neglected compared to the Coulomb part (\ref{eq:Hee00}).
We also mention that the trigonal warping effect $\propto \Tc_{xz},\Tc_{yz}$
does not couple to  $n=0$ states at the perturbative level.

\section{$\nu=0$ quantum Hall ferromagnet in graphene \label{sec:QHFM}}
\subsection{Basic concept and exact result\label{sec:basicQHFM}}

At integer filling factors $\nu$, interacting  multi-component quantum Hall systems are
described by the general theory of the quantum Hall ferromagnetism~\cite{Arovas_etal,QHB,Ezawa_etal,NM,YDM}.
Its central point is that, as long as electron dynamics may be effectively restricted to the partially filled LL
(sufficient conditions for this will be discussed in Sec.~\ref{sec:renorm})
and the interactions are symmetric in the ``spin'' space of discrete degrees of freedom,
the family of the many-body bulk ground states
can be found exactly as follows.
In order to minimize the energy of the Coulomb repulsion, one makes the orbital part of the wave-function totally antisymmetric,
thus putting electrons, for a given density,  as far apart from each other as possible.
Since one has on average an integer number of electrons per orbital, such wave-function can be realized
if electrons occupy the discrete states of all orbitals in exactly the same fashion.

Specifically for the $\nu=0$ state in graphene, which hosts two electrons
per four-fold degenerate orbital of the $n=0$ LL (Fig.~\ref{fig:LLs}), the many body wave-function can be written as
\beq
    \Psi= \lt[ \prod_p \lt(\sideset{}{'}\sum_{\la\sig,\la'\sig'}\Phi_{\la\sig,\la'\sig'}^* \, c^\dg_{0p\la\sig} c^\dg_{0p\la'\sig'}\rt) \rt] |0\ran.
\label{eq:PsiQHFM}
\eeq
Here, $|0\ran$ is the ``vacuum'' state with completely empty $n\geq 0$ LLs and completely filled $n<0$ LLs.
Each factor in the product $\prod_p$ creates a pair of electrons in the state $\Phi=\{\Phi_{\la\sig,\la'\sig'}\}$
($\la,\la'=A,B$ and $\sig,\sig'=\ua,\da$) at orbital $p$ of the $n=0$ LL,
see Eq.~(\ref{eq:psibasis}) and we identify $K\leftrightarrow A$ and $K' \leftrightarrow B$;
the antisymmetric two-particle spinor  $\Phi$ describes the occupation
of the 4D $KK' \otimes s$ isospin-spin space of each orbital by two electrons.
The sum $\sideset{}{'}\sum_{\la\sig,\la'\sig'}$ in Eq.~(\ref{eq:PsiQHFM}) goes over the upper-right off-diagonal elements of $\Phi$
and we normalize the spinor according to the number of particles per orbital,
\beq
    \sum_{\la\sig,\la'\sig'} |\Phi_{\la\sig,\la'\sig'}|^2=2,
\label{eq:Phinorm}
\eeq

For the purpose of illustrating the said exact property, in this subsection, we simply neglect the other $n\neq 0$ LLs.
To make this justified, one may temporarily assume here that the Coulomb interactions are weak, $e^2/v \ll 1$,
and in Sec.~\ref{sec:renorm} we demonstrate how the LL mixing effects can systematically be taken into account for stronger interactions $e^2/v \sim 1$ within the large-$N$ approximation.

Acting with the Coulomb interaction Hamiltonian (\ref{eq:Hee00}), symmetric in $KK'\otimes s$ space,
on the state~(\ref{eq:PsiQHFM}), we obtain that $\Psi$ is an eigenstate,
\[
    \Hh_\text{e-e,0}^{(0)} \Psi = E_0 \Psi,
\]
if and only if $\Phi$ satisfies the constraint
\beq
    \Phi_{K\ua,K'\da} \Phi_{K\da,K'\ua} + \Phi_{K\ua,K\da} \Phi_{K'\ua,K'\da} = \Phi_{K\ua,K'\ua} \Phi_{K\da,K'\da}.
\label{eq:Pluecker}
\eeq
The energy of the state equals
\beq
    E_0 = \frac{1}{2} \sum_{k,k'}  \lt[ 4 V^\dt_0 (k, k')- 2 V^\etxt_0 (k, k') \rt],
\eeq
where
\[
    V^\dt(k,k') =
    \frac{1}{L_y} \int \frac{\dt q_x}{2\pi} \, \etxt^{\lt[-\frac{q_x^2}{2}+ \itxt q_x (k-k') \rt] l_B^2} V_0(q_x,q_y=0),
\]
\[
   V^\etxt(k,k') =
         \frac{1}{L_y} \int \frac{\dt q_x}{2\pi} \, \etxt^{\lt[-\frac{q_x^2+ (k-k')^2}{2}\rt] l_B^2} V_0(q_x,k-k')
\]
are the ``direct'' (Hartree) and ``exchange'' (Fock) matrix elements, respectively,
and $V_0(\qb) = 2 \pi e^2/|\qb|$, $\qb=(q_x,q_y)$, is the Fourier transform of the Coulomb potential.
The energy  $E_0$ can easily be calculated explicitly.

Equation (\ref{eq:Pluecker}) is a necessary and sufficient condition
for the two-particle spinor $\Phi$ to be a Slater-determinant state,
\beq
    \Phi= \chi_a \circ \chi_b - \chi_b \circ \chi_a,
\label{eq:PhiSlater}
\eeq
described by two orthogonal single-particle spinors $\chi_{a,b}$ in $KK' \otimes s$ space;
the symbol $\circ$ denotes the direct product of $KK' \otimes s$ spaces of two electrons.
We see that not every antisymmetric spinor $\Phi$ delivers a many-body eigenstate
of the interaction Hamiltonian $\Hh_\text{e-e,0}^{(0)}$:
for example, the spin-singlet isospin-triplet state with zero isospin projection is not an eigenstate.
However, any state (\ref{eq:PsiQHFM}) with $\Phi$ in the form of a Slater determinant (\ref{eq:PhiSlater}) is an eigenstate with
the energy $E_0$ and the ground state~\cite{QHFMgroundstate} is, therefore, degenerate.

Let us count the number of degrees of freedom parameterizing the ground state $\Psi$.
An arbitrary antisymmetric spinor $\Phi$ has (six complex)=(twelve real) degrees of freedom. Fixing its norm and inconsequential overall phase factor leaves
ten real parameters, and the complex constraint (\ref{eq:Pluecker}) reduces this to the final {\em eight} real parameters.

Each Slater-determinant state $\Phi$ is uniquely specified
by the  2D subspace of the $KK' \otimes s$ space, occupied by two electrons and generated by the vectors $\chi_{a,b}$.
This establishes a one-to-one correspondence between the  states (\ref{eq:PhiSlater}) and
the elements of the Grassmannian manifold $\text{Gr}(2,4)$,
known as the Pl\"{u}cker embedding in mathematical literature~\cite{Grassmannian};
the constraint (\ref{eq:Pluecker}) is called the Pl\"{u}cker relation.
The parametrization of the occupied subspaces, generated by $\chi_{a,b}$,
and therefore of the Grassmannian $\text{Gr}(2,4)$,
is efficiently realized by the matrix
\beq
    P_{\la\sig,\la'\sig'}= \lan \Psi|\ch^\dg_{0p \la' \sig'} \ch_{0p \la \sig} |\Psi \ran.
\label{eq:Pdef}
\eeq
Using Eqs.~(\ref{eq:PsiQHFM}) and (\ref{eq:PhiSlater}), this gives
\beq
    P=    \chi_a \chi_a^\dg + \chi_b \chi_b^\dg
\label{eq:Pchi}
\eeq
in the matrix form in the $KK'\otimes s$ space.
The single-particle density matrix $P$
satisfies the properties of a hermitian projection operator
\beq
     P^\dg= P, \mbox{ } P^2 = P,
\label{eq:Pcond1}
\eeq
and also, for the doubly-filled $\nu=0$ state,
\beq
    \tr P = 2.
\label{eq:Pcond2}
\eeq

The matrix $P$ [Eq.~(\ref{eq:Pdef})] plays the role of the order parameter of the
broken-symmetry state (\ref{eq:PsiQHFM});
the observables and coupling of the state (\ref{eq:PsiQHFM}) to various perturbations can
be expressed through it.
The matrix $P$ is related to the matrix $Q$ ($R$ in Ref.~\onlinecite{Arovas_etal}), commonly used~\cite{Arovas_etal,YDM} in the QHFM theory,
as $P = \frac{1}{2}(\hat{1}+Q)$.

Speaking of symmetries, since $\Psi$ [Eq.~(\ref{eq:PsiQHFM})] is a ground state for any Slater-determinant state
(\ref{eq:PhiSlater}), any $\Stxt\Utxt(4)$ transformation in the single-particle $KK'\otimes s$ space keeps the energy $E_0$ invariant.
However, any $\Stxt \Utxt(2) \times \Stxt \Utxt (2) \times \Utxt(1) $
transformation, corresponding to independent rotations
within the subspace of the occupied states, generated by $\chi_a$ and $\chi_b$,
and its orthogonal complement -- the subspace of empty states, not only does not change the energy $E_0$,
but also leaves the state $\Phi$ intact. Therefore, the symmetry of the $\nu=0$ QHFM state $\Psi$ is described~\cite{Arovas_etal,YDM}
by the factor group
$\Stxt \Utxt(4)/[\Stxt \Utxt(2) \times \Stxt \Utxt (2) \times \Utxt(1)] = \Utxt(4) / [\Utxt(2) \times \Utxt(2)] $.
This group also determines the transformation properties of the order parameter $P$.
The dimensionality of the space of matrices $P$ as an $\Utxt(4) / [\Utxt(2) \times \Utxt(2)] $ manifold is $4^2- 2^2-2^2=8$,
which agrees with the number of the physical degrees of freedom of the Slater-determinant  states $\Phi$ [Eq.~(\ref{eq:PhiSlater})].

\subsection{Energy functional of the $\nu=0$ QHFM}

The exact result of the previous section lays down the basis of the QHFM theory.
In the presence of the isospin-asymmetric interactions  (\ref{eq:Hee10}) and (\ref{eq:Heph0}),
the state (\ref{eq:PsiQHFM}) will generally no longer be an exact ground state.
Besides, it is desirable to know not only the ground state of the system,
but also the excitations.
Provided the energy scales (per orbital) of these perturbations and excitations
are small compared to the energy $\min(\frac{e^2}{l_B}, \frac{v}{N l_B})$ of the screened Coulomb interactions (see Sec.~\ref{sec:renormrhos} below),
the local deviations of the actual many-body eigenstate from the QHFM state (\ref{eq:PsiQHFM}) are also minor.
This makes possible to develop a systematic low-energy quantum field theory
that describes the dynamics of the system.
Such theory has the form of a $\Utxt(4)/[\Utxt(2)\times\Utxt(2)]$ sigma-model for the order parameter $P(t,\rb)$,
which acquires time and coordinate dependence.

The rigorous derivation~\cite{Arovas_etal,QHB}  of the sigma-model
involves a procedure of projecting onto the QHFM state with a given order parameter $P(t,\rb)$.
Proceeding along the standard steps~\cite{Arovas_etal,QHB}, we arrive at the following Lagrangian of the $\nu=0$ QHFM
\beq
    L[P(t,\rb)] =  K[P(t,\rb)] - E[P(t,\rb)].
\label{eq:L}
\eeq
Here, $K[P(t,\rb)]$ is the kinetic term containing the time derivative of $P(t,\rb)$;
it is most simply expressed in terms of the single-particle spinors $\chi_{a,b}(t,\rb)$ [Eq.~(\ref{eq:Pchi})]
\[
    K[P(t,\rb)] =  \itxt  \int \frac{\dt^2 \rb}{2 \pi l_B^2} (\chi_a^\dg \pd_t \chi_a + \chi_b^\dg \pd_t \chi_b).
\]

The energetics  of the $\nu=0$ QHFM is described by the energy functional
\beq
    E[P(t,\rb)] = \int \frac{\dt^2 \rb}{2 \pi l_B^2} [ \Ec_\circ(P)+\Ec_\dm(P) +\Ec_Z(P)].
\label{eq:E}
\eeq
Here,
\beq
    \Ec_\circ(P) = \rho_s \tr [ \nabla P \nabla P ]
\label{eq:E0}
\eeq
is the gradient term characterized by the stiffness $\rho_s$ and
\beq
    \Ec_Z(P) = - \e_Z \,\tr [ S_z P ]
\label{eq:EZ}
\eeq
is the Zeeman term characterized by the energy $\e_Z =\mu_B B$.
Most importantly,
\beq
    \Ec_\dm(P) = \frac{1}{2}\sum_{\al=x,y,z} u_\al t_\al(P),
\label{eq:Ea}
\eeq
\beq
    t_\al (P) = \tr [\Tc_\al P ] \,\tr [\Tc_\al P ] - \tr [\Tc_\al P \Tc_\al P ],
\label{eq:t}
\eeq
is the  isospin anisotropy energy arising from the short-range e-e [Eq.~(\ref{eq:Hee})] and e-ph [Eq.~(\ref{eq:Heph})] interactions,
asymmetric in the valley-sublattice space.
The $KK'$-isospin matrices  $\Tc_\al$ were introduced in Eq.~(\ref{eq:Tal}).
As we will see below, due to symmetries of the e-e and e-ph coupling constants, the anisotropy
energies $u_\al$ for $\al=x,y$ isospin channels are equal,
\[
    u_\perp  \equiv u_x =u_y.
\]
Thus, the isospin anisotropy is fully characterized by two energies, $u_\perp$ and $u_z$.

The expressions (\ref{eq:Ea}) and (\ref{eq:t}) for the isospin anisotropy energy of the $\nu=0$ QHFM
constitute one of the key results of the present work. This is a generic form of the anisotropy,
arising from spin-symmetric two-particle electron interactions (including phonon-mediated interactions)
with arbitrary structure in the valley-sublattice space.

In Sec.~\ref{sec:ground}, we minimize the energy functional (\ref{eq:E})
to obtain a phase diagram of the $\nu=0$ QHFM in the presence of the isospin anisotropy and Zeeman effect
in the space of parameters $(u_\perp,u_z,\e_Z)$. In the rest of this and in the whole next section,
we discuss the expressions  for the stiffness $\rho_s$ and anisotropy energies $u_{\perp,z}$
in terms of the microscopic parameters of the Dirac Hamiltonian.
Their bare values $\rho_s^{(0)}$ and $u_{\perp,z}^{(0)}$, obtained in the lowest order in interactions,
can be determined from the projected Hamiltonian, Eqs.~(\ref{eq:Hn0})-(\ref{eq:Heph0}).

\begin{figure}
\includegraphics[width=.42\textwidth]{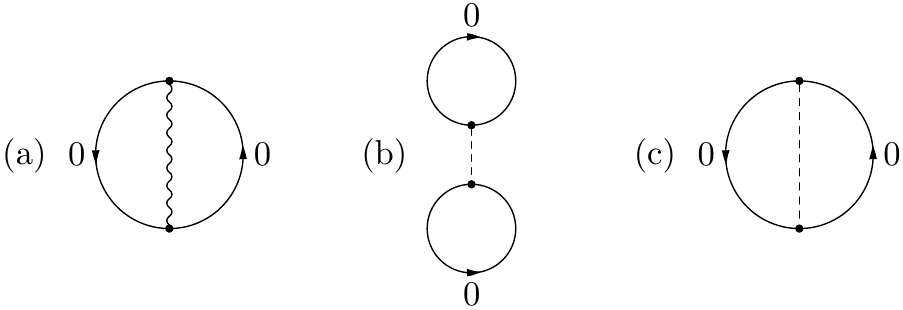}
\caption{
Diagrammatic
representation of the terms in the energy functional (\ref{eq:E}) of the $\nu=0$ QHFM, defining
the bare (lowest order in interactions) values of parameters.
(a) Diagram for the gradient term $\Ec_\circ(P)$ [Eq.~(\ref{eq:E0})], determining the bare stiffness $\rho_s^{(0)}$ [Eq.~(\ref{eq:rhos0})];
the wavy line stands for the Coulomb interaction [Eq.~(\ref{eq:Hee00})].
(b) and (c) Diagrams for the isospin anisotropy term $\Ec_\dm(P)$ [Eqs.~(\ref{eq:Ea}) and (\ref{eq:t})],
determining the bare anisotropy energies $u_{\perp,z}^{(0)}$ [Eqs.~(\ref{eq:u0}), (\ref{eq:uee0}), and (\ref{eq:ueph0})].
The dashed line represents either the short-range e-e [Eq.~(\ref{eq:Hee10})] or e-ph [Eq.~(\ref{eq:Heph0})] interactions.
The Hartree (b) and Fock (c) contributions produce the first and second terms in Eq.~(\ref{eq:t}), respectively.
}
\label{fig:bare}
\end{figure}

The gradient term $\Ec_\circ(P)$ arises from the Fock free-energy diagram in Fig.~\ref{fig:bare}(a), in which
one needs to take the spatial inhomogeneity
of order parameter $P(t,\rb)$ into account.
This yields the standard expression~\cite{QHB,Arovas_etal,YDM}
\beq
    \frac{\rho_s^{(0)}}{2\pi l_B^2} = \frac{1}{16 \sqrt{2 \pi}} \frac{e^2}{l_B}
\label{eq:rhos0}
\eeq
for the bare stiffness.

The isospin anisotropy term $\Ec_\dm(P)$ can be represented by the free-energy diagrams in Figs.~\ref{fig:bare}(b) and \ref{fig:bare}(c);
the first and second terms in Eq.~(\ref{eq:t}) arise from the Hartree (b) and Fock (c) diagrams, respectively.
The diagrams for the short-range e-e [Eq.~(\ref{eq:Hee10})] and e-ph [Eqs.~(\ref{eq:Dph}) and (\ref{eq:Heph0})]
interactions have the same form and the dashed line in the figures stands for either the short-range e-e or e-ph interactions.
This way, for the bare anisotropy energies in terms of the valley-sublattice asymmetric couplings, we obtain
\beq
    u_\al^{(0)} = u_\al^{(\text{e-e},0)} + u_\al^{(\text{e-ph},0)} , \mbox{ } \al = \perp,z,
\label{eq:u0}
\eeq
where
\beq
    u_\al^{(\text{e-e},0)} = \frac{1}{2\pi l_B^2} (g_{\al 0}+g_{\al z}), \mbox{ } \al = \perp,z,
\label{eq:uee0}
\eeq
and
\beq
    u_\perp^{(\text{e-ph},0)} = - \frac{f_{\perp z}}{2\pi l_B^2},
\mbox{ }
    u_z^{(\text{e-ph},0)} =0
\label{eq:ueph0}
\eeq
are the anisotropy energies due to short-range e-e and e-ph interactions, respectively.

In Eq.~(\ref{eq:ueph0}),
\beq
    f_{\perp z}=\frac{F_{A_1}^2 s_0}{M \om_{A_1}^2}
\label{eq:fperpzdef}
\eeq
is the coupling constant of the phonon-mediated interactions between the electrons.
Note that the combination $M \om_{A_1}^2$ is the curvature of the interaction potential between the carbon atoms,
which has electrostatic origin;
therefore, it does not depend on the carbon mass $M$ and scales as $M \om_{A_1}^2 \sim e^2/a_0^3$.
Together with Eq.~(\ref{eq:Fest}), this leads to the order-of-magnitude estimate
\beq
    f_{\perp z}\sim e^2 a_0.
\label{eq:fest}
\eeq
Comparing with Eq.~(\ref{eq:gest}), we see, that the bare anisotropies (\ref{eq:uee0}) and (\ref{eq:ueph0})
due to short-range e-e and e-ph interactions are actually parametrically the same and can differ only numerically.

The bare expressions (\ref{eq:rhos0})-(\ref{eq:ueph0}) determine the parameters of the $\nu=0$ QHFM,
provided one may neglect the effects of the interaction-induced electron transitions between different LLs,
also known as ``Landau level mixing''.
As we find in the next section, this is not the case and, in fact,
the parameters are drastically affected by the LL mixing effects.

\section{Renormalizations of parameters of the  $\nu=0$ QHFM \label{sec:renorm}
by the Landau level mixing effects}

The weakness of the Coulomb interactions is
generally a sufficient condition for the applicability of the QHFM theory.
For graphene, this is formulated as $e^2/v \ll 1$,
in which case, since the typical Coulomb energy $e^2/l_B$ per orbital is much smaller than the LL spacing $\e_1 = \sqrt{2} v/l_B$,
one could expect LL mixing effects to be inefficient and operate within the $n=0$ LL only.
However, interactions are not weak in graphene: for suspended samples, $e^2/v \approx 2.2 $ (taking $v=10^8 \text{cm/s}$), which may be regarded as moderate strength.
Besides, Coulomb interactions are known to be marginal~\cite{GGV} in graphene:
they produce large logarithmic contributions in the diagrammatic series
regardless of the their strength, even if weak.
These logarithms come from the wide range of energies  $|\e | \lesssim v/a_0$ up to the bandwidth and involve many LLs.
Therefore, taking the nonzero LLs into account is essential
and the question arises whether the QHFM theory still holds
for a realistic model of graphene, with $e^2/v \sim 1$.

In this respect, the large-$N$ expansion in the number of ``flavors'' has gained popularity~\cite{AKT,Son,FA,MFS} for graphene.
This approach allows  one to single out the leading in $N$ diagrams, in each order in the bare Coulomb interactions,
and perform partial summation of diagrams, formally analogous to the random-phase-approximation (RPA) series.
The physical justification of the method is that large $N$ makes screening of the interactions
especially efficient.
This reduces the coupling constant from its bare value $e^2/v\sim 1 $ to that of the screened interactions $1/N \ll 1$,
resulting in the effectively weak-coupling theory.

In this section, we will use the large-$N$ approach to systematically take into account the effects of LL mixing.
In reality,  $N=4$ in graphene due to two valleys and two projections of spin.
Although this value is not particularly large, one can still expect the large-$N$ approach to adequately
describe the correlated physics in graphene for moderate Coulomb coupling $e^2/v \sim 1$.

\begin{figure}
\includegraphics[width=.42\textwidth]{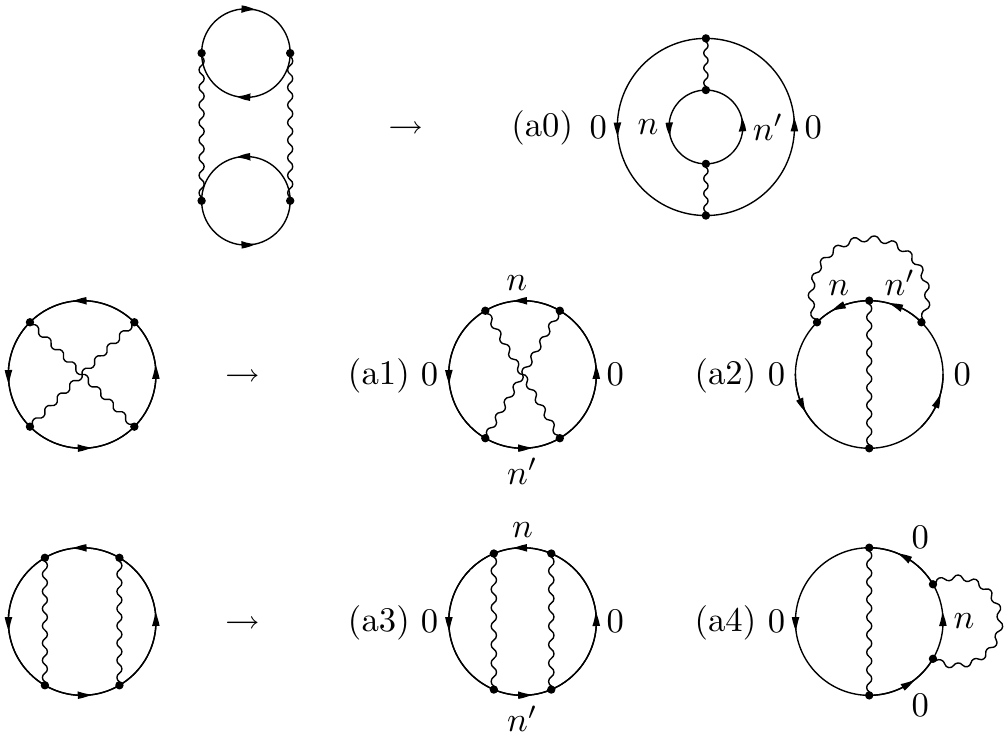}
\caption{
Diagrams for the energy functional  (\ref{eq:E}), second order in the Coulomb interactions (wavy lines).
Diagram (a0) represents the first-order large-$N$ correction to the Coulomb propagator.
Diagrams (a1)-(a4) diverge logarithmically, but cancel each other within the logarithmic accuracy.
}
\label{fig:2nda}
\end{figure}

\begin{figure}
\includegraphics[width=.42\textwidth]{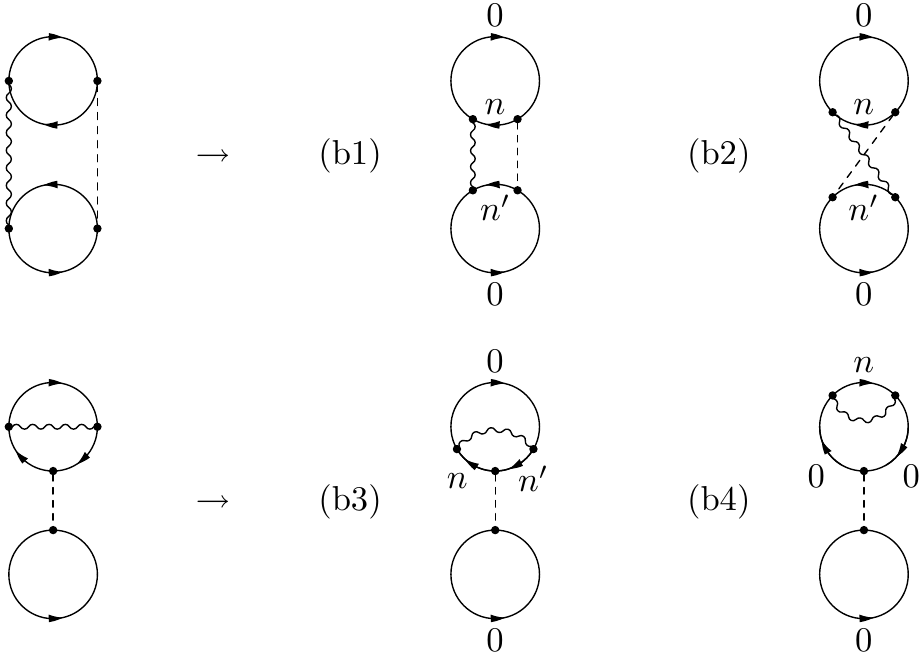}
\caption{Diagrams for the energy functional (\ref{eq:E}), first order in the Coulomb interactions (wavy line)
and in either the short-range e-e or e-ph interactions (dashed line).
Diagrams (b1)-(b4) diverge logarithmically and represent the lowest-order correction
to the Hartree contribution [Fig.~\ref{fig:bare}(b)] to the anisotropy energy.
}
\label{fig:2ndb}
\end{figure}

\begin{figure}
\includegraphics[width=.42\textwidth]{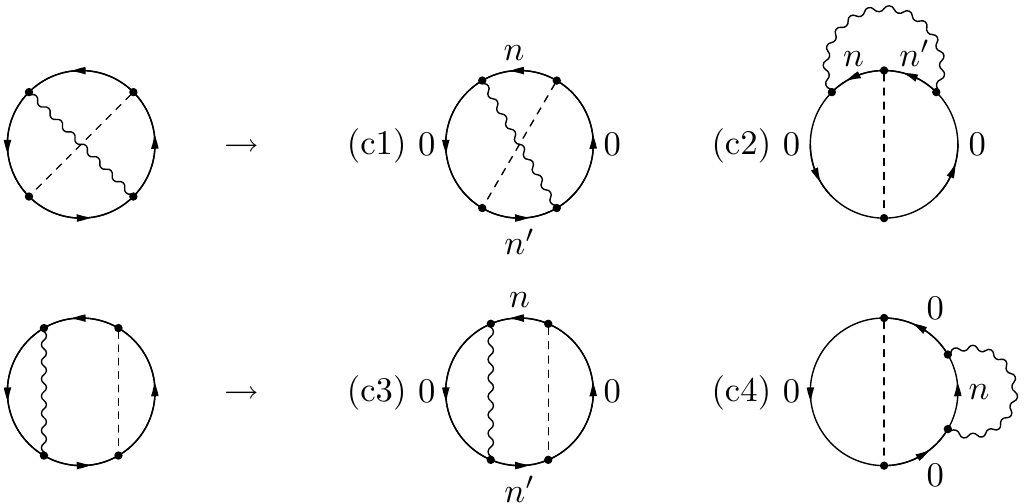}
\caption{
Same as in Fig.~\ref{fig:2ndb}, but with respect to the Fock contribution [Fig.~\ref{fig:bare}(c)].
}
\label{fig:2ndc}
\end{figure}

The effects produced by LL mixing can be identified already in the second order
in the interactions (\ref{eq:Hee0}), (\ref{eq:Hee1}), and (\ref{eq:Heph}).
Fig.~\ref{fig:2nda} shows the free-energy diagrams for the energy functional (\ref{eq:E}), second order in the Coulomb interactions (\ref{eq:Hee0}),
while Figs.~\ref{fig:2ndb} and \ref{fig:2ndc} shows the diagrams, first order in the Coulomb and in either short-range e-e (\ref{eq:Hee1}) or
e-ph (\ref{eq:Heph}) interactions.
As a starting point, one includes the contributions from all LLs in these diagrams,
using the full field operators (\ref{eq:psiexpr}).
One then separates the contributions from $n=0$ and $n\neq 0$ LLs in the electron Green's functions (solid lines),
at which point the following observations can be made.

The diagrams maintain the structure of the lowest-order diagrams in Fig.~\ref{fig:bare} with only $n=0$ LL present,
but contain the blocks involving $n\neq 0$ LLs  that represent
corrections to either
(i) the Coulomb propagator -- diagram in Fig.~\ref{fig:2nda}(a0);
(ii)  two-particle vertex functions -- diagrams in
Figs.~\ref{fig:2nda}[(a1), (a3)],
\ref{fig:2ndb}[(b1), (b2), (b3)]
and \ref{fig:2ndc}[(c1), (c2), (c3)];
(iii) or the self-energy of the  $n=0$ LL Green's function -- diagrams in Figs.~\ref{fig:2nda}(a4), \ref{fig:2ndb}(b4), and \ref{fig:2ndc}(c4).
Without the magnetic field, the blocks (ii) and (iii) diverge logarithmically~\cite{AKT}, since Coulomb interactions are marginal;
this remains true in the magnetic field.

This pattern persists in higher orders, which allows one to formulate the general recipe for taking the LL mixing effects into account
within the large-$N$ approach.
One first performs a partial summation of the RPA-type series of blocks (i).
This produces the ``dressed'' propagator of the screened Coulomb interactions.
One then performs the summation of the blocks (ii) and (iii).
The leading log-divergent diagrams can efficiently be summed up using the renormalization group (RG) procedure.
One notices that, when combined together, the blocks (ii) and (iii)
describe the critical renormalizations of the short-range e-e and e-ph interactions~\cite{AKT,BA} by the screened Coulomb interactions.
At the same time, the Coulomb interactions themselves are not
renormalized~\cite{AKT}: e.g., the diagrams (a1)-(a4) in Fig.~\ref{fig:2nda} cancel each other within log-accuracy.

With these steps performed, we find that, within the large $N$-approach,
the $\nu=0$ QHFM theory [Eq.~(\ref{eq:L}) - (\ref{eq:t})]
does hold as a controlled approximation for graphene with moderate strength Coulomb interactions $e^2/v\sim 1$,
but LL mixing effects result in crucial renormalizations of its parameters $\rho_s$  and $u_{\perp,z}$.
These renormalizations are considered in the remaining subsections.

We emphasize that the separation of the contributions from $n=0$ and $n\neq 0$ LLs in the diagrams,
with subsequent classification of their blocks according to the types (i), (ii), (iii), is justified only
in the weak-coupling large-$N$ limit of the screened Coulomb interactions,
when the {\em real} occupancy of the $n\neq 0$ LLs is close to full or zero, i.e.,
$\lan c^\dg_{n p \mu \sig} c_{n p \mu \sig} \ran \approx 1$ or $0$, for $n<0$ or $n>0$, respectively.
For example, for the general form of the second-order free-energy diagrams,
with all LLs involved (leftmost diagrams in Figs.~\ref{fig:2nda}, \ref{fig:2ndb}, and \ref{fig:2ndc}),
one cannot meaningfully attribute each diagram to just one of the (i), (ii), (iii) classes.
At the same time, {\em virtual} electron transitions between different LLs are quite efficient and
result in the strong screening of the Coulomb interactions.

\subsection{Stiffness for screened Coulomb interactions\label{sec:renormrhos}}

The first consequence of the LL mixing effects
is that the stiffness $\rho_s$ becomes suppressed due to screening, as compared to its bare value (\ref{eq:rhos0}).
The stiffness is obtained from the Fock diagram in Fig.~\ref{fig:bare} (a),
in which the bare Coulomb potential $V_0(q) = 2 \pi e^2 /q$
should be substituted by the ``dressed'' propagator
\beq
    V(\om, q) = \frac{V_0(q)}{1+V_0(q) \Pi(\om,q)}
\label{eq:V}
\eeq
of the screened interaction (the second term of this series is shown in Fig.~\ref{fig:2nda}(a0)).
The stiffness is given by the standard~\cite{Arovas_etal,QHB,YDM} expression
\beq
    \rho_s = \frac{l_B^4}{16 \pi } \int_0^{\infty} \dt q \, q^3 \etxt^{-\frac{q^2 l_B^2}{2}} V(\om=0,q)
\label{eq:rhosexpr}
\eeq
in terms of an arbitrary potential.
In Eq.~(\ref{eq:V}), $\Pi(\om,q)$ is the polarization operator of graphene
in the presence of the magnetic field.
Since in Eq.~(\ref{eq:rhosexpr}) the frequency is constrained  to $\om =0$ (the typical energy scales of the QHMF theory are $\ll v/l_B$)
and the relevant momenta are $q  \sim 1/l_B$,
one has to use the exact expression for the polarization operator~\cite{Pi,Goerbig}.
At filling factor $\nu=0$, zero temperature,
and neglecting the minor corrections from the Zeeman effect,
the polarization operator reads
\beq
    \Pi(0,q) =
    \frac{ N}{2 \pi l_B^2} \sum_{\substack{n>0\\n' \leq 0}} \frac{2}{ \e_n + |\e_{n'}|} |\Kbr_{nn'}(\qb)|^2.
\label{eq:Pi}
\eeq
Here, $\Kbr_{nn'}(\qb)$ [$\qb=(q_x,q_y)$, $q= |\qb|$] are the graphene magnetic form-factors;
they are expressed in terms of the conventional form-factors
\beq
    K_{n n'}(\qb) = \int_{-\infty}^{+\infty} \dt x\,
    \etxt^{\itxt q_x x} \phi_n \lt(x- \frac{q_y}{2} l_B^2\rt ) \phi_{n'}\lt (x + \frac{q_y}{2} l_B^2\rt)
\label{eq:K}
\eeq
for the quadratic spectrum as
\[
    \Kbr_{nn'}(\qb) = \frac{1}{2} [K_{|n|,|n'|}(\qb) + \sgn n \, \sgn n' K_{|n|-1,|n'|-1}(\qb)],
\]
if both $n \neq 0$ and  $n' \neq 0$, and
$
    \Kbr_{n 0} (\qb) = K_{|n|,0}(\qb).
$

The polarization operator $\Pi(0,q)$ depends only on the combination $q l_B$.
For $q l_B \gg 1$, $\Pi(0,q) \rtarr N q /(16 v)$ approaches its expression in the absence of the magnetic field.
At arbitrary $q l_B$, $\Pi(0,q)$ can be calculated only numerically.

Since the Coulomb potential $V_0(q)$ has no scale and $\Pi(0,q)$ depends solely on $q l_B$,
the stiffness (\ref{eq:rhosexpr}) scales as
\beq
    \frac{\rho_s}{2 \pi l_B^2} = \frac{v}{N \l_B}
    R  (e^2 N /v),
\label{eq:rhos}
\eeq
where the dimensionless function $R(e^2 N/v)$ of the coupling strength $e^2 N/ v$ is
defined by Eqs.~(\ref{eq:rhosexpr}), (\ref{eq:V}), and (\ref{eq:Pi}).
In the limit $e^2 N/v \ll 1$ of negligible screening, $V(0,q) \approx V_0(q)$,
one obtains
\[
    R(e^2 N/v \ll 1)
    \approx \frac{1}{8}\sqrt{\frac{\pi}{2}} \frac{e^2 N}{v},
\]
recovering the expression (\ref{eq:rhos0}) for the bare stiffness.
The function $R(e^2 N/v )$ grows with increasing $e^2 N/v$ and  saturates to the  maximum value $R(\infty) \sim 1$
in the limit $e^2 N /v \gg 1$ of complete screening, when $V(0,q) \approx 1/\Pi(0,q)$.

We see that, upon taking the screening effects of LL mixing into account, the stiffness (\ref{eq:rhos})
retains its square-root scaling $\rho_s(B_\perp)/(2 \pi l_B^2) \propto \sqrt{B_\perp}$ with the magnetic field,
but the numerical prefactor of the dependence becomes suppressed.
The main practical implication of this concerns the activation transport through the bulk of the sample,
since the gaps of the charge excitations are determined by the typical energy of symmetric interactions;
e.g., the energy of the unit charge Skyrmions equals $E_\text{Sk} =  2 \rho_s/l_B^2$.

\subsection{Renormalization of the anisotropy energies\label{sec:renormu}}

The second and a more physically significant effect arising from nonzero LLs
is the renormalization of the isospin anisotropy energies $u_{\perp,z}$ [Eq.~(\ref{eq:Ea})] by the Coulomb interactions.
The bare energies $u_{\perp,z}^{(0)}$ [Eqs.~(\ref{eq:u0}), (\ref{eq:uee0}), and (\ref{eq:ueph0})]
are  determined by the diagrams (b) and (c) in Fig.~\ref{fig:bare}, while
the diagrams in Figs.~\ref{fig:2ndb} and \ref{fig:2ndc} represent the lowest-order corrections to them.
Since the latter diverge logarithmically, one is forced to sum up the whole series of log-divergent diagrams.
Within the logarithmic accuracy, this can be achieved by means of the RG procedure.
It proves more efficient to consider the renormalizations of the bare couplings $g_{\al\be}$ and $f_{\perp z}$,
rather than $u_{\perp,z}^{(0)}$ directly,
and then express the anisotropy energies $u_{\perp,z}$ in terms of the renormalized couplings.

In the RG procedure for monolayer graphene~\cite{AKT,GGV,Son}, one starts with the Hamiltonian (\ref{eq:H})
of massless weakly interacting Dirac fermions.
Integrating out the high-energy fermionic modes in the frequency-momentum space yields the RG equations for the involved couplings.
One finds~\cite{AKT,GGV,Son} that the Coulomb interactions are marginally irrelevant
and, consequently, weakly interacting electrons in monolayer graphene flow to a fixed point of noninteracting massless Dirac fermions.

The RG analysis of the renormalizations of the short-range e-e and e-ph interactions by the Coulomb interactions
was carried out in Refs.~\onlinecite{AKT,BA}.
Here, we recover the essential results and concentrate on the properties of key relevance to the $\nu=0$ QHFM theory.
Our notation differs from that of Refs.~\onlinecite{AKT,BA}.

\subsubsection{Preparatory remarks \label{sec:preparatory}}

Several comments are in order before we proceed.

(a) In the perturbation theory diagrams, such as in Figs. \ref{fig:2ndb} and \ref{fig:2ndc},
the large logarithms arise from the divergent sums over LLs $\e_n$ (if one first integrates over the frequency $\om$),
which have to be cut by the bandwidth at high energies, $|\e_n| \lesssim v/a_0$.
Since such sums involve many LLs,
the discreteness of the spectrum due to the magnetic field may be neglected:
one may use the expressions for the Green's functions and polarization operators in the absence of the magnetic field,
substituting the sums over LLs by the integrals over momenta $q$.
At the lower limit, these integrals have to be cut by the inverse magnetic length $q\sim 1/l_B$,
once the influence of the magnetic field becomes important.
Hence, the arising logarithms are $\int_{1/l_B}^{1/a_0} \dt q/q \sim \ln (l_B/a_0)$.

Therefore, the magnetic field does not affect the very structure of the RG equations of Refs.~\onlinecite{AKT,BA},
yet defines a natural scale, at which the RG flow stops.
In the RG approach, the coupling constants $g_{\al\be}(l)$ and $f_{\perp z}(l)$ acquire a dependence on
the running length scale $l$.
The RG flow starts at the atomic scale $l=a$,
where the couplings are equal to their bare values [Eqs.~(\ref{eq:Hee1}), (\ref{eq:Heph}), and (\ref{eq:fperpzdef})],
\[
    g_{\al\be}(a) = g_{\al\be}, \mbox{ } f_{\perp z}(a) = f_{\perp z},
\]
and stops at the magnetic length $l=l_B$. The magnetic length $l_B$ defines the scale, at which
the renormalized anisotropy energies $u_{\perp,z}$ are to be determined.
We define the atomic scale as $a\sim a_0$, absorbing the ambiguity of the cutoffs in it.

(b) Besides the
renormalizations  arising from the interactions in the process of ``integrating out'' higher energy degrees of freedom
(``mode elimination'' part, in the terminology of Ref.~\onlinecite{Shankar}),
the full RG scheme also includes the renormalizations due to rescaling of frequencies, momenta, and quantum fields
(scaling, or ``tree-level'', renormalization).
While the former represent actual physical processes, described explicitly by the diagrams
(such as in Figs.~\ref{fig:2nda}, \ref{fig:2ndb}, and \ref{fig:2ndc}),
the latter are introduced in order to restore the original phase space
and thus make comparison of the theories with different bandwidths meaningful.
An important question arises, whether the latter, tree-level, renormalizations
also have to be taken into account when determining the renormalized anisotropy energies $u_{\perp,z}$.
Our understanding is the answer is negative, as they do not correspond to any physical processes.
The  following arguments can be given.

(b.1) Suppose the logarithmic contributions due to the Coulomb interactions were absent or negligible:
for the sake of argument, one may certainly consider a model with well-behaved finite-range symmetric interactions (e.g., Gaussian potential),
or the Coulomb interactions so weak or the number of flavors $N$ so large that $\min(e^2/v,1/N) \ln (l_B/ a_0) \ll 1$.
Then, all higher-order contributions to the anisotropy energy would be small compared to the first-order contribution,
which is given by the diagrams (b) and (c) in Fig.~\ref{fig:bare}
(the ultraviolet-divergent contributions, mentioned in Sec.~\ref{sec:Hee},
from the short-range e-e interactions  may also be assumed small, $g_{\al\be}/(v a_0) \ll 1$).
The renormalized anisotropy energies $u_{\perp,z}$
would then be determined just by the bare expressions~(\ref{eq:u0}), (\ref{eq:uee0}), and (\ref{eq:ueph0}),
\beq
    u_{\perp,z} \rtarr u_{\perp,z}^{(0)}.
\label{eq:uargument}
\eeq
At the same time, the full RG procedure would consist just of the tree-level (TL) renormalization
and the short-range e-e and e-ph couplings would renormalize as
$g_{\al\be}^\text{TL}(l) = g_{\al\be} \frac{a}{l}$
and
$f_{\perp z}^\text{TL}(l) = f_{\perp z} \frac{a}{l}$,
since their scaling dimension  is $-1$ (see paragraph (c) below regarding e-ph coupling).
Therefore, taking the scaling renormalization into account, i.e.,
using the expression (\ref{eq:uee0}) for the renormalized energies $u_{\perp,z}$ in terms of
the couplings $g_{\al\be}^\text{TL}(l_B)$ and
$f_{\perp z}^\text{TL}(l_B)$,
one would have to multiply the bare energies $u_{\perp,z}^{(0)}$ by the factor $a/l_B$,
\[
    u_{\perp,z} \xrightarrow{?} \frac{a}{l_B} u_{\perp,z}^{(0)}.
\]
This would, in apparent disagreement with Eq.~(\ref{eq:uargument}),
both drastically decrease the magnitude of the anisotropy energies and alter their dependence on the magnetic field.

(b.2) Alternatively,
such question and the extra factor $a/l_B$ in $u_{\perp,z}$ never arise,
if one starts with the ``poor man's'' formulation of the problem,
as to sum up the complicated parquet series of log-divergent diagrams (Figs.~\ref{fig:2ndb} and \ref{fig:2ndc} and all higher orders).
One can then invoke the RG procedure, or rather, just its ``mode elimination'' part, as the latter
is known to be an elegant way of accomplishing this task.

We, therefore, conclude that the renormalized anisotropy energies $u_{\perp,z}$
must be expressed through the couplings  $\gb_{\al\be}(l_B)$ and $\fb_{\perp z}(l_B)$,
in which only the renormalizations arising from the ``mode elimination'' part of the RG procedure are taken into account:
\beq
    u_\al = u_\al^{(\text{e-e})} + u_\al^{(\text{e-ph})} , \mbox{ } \al = \perp,z
\label{eq:u}
\eeq
\beq
    u_\al^{(\text{e-e})} = \frac{1}{2\pi l_B^2} [\gb_{\al 0}(l_B)+\gb_{\al z}(l_B)], \mbox{ } \al = \perp,z
\label{eq:uee}
\eeq
\beq
    u_\perp^{(\text{e-ph})} = - \frac{\fb_{\perp z}(l_B)}{2\pi l_B^2},
\mbox{ }
    u_z^{(\text{e-ph})} =0.
\label{eq:ueph}
\eeq
The couplings $\gb_{\al\be}(l)$ and  $\fb_{\perp z}(l)$ differ from the couplings $g_{\al\be}(l)$ and $f_{\perp z}(l)$ of the full RG scheme
by the tree-level renormalization,
\[
    g_{\al\be}(l) = \gb_{\al\be}(l) \frac{a}{l}, \mbox{ } f_{\perp z}(l) = \fb_{\perp z}(l) \frac{a}{l}.
\]

(c) Finally, we comment on the peculiarities of e-ph interactions in graphene.
Since the phonon dynamics is characterized by the phonon frequencies $\om_\mu$ [$\mu=E_2,A_1$, Eq.~(\ref{eq:Dph})],
the properties of e-ph interactions depend on the energy scale $\e$ at which considered.
At $\e \ll \om_\mu$, the phonon-mediated electron interactions
are effectively instantaneous and, being also short-ranged, are quite analogous
to the short-range e-e interactions; in particular, they are irrelevant in the RG sense with the scaling dimension $-1$.
At energies $\e \gg \om_\mu$,
retardation is strong and phonon-mediated interactions become marginal~\cite{BA}: they produce log-divergencies,
which have to be cut by  $\om_\mu$ at the lower limit, yielding the logarithms $\ln [v/ (a_0 \om_\mu)]$.
As a result, in principle, e-ph interactions renormalize both themselves and the  short-range e-e interactions.
In practice, however, for the typical values of parameters in graphene,
these renormalizations turn out to be numerically smaller that those due to the Coulomb interactions,
as can be inferred from the analysis of Ref.~\onlinecite{BA}.
Therefore, here, we neglect the renormalizations due to e-ph interactions.
On the one hand, since e-ph interactions couple different valley channels,
including these renormalizations would significantly complicate the RG equations.
On the other hand, this should not alter the main conclusions
of the RG analysis, which imply that essentially any algebraic possibility (i.e., including signs)
for the isospin anisotropy energies $(u_\perp,u_z)$ could be realized in graphene.

\subsubsection{RG analysis\label{sec:RGanalysis}}

We can now proceed with the RG analysis~\cite{AKT,BA}.
The one-loop renormalizations of the short-range e-e and e-ph couplings
by the screened Coulomb interactions are actually described by the diagrams in Figs.~\ref{fig:2ndb} and \ref{fig:2ndc},
upon ``dressing'' the bare Coulomb lines.
The corresponding RG equations for the coupling constants $\gb_{\al\be}(l)$ read~\cite{AKT}
\beq
    \frac{\dt \gb_{\al 0}(l)}{\dt \xi} = 0 \mbox{ for } \al=x,y,z,
\label{eq:gal0eq}
\eeq
and
\beq
    \frac{\dt \gb_{\al \perp}(l)}{\dt \xi} = 2 F(w(l))[\gb_{\al \perp}(l) - \gb_{\al z}(l)],
\label{eq:galperpeq}
\eeq
\beq
    \frac{\dt \gb_{\al z}(l)}{\dt \xi} = 4 F(w(l))[\gb_{\al z}(l) - \gb_{\al \perp}(l)],
\label{eq:galzeq}
\eeq
for $\al =0,x,y,z$. Here, $\xi = \ln l/a$.

One observes that the couplings
\beq
    \gb_{\al 0}(l) = g_{\al 0}, \mbox{ } \al=x,y,z,
\label{eq:gal0}
\eeq
are not renormalized, whereas  $\gb_{\al x}(l) = \gb_{\al y}(l) \equiv \gb_{\al \perp}(l)$ and $\gb_{\al z}(l)$, $\al=0,x,y,z$, are.
Different valley channels $\al$ are not coupled,
but the $\perp$ and $z$ sublattice channels do couple to each other within each valley channel.
The reason for this are the properties of the electron Green's function:
while a unit matrix in the $KK'$ valley space, it has nontrivial matrix structure in the $\Abr \Bbr$ sublattice space.
The symmetric Coulomb interactions themselves couple neither valley nor sublattice channels.

\begin{figure}
\includegraphics[width=.35\textwidth]{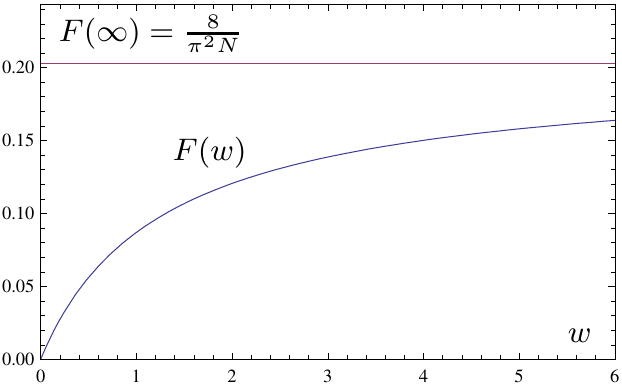}
\caption{
The function $F(w)$ [Eq.~(\ref{eq:F})] of the Coulomb coupling constant $w$ [Eq.~(\ref{eq:gC})]
describes the renormalizations of the Dirac velocity $v(l)$, $w(l)$ itself, and the short-range e-e and e-ph couplings.
}
\label{fig:Fw}
\end{figure}

In Eqs.~(\ref{eq:galperpeq}) and (\ref{eq:galzeq}).
\beq
    F(w) = \frac{8}{\pi^2 N} \lt(1 - \frac{\pi}{2 w} + \frac{\arccos w}{w \sqrt{1-w^2}} \rt)
\label{eq:F}
\eeq
is a function of the dimensionless coupling constant
\beq
    w(l) = \frac{\pi N}{8} \frac{e^2}{v(l)}
\label{eq:gC}
\eeq
of the Coulomb interactions, plotted in Fig.~\ref{fig:Fw}.
The RG equation~\cite{GGV,AKT,Son} for $w(l)$
is also determined by Eq.~(\ref{eq:F}),
\beq
    \frac{\dt w(l)}{\dt \xi}  = - F(w(l)) w(l).
\label{eq:geq}
\eeq
The renormalization of $w(l)$ comes from the renormalization of the Dirac velocity $v(l)$, while the charge $e^2$ is not renormalized
(the diagrams (a1)-(a4) in Fig.~\ref{fig:2nda} cancel each other within the log-accuracy).

\begin{figure}
\includegraphics[width=.35\textwidth]{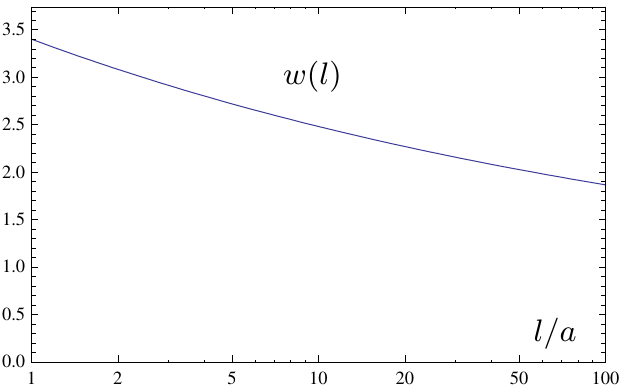}
\caption{
Renormalization of the Coulomb coupling constant $w(l)$ [Eq.~(\ref{eq:gC})]. The bare value $w(a) =3.4$ was used.
}
\label{fig:g}
\end{figure}

Equation (\ref{eq:geq}) can be  solved analytically only the limits of
weak~($w(l)\ll 1$, interactions are not screened, $F(w) \approx \frac{2}{\pi N} w$ ) or strong~($w(l)\gg 1$, interactions are fully screened,
$F(w) \approx F(\infty) = \frac{8}{\pi^2 N}$) coupling.
In Fig.~\ref{fig:g}, we plot the numerical solution $w(l)$ to Eq.~(\ref{eq:geq})
for the bare value $w(a) = w \approx 3.4$ obtained at  $v(a) = v  \approx 10^8\text{cm}/\text{s}$,
which should be typical for suspended graphene.
As a general property, $w(l)$ decreases [the velocity $v(l)$ grows] with increasing the length scale $l$;
therefore, Coulomb interactions flow towards weak coupling.
At  $l_B /a =40$ (taking $l_B \approx 10 \text{nm}$ at
$B_\perp \approx 10 \Ttxt$ and $a=a_0 \approx 2.5 \AA$),
we obtain $w(l_B) \approx  2.1$, i.e., for experimentally relevant values,
Coulomb interactions remain in the intermediate, moderate strength, regime.

Since the renormalization of the Coulomb coupling $w(l)$ [Eq.~(\ref{eq:geq})]
does not involve the short-range e-e or e-ph interactions (we neglect the later, see paragraph (c) in Sec.~\ref{sec:preparatory} and Ref.~\onlinecite{BA}), one may regard $F(w(l))$ in Eqs.~(\ref{eq:galperpeq}) and (\ref{eq:galzeq}) as a known function of $l$.
The system of the coupled equations (\ref{eq:galperpeq}) and (\ref{eq:galzeq})
can then easily be solved by making a linear transformation to the couplings~\cite{AKT}
\[
    \gb_{\al +}(l) = \frac{1}{3}[\gb_{\al z}(l) + 2 \gb_{\al \perp}(l)] , \mbox{ } \gb_{\al -}(l) = \frac{1}{3}[\gb_{\al z}(l) - \gb_{\al \perp}(l)],
\]
which translates the equations to
\[
    \frac{\dt \gb_{\al +}(l) }{\dt \xi} = 0, \mbox{ }
    \frac{\dt \gb_{\al -}(l) }{\dt \xi}  = 6 F(w(l)) \gb_{\al -}(l).
\]
The solution of these equations is straightforward and, coming back to $\gb_{\al\perp}(l)$ and $\gb_{\al z}(l)$, one obtains
\beq
    \gb_{\al \perp}(l) = \frac{1}{3} (g_{\al z} + 2  g_{\al \perp })
    - \frac{1}{3} (g_{\al z} - g_{\al \perp }) \Fc_\text{e-e}(l/a,w),
\label{eq:galperp}
\eeq
\beq
    \gb_{\al z}(l) = \frac{1}{3} (g_{\al z} + 2  g_{\al \perp }) + \frac{2}{3} (g_{\al z} - g_{\al \perp }) \Fc_\text{e-e}(l/a,w),
\label{eq:galz}
\eeq
\beq
    \Fc_\text{e-e}(l/a,w)=  \exp\lt[ \int_a^l \frac{\dt l'}{l'} \, 6 F(w(l'))\rt],
\label{eq:Fcee}
\eeq
for $\al=0,x,y,z$.
The function $\Fc_\text{e-e}(l/a,w)$ determines the strength of renormalization. It depends on the bare Coulomb coupling $w$
as parameter and reaches its maximum
\[
    \Fc_\text{e-e}^\text{max}(l/a) = \Fc_\text{e-e}(l/a,\infty) = \lt( \frac{l}{a} \rt)^{\frac{48}{\pi^2 N}}
\]
in the strong coupling regime $w\rtarr \infty$, when the interactions are fully screened.
This provides an upper bound for the strength of renormalization.

\begin{figure}
\includegraphics[width=.30\textwidth]{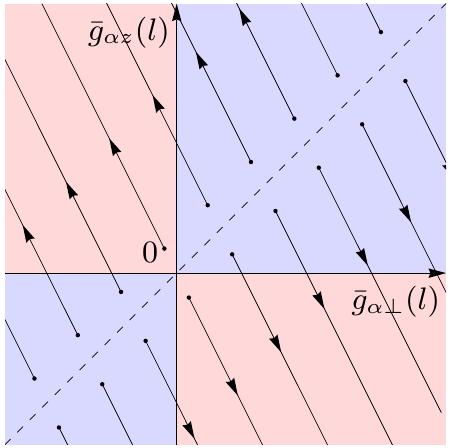}
\caption{RG flows [Eqs.~(\ref{eq:galperp}), (\ref{eq:galz}), and (\ref{eq:Fcee})] of the coupling constants $\gb_{\al \perp}(l)$ and $\gb_{\al z}(l)$ of the short-range electron-electron interactions (\ref{eq:Hee1}).
}
\label{fig:RGflows}
\end{figure}

The RG flows in the space of couplings $(\gb_{\al \perp}(l),\gb_{\al z}(l))$, defined by Eqs.~(\ref{eq:galperp}) and (\ref{eq:galz}), are plotted in Fig.~\ref{fig:RGflows}.
Their key properties are as follows.
Starting from the point $(\gb_{\al \perp}(a),\gb_{\al z}(a)) = (g_{\al \perp},g_{\al z})$
at the lattice scale $l=a$,
the RG flow $(\gb_{\al \perp}(l),\gb_{\al z}(l))$ follows the straight line
\beq
    2\gb_{\al \perp}(l) + \gb_{\al z}(l) = 2 g_{\al \perp} + g_{\al z}.
\label{eq:gconserve}
\eeq
This constraint implies that the sum of couplings
$\sum_{\be=x,y,z} \gb_{\al\be}(l)= 2\gb_{\al \perp}(l) + \gb_{\al z}(l)$ in different sublattice channels $\be$
is conserved under renormalization.

Further, depending on the relation between the bare values $g_{\al \perp}$ and $g_{\al z}$,
eventually, for strong enough renormalization, the flow line ends up either in the quadrant
\beq
    (\gb_{\al \perp}(l)>0,\mbox{ } \gb_{\al z}(l)<0), \mbox{ if } g_{\al \perp}> g_{\al z},
\label{eq:gprop1}
\eeq
or in the quadrant
\beq
    (\gb_{\al \perp}(l)<0,\mbox{ } \gb_{\al z}(l)>0), \mbox{ if }  g_{\al \perp} < g_{\al z},
\label{eq:gprop2}
\eeq
light red regions in Fig.~\ref{fig:RGflows}.
That is, the initially algebraically larger (smaller) coupling always becomes or stays positive (negative).
Deeper reasons for this interesting behavior could be sought in the chiral nature of the Dirac quasiparticles,
which is the ultimate cause of coupling between $\perp$ and $z$ sublattice channels.

What concerns the renormalizations of e-ph interactions by the Coulomb interactions,
the RG equations for the couplings $\fb_{z\perp}(l) = \frac{F_{E_2}^2 s_0}{M \om_{E_2}^2}$ ($E_2$ phonons)
and $\fb_{\perp z}(l)$ ($A_1,B_1$ phonons) read~\cite{BA}
\[
    \frac{\dt \fb_{z\perp}(l)}{\dt \xi} = 2 F(w(l)) \fb_{z \perp}(l),
\]
\beq
    \frac{\dt \fb_{\perp z}(l)}{\dt \xi} = 4 F(w(l)) \fb_{\perp z}(l).
\label{eq:fperpzeq}
\eeq
For e-ph interactions, different sublattice channels do not couple in the renormalization process.
The solution to Eq.~(\ref{eq:fperpzeq}) is
\beq
    \fb_{\perp z} (l)  = f_{\perp z} \Fc_\text{e-ph}(l/a,w),
\label{eq:fperpz}
\eeq
\beq
    \Fc_\text{e-ph}(l/a,w) = \exp\lt[\int_a^l \frac{\dt l'}{l'} 4 F(w(l')) \rt].
\label{eq:Fceph}
\eeq
Comparing Eqs.~(\ref{eq:Fcee}) and (\ref{eq:Fceph}),
we see that the renormalization of the e-ph coupling is parametrically weaker
than that of the short-range e-e couplings:  $\Fc_\text{e-ph}(l/a,w) =  [\Fc_\text{e-e}(l/a,w)]^{2/3}$
due to the factor 4 instead of 6 in the exponential.
The maximum, as a function of $w$, equals
\[
    \Fc_\text{e-ph}^\text{max}(l/a) = \Fc_\text{e-ph}(l/a,\infty) =  \lt(\frac{l}{a}\rt)^\frac{32}{\pi^2 N}.
\]

Concluding this section, the renormalizations of the short-range e-e and e-ph couplings
are described by Eqs.~(\ref{eq:gal0}), (\ref{eq:galperp}), (\ref{eq:galz}), (\ref{eq:Fcee}), (\ref{eq:fperpz}), and (\ref{eq:Fceph}).
Since  $\gb_{\al 0}(l_B)=g_{\al 0}$, $\al=\perp,z$, are not renormalized and,
according to Eq.~(\ref{eq:uee}),
$\gb_{0 \be}(l_B)$, $\be=\perp,z$, do not affect the anisotropies,
the main contribution to the anisotropy energies $u_\al^{(\text{e-e})}$
due to short-range e-e interactions comes from the couplings $\gb_{\al z}(l_B)$.
In the regime $\Fc_\text{e-e}(l_B/a, w) \gg 1$ of strong renormalization,
one may neglect the first term in Eq.~(\ref{eq:galz})
(except for the special case, when the bare values are close to the
unstable line  $g_{\al \perp} =g_{\al z}$) to obtain
\beq
    u^{(\text{e-e})}_\al \approx \frac{1}{2\pi l_B^2} \frac{2}{3} (g_{\al z} - g_{\al \perp}) \Fc_\text{e-e}(l_B/a,w), \mbox{ } \al =\perp,z.
\label{eq:ueestrong}
\eeq
From Eqs.~(\ref{eq:ueph}), (\ref{eq:fperpz}), and (\ref{eq:Fceph}), the anisotropy energies due to e-ph interactions
equal
\beq
    u^{(\text{e-ph})}_\perp = -\frac{f_{\perp z}}{2\pi l_B^2} \Fc_\text{e-ph}(l_B/a,w), \mbox{ }  u^{(\text{e-ph})}_z=0.
\label{eq:uephstrong}
\eeq
Finally, we also mention that the Zeeman energy $\e_Z$ [Eq.~(\ref{eq:EZ})] is not renormalized~\cite{AKT} by the Coulomb interactions,
since the Zeeman splitting term in Eq.~(\ref{eq:H0}) is scalar in $KK' \otimes \Abr\Bbr$ space.

\subsection{Consequences of renormalizations of the anisotropy energies $u_{\perp,z}$ for the $\nu=0$ QHFM\label{sec:implications}}

The properties of the renormalized isospin anisotropy energies, obtained in Sec.~\ref{sec:renormu},
have very important consequences for the physics of the $\nu=0$ QHFM.
The conclusions below constitute some of the key results of the present  work.

\subsubsection{Magnitude of the anisotropy energies $u_{\perp,z}$}

The first consequence concerns the absolute values of the anisotropy energies $u_{\perp,z}$.
As follows from Eqs.~(\ref{eq:gest}), (\ref{eq:uee0}), (\ref{eq:ueph0}), and (\ref{eq:fest}),
the bare anisotropy energies
can be roughly estimated as
\beq
    | u_{\perp,z}^{(\text{e-e},0)}(B_\perp)|, \mbox{ } |u_\perp^{(\text{e-ph},0)}(B_\perp)|
        \sim \frac{e^2}{a_0} \lt(\frac{a_0}{l_B}\rt)^2 \sim B_\perp[\Ttxt] \text{K}.
\label{eq:u0est}
\eeq
They scale linearly with the perpendicular magnetic field $B_\perp$
and are on the same order as the Zeeman energy $\e_Z = \mu_B B \approx 0.7 B[\Ttxt] \text{K}$ (for moderate tilt angles),
as obtained earlier in Refs.~\onlinecite{AF,Goerbig,Nomura,Chamon}.

According to Eqs.~(\ref{eq:ueestrong}) and (\ref{eq:uephstrong}),
\beq
    |u_{\perp,z}^{(\text{e-e})}(B_\perp) | \sim  |u_{\perp,z}^{(\text{e-e},0)}(B_\perp) | \Fc_\text{e-e}(l_B/a,w)
\label{eq:ueeest}
\eeq
and
\beq
    u_\perp^{(\text{e-ph})}(B_\perp)  = u_\perp^{(\text{e-ph},0)}(B_\perp)  \Fc_\text{e-ph}(l_B/a,w),
\label{eq:uephest}
\eeq
i.e.,  the renormalized anisotropy energies are enhanced by the factors $\Fc_\text{e-e}(l_B/a,w)$ and $\Fc_\text{e-ph}(l_B/a,w)$,
compared to the bare values and their functional dependence on  $B_\perp$ is altered.

\begin{figure}
\includegraphics[width=.35\textwidth]{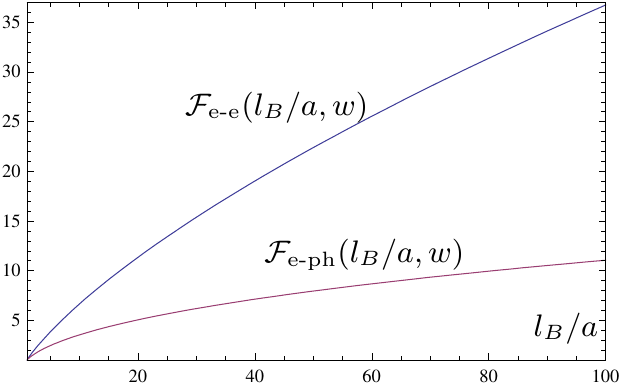}
\caption{
The functions $\Fc_\text{e-e}(l_B/a,w)$ and $\Fc_\text{e-ph}(l_B/a,w)$ [Eqs.~(\ref{eq:Fcee}) and (\ref{eq:Fceph})],
characterizing the strength of renormalizations of the isospin anisotropy energies
$u_\al^{\text{(e-e)}}$ and $u_\al^{\text{(e-ph)}}$ [Eqs.~(\ref{eq:ueestrong}) and (\ref{eq:uephstrong})]
due to short-range e-e and e-ph interactions.
The value $w=3.4$ for the bare Coulomb coupling [Eq.~(\ref{eq:gC})] was used.
In a typical experimental range $l_B/a\sim 10-100$, the anisotropy energies
are enhanced by about one order compared to their bare values $u_\al^{(\text{e-e},0)}$ and $u_\al^{(\text{e-ph},0)}$
[Eqs.~(\ref{eq:u0})-(\ref{eq:fperpzdef}), and (\ref{eq:u0est})].}
\label{fig:FeeFeph}
\end{figure}

In Fig.~\ref{fig:FeeFeph}, we plot $\Fc_\text{e-e}(l_B/a,w)$ and $\Fc_\text{e-ph}(l_B/a,w)$ as functions of $l_B/a$
for the bare Coulomb strength $w=3.4$. Some values are $(\Fc_\text{e-e}(l_B/a,w), \Fc_\text{e-ph}(l_B/a,w)) \approx
(6.6,3.5)$, $(19,7)$, $(37,11)$ at $l_B/a = 10,40,100$, respectively.
We see that, in a typical experimental situation ($l_B \sim 10\text{nm}$ and $a \approx 2.5 \AA$ gives $l_B/a \approx 40$),
the renormalization of the anisotropy energy is expected to be very strong.
Since the Zeeman energy $\e_Z = \mu_B B $ is not renormalized,
$u_{\perp,z}$ can easily exceed $\e_Z$ by one order.
This implies that the isospin anisotropies play a major, perhaps,
more significant role than the Zeeman effect in the physics of the $\nu=0$ QHFM.

\subsubsection{Signs of the anisotropy energies $u_{\perp,z}$}
The second consequence concerns the possible signs of the anisotropy energies $u_{\perp,z}$.
According to Eq.~(\ref{eq:ueestrong}), in the regime of strong renormalization $\Fc_\text{e-e}(l_B/a,w) \gg 1$,
the sign of the anisotropy energy $u^{(\text{e-e})}_\al$, $\al=\perp,z$,
is determined by the relation $g_{\al z} \gtrless g_{\al \perp}$ between the bare couplings.
This follows from the discussed peculiar behavior of the RG flows, see Fig.~\ref{fig:RGflows} and Eqs.~(\ref{eq:gprop1}) and (\ref{eq:gprop2}).
This property suggests that,

{\em essentially, any sign combination}
\beq
    u_\perp^{(\text{e-e})} \gtrless 0, \mbox{ }u_z^{(\text{e-e})} \gtrless 0
\label{eq:ueesigns}
\eeq
{\em of the anisotropy energies due to short-range e-e interactions could be realized in a real graphene system,
regardless of the potential restrictions on the signs of the bare couplings $g_{\al\be}$, $\al,\be = \perp,z$.}

To elaborate on this statement, it is instructive to consider the weak-coupling limit $e^2/v \ll 1$.
In this regime, the bare couplings $g_{\al\be}  = g_{\al\be}^{(1)}$, $\al,\be = \perp,z$,
obtained in the first order in the Coulomb interaction, are given by Eq.~(\ref{eq:g1expr}).
The expression~(\ref{eq:g1expr}) in the form of the electrostatic energy leads us to conclude [Eq.~(\ref{eq:g1sign})]
that the couplings $g_{\al\be}^{(1)}>0$, $\al,\be=\perp,z$,  are positive, i.e., e-e interactions in these channels are repulsive.
Then, according to Eq.~(\ref{eq:uee0}),
the bare anisotropy energies $u^{(\text{e-e},0)}_\al = g_{\al z}^{(1)} / (2 \pi l_B^2) >0 $, $\al=\perp,z$, can also only be positive
($g_{\al 0}=g_{\al 0}^{(2)} \ll g_{\al z}^{(1)}$, for $e^2/v \ll 1$).

If the bare expressions $u^{(\text{e-e},0)}_{\perp,z}$ provided correct values for  the anisotropy energies,
this would significantly restrict the variety of possible orders of the $\nu=0$ QHFM that could be realized in the system.
Anticipating the results of Sec.~\ref{sec:ground}, the fully spin-polarized ferromagnetic phase
would be the only phase favored by the short-range e-e interactions.
However, the positiveness of the bare couplings $g_{\al\be}^{(1)}>0$
does not impose any sign restrictions on the renormalized energies $u_{\perp,z}^{(\text{e-e})}$:
according to Eq.~(\ref{eq:ueestrong}), the signs of  $u^{(\text{e-e})}_\al$ ($\al=\perp,z$)
are determined by the relations $g_{\al z}^{(1)} \gtrless g_{\al \perp}^{(1)}$ between the couplings,
and not their signs, and there seems to be no physical restriction on the relative values $g_{\al z}^{(1)} /g_{\al \perp}^{(1)}$.

This demonstrates that, even if the signs of the bare couplings $g_{\al \be}$ ($\al,\be=\perp,z$)
were fixed to positive by the repulsive nature of  the Coulomb interactions,
the renormalized anisotropy energies $u_{\perp,z}^{(\text{e-e})}$
can still come in any possible sign combination, which brings us to the above statement.
We also do not expect the sign restriction on
$g_{\al\be}$ ($\al,\be=\perp,z$)
to necessarily remain in the regime of stronger interactions, $e^2/v \sim 1$:
the sign change of some couplings could occur already due to renormalizations at the lattice scale, see Sec.~\ref{sec:Hee}.

What concerns the anisotropies due to electron interactions with in-plane $A_1,B_1$ phonons,
as follows from Eqs.~(\ref{eq:ueph0}) and (\ref{eq:uephstrong}), the energy
\beq
    u_\perp^{(\text{e-ph})} < 0
\label{eq:uephsign}
\eeq
always remains negative. The negative sign is a consequence of the attractive nature of the phonon-mediated interactions
and appears to stay preserved in the renormalization process.
We also mention that electron interactions with the out-of-plane phonons result in a negative anisotropy energy $ u_z^{(\text{e-ph})} < 0$,
which is, however, small, $ |u_z^{(\text{e-ph})}| \ll | u_\perp^{(\text{e-ph})}|$ and neglected here.

The discussion of the implications of the properties (\ref{eq:ueesigns}) and (\ref{eq:uephsign})
for the physics of the $\nu=0$ QHFM will be continued in Sec.~\ref{sec:phasesnoZ}.

\section{Phase diagram for the  $\nu=0$ quantum Hall ferromagnet \label{sec:ground}}

\begin{table*}

\begin{tabular}{|c|c|c|c|}
    \hline $\nu=0$ QHFM state
        &   single-particle spinors $\chi_{a,b}$ & order parameter $P$ & anisotropy energy $\Ec_\dm(P)$ \\
    \hline spin-polarized isospin-singlet  & $|\nb \ran \otimes |\s\ran$, $|-\nb \ran \otimes |\s\ran$ &
                            $\hat{1} \otimes P_\s$  & $-2 u_\perp -u_z$  \\
   \hline  isospin-polarized spin-singlet  &   $|\nb \ran \otimes |\s\ran$, $|\nb \ran \otimes |-\s\ran$ & $ P_\nb \otimes \hat{1}$ &
                                                                                $u_\perp (n_x^2+n_y^2)+u_z n_z^2$ \\
   \hline  $P^{ns}$ state   & $|\nb \ran \otimes |\s\ran$, $|-\nb \ran \otimes |-\s\ran$
                                & $P_\nb \otimes P_\s + P_{-\nb} \otimes P_{-\s}$
                                & $- u_\perp (n_x^2+n_y^2)-u_z n_z^2$ \\
    \hline
\end{tabular}
\caption{Relevant states of the $\nu=0$ QHFM. For given values of $(u_\perp,u_z)$, the anisotropy energy $\Ec_\dm(P)$ is minimized
by one of these states, see Tab.~\ref{tab:Eamin}}
\label{tab:states}
\end{table*}

In this section, we obtain the phase diagram of the $\nu=0$ QHFM in graphene in the presence
of the isospin anisotropy and Zeeman effect, as described by the energy functional $E[P(t,\rb)]$
[Eqs.~(\ref{eq:E})-(\ref{eq:t})].
We will consider only the ``classical'' ground states, i.e., time-independent configurations $P(\rb)$
that minimize the energy functional;
effects of thermal fluctuations are beyond the scope of this paper.
In the bulk of the sample,
the functional $E[P(\rb)]$ is minimized by a spatially homogeneous configuration $P(\rb)=P$
that minimizes the sum
\beq
    \Ec(P)= \Ec_\dm(P)+\Ec_Z(P)
\label{eq:Etotal}
\eeq
of the anisotropy and Zeeman energies.
Which  $P$ delivers the minimum of $\Ec(P)$ depends
on the signs of the anisotropy energies $u_{\perp,z}$
and their relations between each other and the Zeeman energy $\e_Z$.
The results of the previous section suggest that essentially any
scenario for $(u_\perp,u_z)$ could take place in real graphene. Therefore, here, we consider all possibilities
and obtain a generic phase diagram in the space of parameters $(u_\perp,u_z,\e_Z)$.

\subsection{Relevant states}

According to Sec.~\ref{sec:basicQHFM},
the order parameter matrices $P$ of the $\nu=0$ QHFM in graphene form a representation of the eight-dimensional
Grassmanian  manifold Gr(2,4).
For the purpose of finding the ground states, however, considering the most general parametrization of $P$
is not necessary, since the anisotropy terms [Eqs.~(\ref{eq:Ea}) and (\ref{eq:t})] act explicitly on the isospin only.
Here, we will consider two simpler six-dimensional submanifolds of Gr(2,4) of physical relevance to the problem.
We will find that all possible ground states belong to these subsets.

First, consider a family of states [see Eqs.~(\ref{eq:PhiSlater}) and (\ref{eq:Pchi}), direct products are in $KK' \otimes s$ space],
\beq
    \chi_a = |\nb_a\ran \otimes |\s \ran, \mbox{ } \chi_b = |\nb_b\ran \otimes | -\s\ran,
\label{eq:chin}
\eeq
\beq
     P^n = P_{\nb_a} \otimes P_{\s}+P_{\nb_b} \otimes P_{-\s},
\label{eq:Pn}
\eeq
in which two electrons occupy the states $\chi_{a,b}$ with opposite spin ($\pm \s$)
and arbitrary isospin ($\nb_{a,b}$) polarizations. Here and below,
\[
    P_\s = |\s \ran \lan \s | = \frac{1}{2}(\hat{1}+ \taub \s)
\]
is the density matrix of the spin (isospin, if in $KK'$ space)
in the state $|\s\ran$; $\nb_{a,b}$, $\s $, and $\nb$, $\s_{a,b}$ below, are unit vectors defining the spin or isospin states.
Calculation of the anisotropy energy [Eqs.~(\ref{eq:Ea}) and (\ref{eq:t})] of the state (\ref{eq:Pn}) gives
\beq
    \Ec_\dm(P^n) = u_\perp (n_{ax} n_{bx}+n_{ay} n_{by}) + u_z n_{az} n_{bz}
\label{eq:En}
\eeq

\begin{figure}
\centerline{
\includegraphics[width=.22\textwidth]{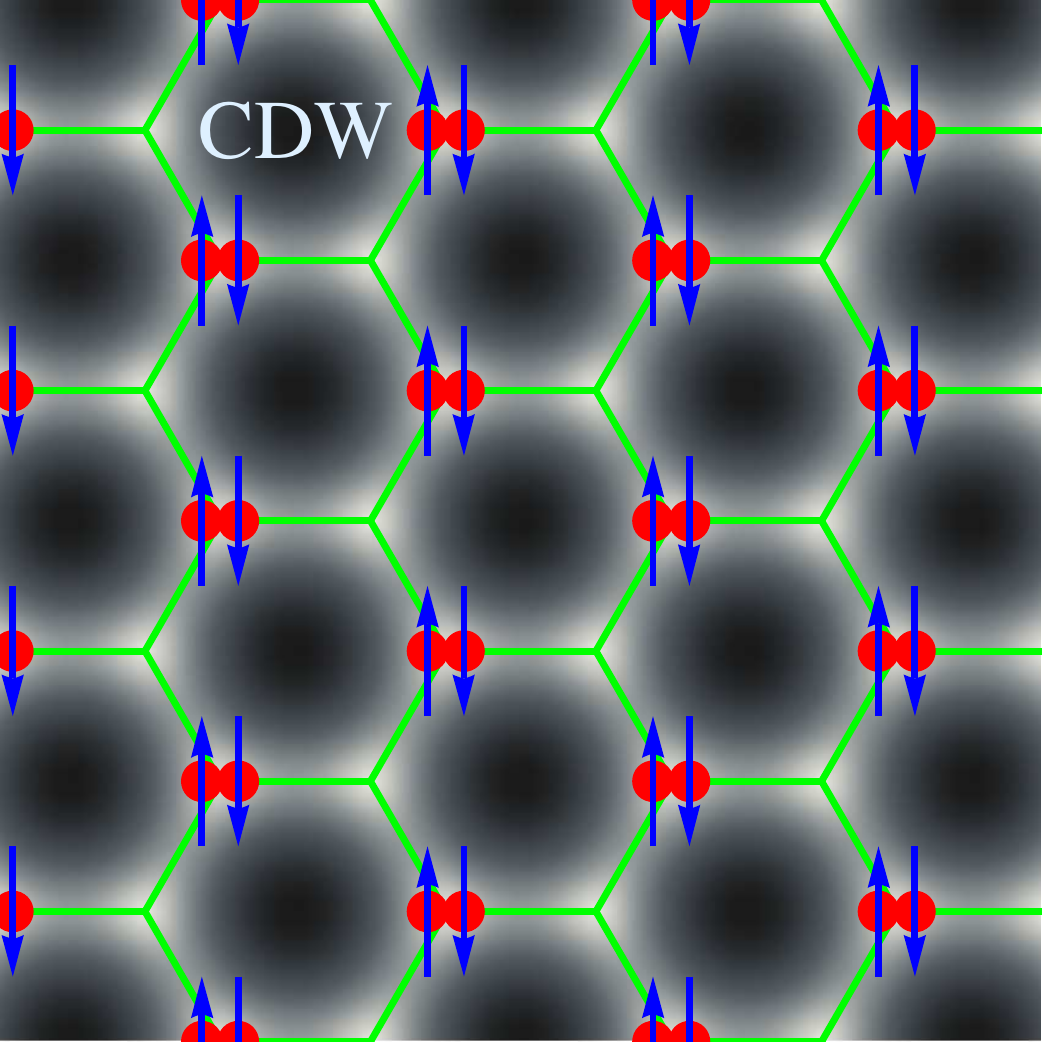}
\hspace{.3cm}
\includegraphics[width=.20\textwidth]{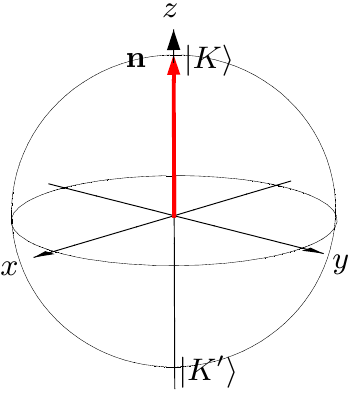}
}
\caption{Charge-density-wave (CDW) phase of the $\nu=0$ QHFM.}
\label{fig:CDW}
\end{figure}

\begin{figure}
\includegraphics[width=.22\textwidth]{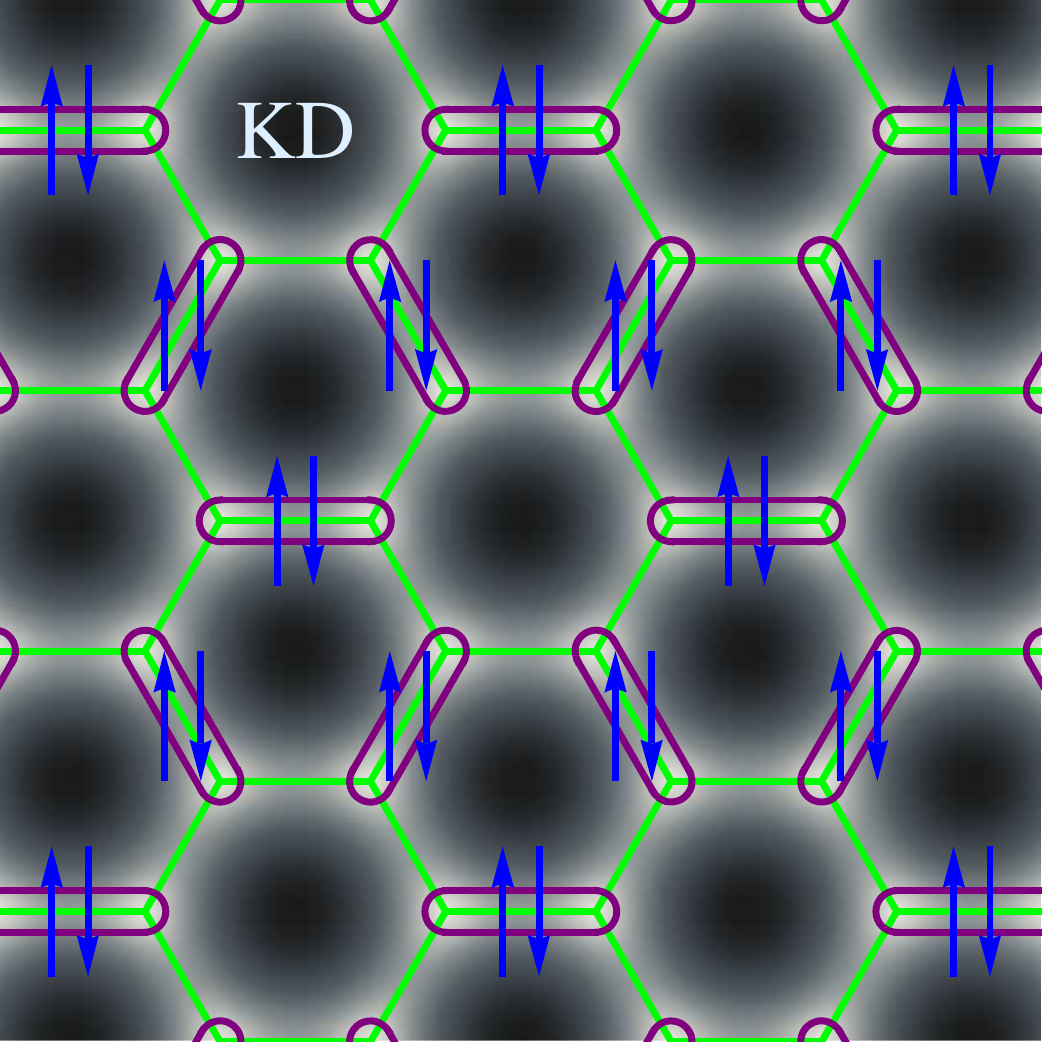}
\hspace{.2cm}
\includegraphics[width=.22\textwidth]{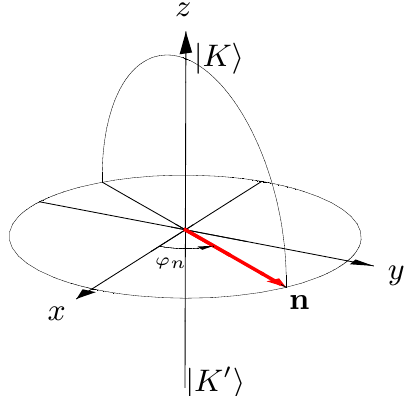}
\caption{Kekul\'{e}-distortion (KD) phase of the $\nu=0$ QHFM.}
\label{fig:KD}
\end{figure}

As an important special case of $P^n$ states, for coinciding isospins $\nb_a=\nb_b\equiv \nb$, one obtains
a fully isospin-polarized (IP) spin-singlet state
\beq
    P^\text{IP} =
    P_\nb \otimes \hat{1}
\label{eq:PS0}
\eeq
with the anisotropy energy
\beq
    \Ec_\dm(P^\text{IP}) = u_\perp n_\perp^2+ u_z n_z^2, \mbox{ } n_\perp^2= n_x^2 +n_y^2.
\label{eq:ES0}
\eeq
The observable order of such state depends on the orientation of the isospin $\nb$.
In particular, when the isospin is at the poles of the Bloch sphere, $\nb= \pm \nb_z$, $\nb_z=(0,0,1)$,
the state has a charge-density-wave (CDW) order (Fig.~\ref{fig:CDW}), with both electrons per orbital occupying the
same sublattice, and energy $\Ec_\dm^\text{CDW} = u_z$.
When the isospin is on the equator of the Bloch sphere, $\nb = \nb_\perp = (\cos \vphi_n,\sin \vphi_n,0)$,
the state has the Kekul\'{e}-distortion (KD) order, represented schematically in Fig.~\ref{fig:KD},
and energy  $\Ec_\dm^\text{KD} = u_\perp$.
The angle $\vphi_n$ is related to the  orientation of the atom displacements in Fig.~\ref{fig:phmodes}(right).

The second important family of states is ``dual'' to $P^n$,
\beq
    \chi_a = |\nb\ran \otimes | \s_a \ran, \mbox{ } \chi_b = |-\nb\ran \otimes  \s_b\ran,
\label{eq:chis}
\eeq
\beq
     P^s = P_{\nb} \otimes P_{\s_a}+P_{-\nb} \otimes P_{\s_b},
\label{eq:Ps}
\eeq
Here, two electrons occupy the states $\chi_{a,b}$ with opposite isospin  ($\pm \nb$)
and arbitrary spin  ($\s_{a,b}$)  polarizations. Calculation of the anisotropy energy gives
\beq
    \Ec_\dm(P^s) =
    -(2 u_\perp+u_z)\frac{1+\s_a \s_b}{2} - (u_\perp n_\perp^2 + u_z n_z^2)\frac{1-\s_a\s_b}{2}.
\label{eq:Es}
\eeq

\begin{figure}
\includegraphics[width=.22\textwidth]{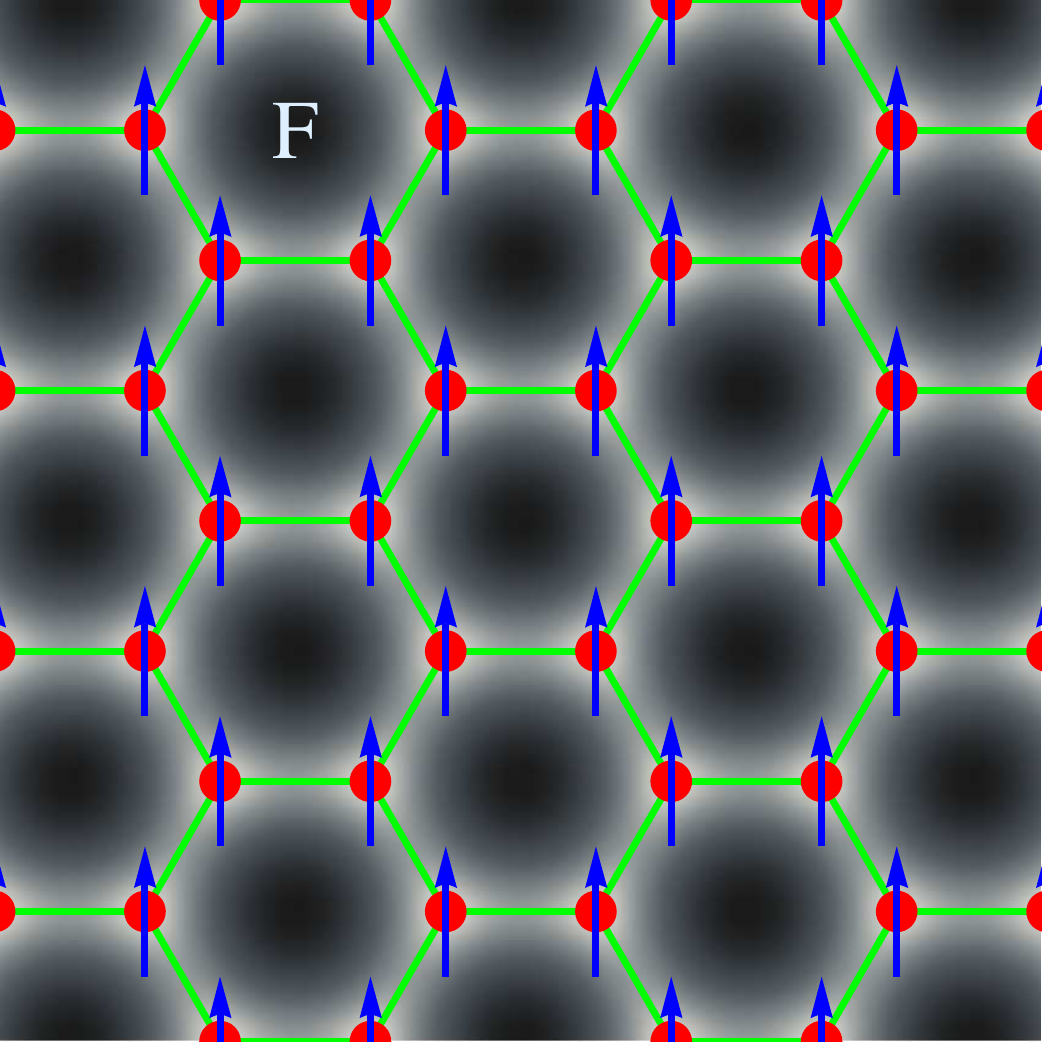}
\caption{Spin-polarized ferromagnetic (F) phase of the $\nu=0$ QHFM.}
\label{fig:F}
\end{figure}

As a special case of $P^s$ states, for coinciding spins $\s_a=\s_b\equiv\s$,
one obtains a fully spin-polarized isospin-singlet state
\beq
    P^\text{F} = \hat{1} \otimes P_\s,
\label{eq:PT0}
\eeq
with the ferromagnetic (F) order (Fig.~\ref{fig:F})
and anisotropy energy
\beq
    \Ec_\dm^\text{F} = - 2 u_\perp - u_z.
\label{eq:ET0}
\eeq
Note that in the isospin-singlet F state,
each isospin channel $\al=x,y,z$ contributes $- u_\al$ to the anisotropy energy~(\ref{eq:Ea}).

\begin{figure}
\includegraphics[width=.22\textwidth]{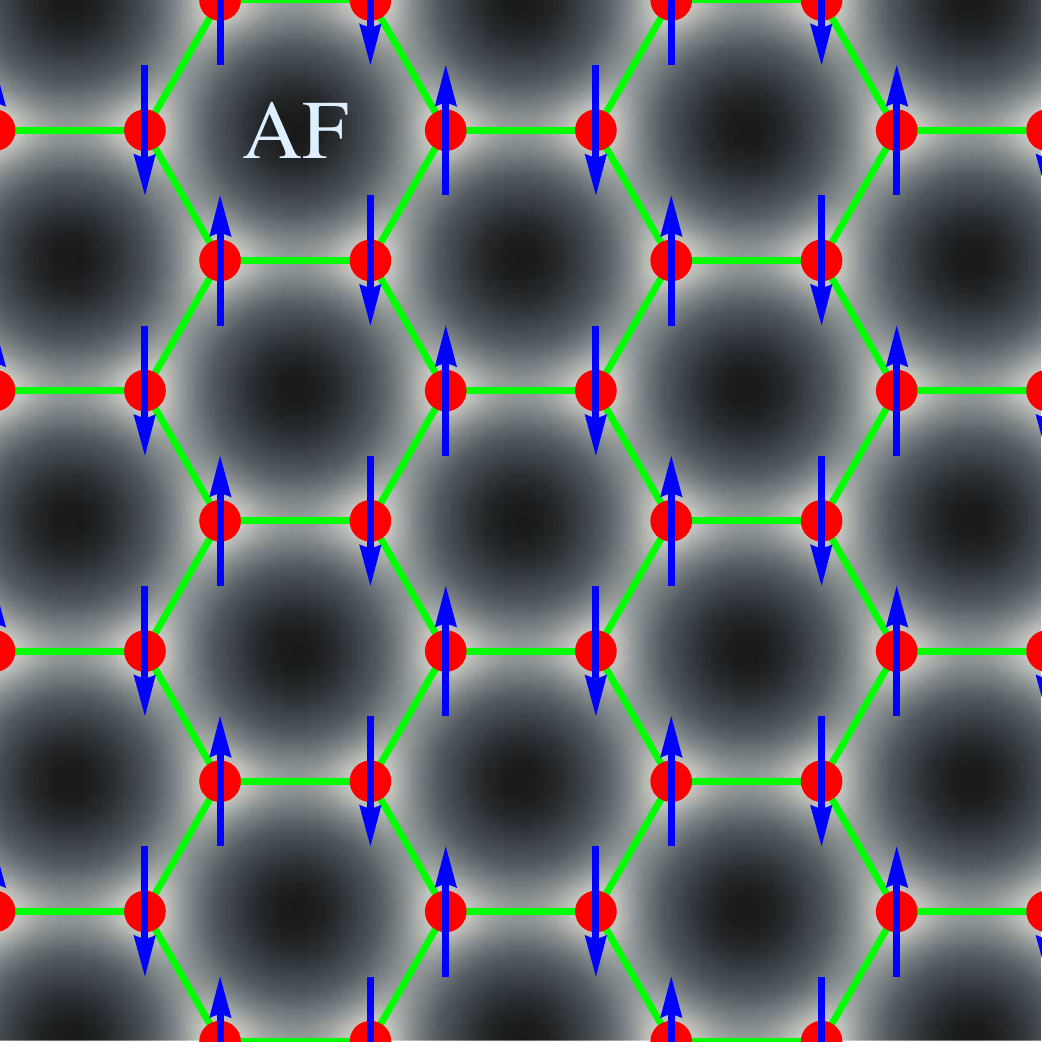}
\caption{Antiferromagnetic (AF) phase of the $\nu=0$ QHFM.}
\label{fig:AF}
\end{figure}

An intersection of the subsets (\ref{eq:Pn}) and (\ref{eq:Ps}) is the family of
states
\[
    \chi_a = |\nb \ran \otimes |\s \ran, \mbox{ }  \chi_b= |-\nb\ran \otimes | -\s\ran
\]
\beq
    P^{ns} = P_{\nb} \otimes P_{\s}+P_{-\nb} \otimes P_{-\s}.
\label{eq:Pns}
\eeq
Here, two electrons have simultaneously opposite spin ($\pm \s$) and isospin ($\pm \nb$) polarizations.
For brevity, we will denote  the states (\ref{eq:Pns}) with  $\nb =\nb_\perp$ and $\nb= \pm \nb_z$ as $P^{ns}_\perp$ and $P^{ns}_z$, respectively.
At $\nb=\pm \nb_z$, electrons with opposite spin polarizations $\pm \s$ reside on different
sublattices and  $P^{ns}_z$ state has an antiferromagnetic (AF) order (Fig.~\ref{fig:AF}).

The state (\ref{eq:Pns}) has the anisotropy energy
\beq
    \Ec_\dm^{ns} = - u_\perp n_\perp^2-u_z n_z^2,
\label{eq:Ens}
\eeq
as can be obtained from either Eq.~(\ref{eq:En}) or (\ref{eq:Es}).
Comparing Eqs.~(\ref{eq:Es}), (\ref{eq:ET0}) and (\ref{eq:Ens}), we also notice that,
for arbitrary orientations of spins $\s_a$ and $\s_b$,
the anisotropy energy (\ref{eq:Es}) of the state (\ref{eq:Ps})
is a linear combination of the energies of the F and $P^{ns}$ states,
with the weights determined by the product $\s_a \s_b$, i.e., by the angle between the spins.

The properties of the states (\ref{eq:PS0}), (\ref{eq:PT0}), and (\ref{eq:Pns}) are summarized in Table~\ref{tab:states}.

\subsection{Phase diagram neglecting the Zeeman effect \label{sec:phasesnoZ}}

\begin{table*}
\begin{tabular}{|c|c|c|c|c|c|c|c|}
    \hline sign combination
        & $u_\perp>0$, $u_z>0$ &  \multicolumn{2}{|c|}{$u_\perp>0$, $u_z<0$}  & \multicolumn{2}{|c|}{$u_\perp<0$, $u_z>0$} &
       \multicolumn{2}{|c|}{$u_\perp<0$, $u_z<0$} \\
    \hline
          subcase &  & $u_\perp> |u_z|$ & $u_\perp<|u_z|$ & $|u_\perp|> u_z$ & $|u_\perp|<u_z$ & $|u_\perp|>|u_z|$ & $|u_\perp|<|u_z|$ \\
    \hline
        state minimizing $\Ec_\dm(P)$  & F & F & CDW & KD & AF & KD & CDW\\
    \hline
        minimal  $\Ec_\dm(P)$ & $- 2 u_\perp - u_z$ & $- 2 u_\perp + |u_z|$ & $-|u_z|$  & $-|u_\perp|$ & $-u_z$ & $-|u_\perp|$ & $-|u_z|$\\
    \hline
\end{tabular}
\caption{Minimization of the anisotropy energy $\Ec_\dm(P)$ [Eqs.~(\ref{eq:Ea}) and (\ref{eq:t})].}
\label{tab:Eamin}
\end{table*}

To get a clear understanding of the orders favored by the isospin anisotropy,
let us first neglect the Zeeman energy $\Ec_Z(P)$  completely and find the states $P$ that minimize the anisotropy energy
$\Ec_\dm(P)$ [Eqs.~(\ref{eq:Ea}) and (\ref{eq:t})] alone.

We accomplish this by noticing the following property. For a given isospin channel $\al =x,y,z$,
the function $t_\al (P)$ [Eq.~(\ref{eq:t})] belongs to the range
\[
    -2 \leq t_\al(P) \leq 2.
\]

The maximum $t_\al(P) =2$ is reached at the IP state~(\ref{eq:PS0})
with $\nb$ parallel to the $\al$ axis, $n_\al = \pm 1$. For the remaining components $\bar{\al}$, one has $t_{\bar{\al}}(P)=0$.

The minimum $t_\al(P)=-2$ is reached within the subset (\ref{eq:Ps}), when
\[
    (1-n_\al^2)(\s_+ \s_- -1) = 0,
\]
i.e., in two cases.

(i) In the first case, $\s_+ = \s_-$, and the minimum is reached at the F state (\ref{eq:PT0}) (hence, $\nb$ can be arbitrary).
For the remaining components $\bar{\al}$, one also has $t_{\bar{\al}}(P) = -2 $, see also the comment after Eq.~(\ref{eq:ET0}).

(ii) In the second case, $n_\al =\pm 1 $. For the remaining components $\bar{\al}$, one has $t_{\bar{\al}}(P) = -(1+\s_+ \s_-) $,
for which the maximum $t_{\bar{\al}}(P) = 0$ is reached at  $ \s_+ = - \s_-$, i.e, at the $P^{ns}_\al$ ($\al=\perp,z$) state,
and the minimum $t_{\bar{\al}}(P) = -2$ --  at the F state. The latter just brings us back to the case (i).

Using these properties,
we arrive at the conclusion that, for given signs and ratio of $u_\perp$ and $u_z$, the anisotropy energy
$\Ec_\dm(P)$ is minimized by one of the above states --  F, IP, or $P^{ns}_\al$ -- that either minimize or maximize $t_\al(P)$.

The four possible cases $(u_\perp  \gtrless 0 , u_z  \gtrless 0)$ of sign combinations are considered below;
three of them split  into subcases, depending on the relative absolute value $|u_\perp/u_z|$.

$(++)$: $u_\perp>0$, $u_z>0$.
The anisotropy energy is minimized by the F state~(\ref{eq:PT0}),
which minimizes $t_\al(P)=-2$ simultaneously for all $\al= x,y,z$, and gives $\Ec_\dm^\text{F} = - 2u_\perp-u_z$.

$(--)$: $u_\perp<0$, $u_z<0$.
The anisotropy energy  is minimized by one of the IP states (\ref{eq:PS0}),
which maximize $t_\al(P) =2 $.
For $|u_\perp| > |u_z|$, $\Ec_\dm(P)$ is minimized by the KD state with $\Ec_\dm^\text{KD} = - |u_\perp|$,
whereas for  $|u_\perp| < |u_z|$ -- by the  CDW state with $\Ec_\dm^\text{CDW} = - |u_z|$.

$(-+)$: $u_\perp<0$, $u_z>0$.
The anisotropy energy is minimized by either the KD state with  $\Ec_\dm^\text{KD} = -|u_\perp|$,
which minimizes the $\Ec_\perp(P) = \frac{1}{2} u_\perp[t_x(P)+t_y(P)]$ part of $\Ec_\dm(P)$, or the AF state [Eq.~(\ref{eq:Pns}) with $\nb=\pm \nb_z$]
with $\Ec_\dm^\text{AF}=- u_z$, which minimizes the $\Ec_z(P) = \frac{1}{2} u_z t_z(P)$ part of $\Ec_\dm(P)$, $\Ec_\dm(P)=\Ec_\perp(P)+\Ec_z(P)$.
Comparing these energies, we obtain that for $ | u_\perp|> u_z$,  the KD state is realized,
whereas for  $ |u_\perp|< u_z  $, the AF state is realized.
The F state, which also minimizes $\Ec_z(P)$,
cannot be realized, since $\Ec_\dm^{\text{F}}= 2 |u_\perp| -  u_z >\Ec_\dm^\text{AF}$ in this regime.

$(+-)$: $u_\perp>0$, $u_z<0$.
In this case, one has to compare the energies of three states: the CDW state with $\Ec_\dm^\text{CDW}=-|u_z|$,
which minimizes the $\Ec_z(P)$; the F state with $\Ec_\dm^\text{F} = - 2 u_\perp+ |u_z|$, which
minimizes $\Ec_\perp(P)$ but maximizes $\Ec_z(P)$ at the same time;
and the $P^{ns}_\perp$ state [Eq.~(\ref{eq:Pns}) with $n_z = 0$] with $\Ec_\dm(P^{ns}_\perp) = - u_\perp$,
which has a higher $\Ec_\perp(P)$ than the F state, but zero $\Ec_z(P)$, on the other hand.
Comparing these three energies, we find that only the first two states are realized:
for $u_\perp> | u_z|$, $\Ec_\dm(P)$ is minimized by the F state,
whereas for $u_\perp < | u_z|$ -- by the  CDW state.
The $P^{ns}_\perp$ state is not realized since $\Ec_\dm(P^{ns}_\perp) > \Ec_\dm^\text{F} $
for  $u_\perp> | u_z|$ and $\Ec_\dm(P^{ns}_\perp) < \Ec_\dm^\text{CDW} $  for $u_\perp< | u_z|$.

\begin{figure}
\includegraphics[width=.30\textwidth]{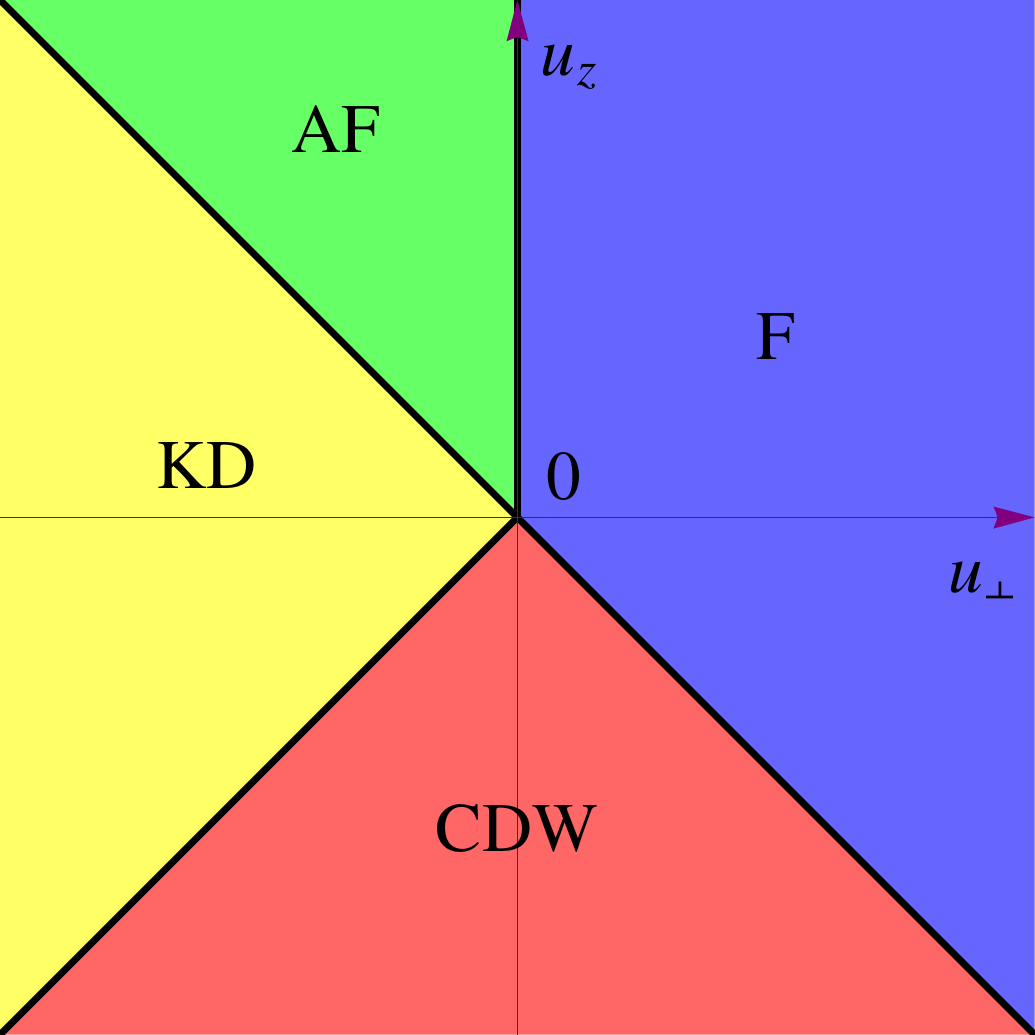}
\caption{Phase diagram of the $\nu=0$ QHFM states minimizing the isospin anisotropy energy $\Ec_\dm(P)$ [Eqs.~(\ref{eq:Ea}) and (\ref{eq:t})],
in the space of the anisotropy energies $(u_\perp,u_z)$. Physical orders of the phases are shown in Figs.~\ref{fig:CDW}-\ref{fig:AF}.
}
\label{fig:phasesnoZ}
\end{figure}

\begin{table}
\begin{tabular}{|c|c|c|c|}
    \hline state &   order parameter $P$ &  $\Ec_\dm(P)$ & symmetry \\
    \hline F & $\hat{1} \otimes P_\s$  & $- 2 u_\perp - u_z $ & $\Stxt \Utxt (2)_s$  \\
   \hline  KD  &    $P_{\nb_\perp} \otimes \hat{1}$ &  $  u_\perp $  & $\Utxt (1)_{KK'}$ \\
   \hline  CDW &   $ P_{\pm \nb_z} \otimes \hat{1}$ &     $  u_z $ & $Z_{2\,KK'}$ \\
   \hline  AF  &  $P_{\nb_z} \otimes P_{\s} + P_{-\nb_z} \otimes P_{-\s}$ & $-  u_z $ & $\Stxt \Utxt (2)_s$ \\
    \hline
\end{tabular}
\caption{States minimizing the isospin anisotropy energy $\Ec_\dm(P)$ [Eqs.~(\ref{eq:Ea}) and (\ref{eq:t})]
and forming the phase diagram in Fig.~\ref{fig:phasesnoZ}.
The last column (symmetry) denotes the symmetries of the ground states in the isospin ($KK'$) and spin ($s$) spaces.}
\label{tab:phasesnoZ}
\end{table}

These cases are summarized in Tab.~\ref{tab:Eamin}.
Together, they combine into the phase diagram of the states of the $\nu=0$ QHFM that minimize the isospin anisotropy energy,
plotted in Fig.~\ref{fig:phasesnoZ}.
The phase diagram consists of four states with the following orders: F, AF, CDW, and KD,
shown in Figs.~\ref{fig:CDW}-\ref{fig:AF}.
The phases are separated by the first-order transition lines:
\[
    u_\perp = -u_z >0
\mbox{ -- between the F and CDW phases;}
\]
\[
    u_\perp=u_z<0
\mbox{ -- between the CDW and KD phases;}
\]
\[
    u_\perp= - u_z <0
\mbox{ -- between the KD and AF phases;}
\]
\[
    u_\perp=0, \mbox{ } u_z >0
\mbox{ -- between the AF and F phases.}
\]
These transition lines come together at the origin $(u_\perp, u_z)=(0,0)$,
where the anisotropy energy vanishes and all orders $P$ have the same energy.

{\em Without} the Zeeman effect, the phases have the following symmetries:
the F and AF phase are SU(2)-symmetric in the spin space [the spin polarization $\s$ in Eq.~(\ref{eq:PT0}) or (\ref{eq:Pns}) can be arbitrary];
the KD phase has U(1) symmetry [$\nb = \nb_\perp= (\cos \vphi_n, \sin \vphi_n,0)$ can have arbitrary orientation $\vphi_n$ in the  $n_z=0$ plane], and the CDW phase has $Z_2$ symmetry ($\nb =\pm \nb_z$,
according to the occupation of either $A$ or $B$ sublattice).
Exactly at the phase transition lines, the symmetry of the ground state becomes higher;
we do not attempt to study the details of the transitions here.
The key properties of the phases are summarized in Tab.~\ref{tab:phasesnoZ}.

Let us come back to the microscopic origins of the isospin anisotropy.
The conclusions of Sec.~\ref{sec:implications} [Eq.~(\ref{eq:ueesigns})] imply that
any possibility for the signs and relative values of the anisotropy energies $u_{\perp,z}^\text{(e-e)}$ could be realized
and, therefore, any phase on the diagram in Fig.~\ref{fig:phasesnoZ}
could be favored by the short-range e-e interactions alone.
The reason for that are peculiar properties of the renormalizations of the short-range e-e interactions in graphene,
which allow for sign changes of the  coupling constants (Fig.~\ref{fig:RGflows}), switching the interactions
from repulsive ($u_\al^\text{e-e}>0$) to attractive ($u_\al^\text{e-e} <0$) in certain channels.

In contrast, according to Eq.~(\ref{eq:uephsign}), the leading e-ph interactions
always favor the Kekul\'{e} distortion order,
in agreement with earlier predictions~\cite{Nomura,Chamon}.
Unlike short-range e-e interactions, different sublattice channels do not couple
in the renormalization process and e-ph couplings retain their negative sign, characteristic of attractive interactions.
We also mention that e-ph interactions with the out-of-plane phonons, neglected here as weak,
result in $u_z^{(\text{e-ph})}<0$ and favor CDW order, in line with Ref.~\onlinecite{FuchsLederer}.

Of course, the values of the bare couplings $g_{\al\be}$ are determined
by the details of e-e interactions and band structure at atomic scale
and should be a robust material property.
Therefore, in real graphene, one can expect one particular situation for the anisotropy energies
$(u_\perp^{(\text{e-e})}, u_z^{(\text{e-e})})$ to be realized
[which, in the experimentally relevant regime of strong renormalizations, cannot be changed by varying $B_\perp$,
see Eq.~(\ref{eq:ueestrong}) and discussion in Sec.~\ref{sec:transitions}]
and one certain order to be favored.
In this sense, a more accurate formulation of the statement (\ref{eq:ueesigns})
is that one cannot theoretically rule out any possibility for $u_{\perp,z}^{(\text{e-e})}$,
based just
on the repulsive nature of the underlying Coulomb interactions.
Doing so requires a reliable numerical estimate for the bare couplings.

In the absence of such an estimate, we point out one case
that may be the most relevant to the real system.
For that, we turn again to the first-order expressions (\ref{eq:g1expr})  for the couplings
$g^{(1)}_{\al\be}$, $\al,\be=\perp,z$.
There, $\rho_{\al\be}(\rv)$, $\al,\be = x,y$, and $\rho_{z z}(\rv)$ are the staggered-type densities
given by the linear combinations of the products  $u_{\mu A}(\rv) u_{\mu' A}(\rv)$ and $u_{\mu B}(\rv) u_{\mu' B}(\rv)$, $\mu=K,K'$,
of the Bloch wave-functions.
On the other hand, the densities $\rho_{\al z}(\rv)$ and $\rho_{z \al}(\rv)$, $\al=x,y$,
are determined by the overlaps $u_{\mu A}(\rv) u_{\mu' KB}(\rv)$ of the Bloch wave-functions
peaked at the atomic sites of different sublattices.
Therefore, it would be reasonable to expect the couplings $g^{(1)}_{\perp z}$ and $g^{(1)}_{z \perp}$ to be smaller
than  $g^{(1)}_{z z}$ and $g^{(1)}_{\perp \perp}$, i.e.,
\[
   g^{(1)}_{z z}, \mbox{ }g^{(1)}_{\perp \perp}>   g^{(1)}_{\perp z}, \mbox{ } g^{(1)}_{z \perp}>0.
\]
(This certainly becomes true in the limit of ``strongly localized'' atomic orbitals, when the size of the atomic wave-function
is much smaller than the bond length.)
In this case, according to Sec.~\ref{sec:renormu}, $ \gb_{z z}(l_B)>0$ will stay positive and grow upon renormalization, while
$\gb_{\perp z}(l_B)$ will turn negative and grow in absolute value. This results in the signs
\[
    u_\perp^{(\text{e-e})}<0, \mbox{ } u_z^{(\text{e-e})}>0
\]
of the anisotropy energies [Eq.~(\ref{eq:uee})].
The same holds for the total anisotropy energies,
\[
    u_\perp<0, \mbox{ } u_z>0,
\]
once e-ph interactions [Eq.~(\ref{eq:uephsign})] are included.
In this case, according to Fig.~\ref{fig:phasesnoZ} only either KD or AF
phase can be favored by the isospin anisotropy (the latter becomes canted in the presence of the Zeeman effect, see Sec.~\ref{sec:phases}).

\subsection{Phase diagram in the presence of the Zeeman effect\label{sec:phases}}

We now take the Zeeman effect into account and find the ground states $P$ that minimize the
the sum $\Ec(P)$ [Eq.~(\ref{eq:Etotal})] of the anisotropy and Zeeman energies.
The question is how the phase diagram in Fig.~\ref{fig:phasesnoZ} is modified by the Zeeman effect.

\begin{figure}
\centerline{
\includegraphics[width=.22\textwidth]{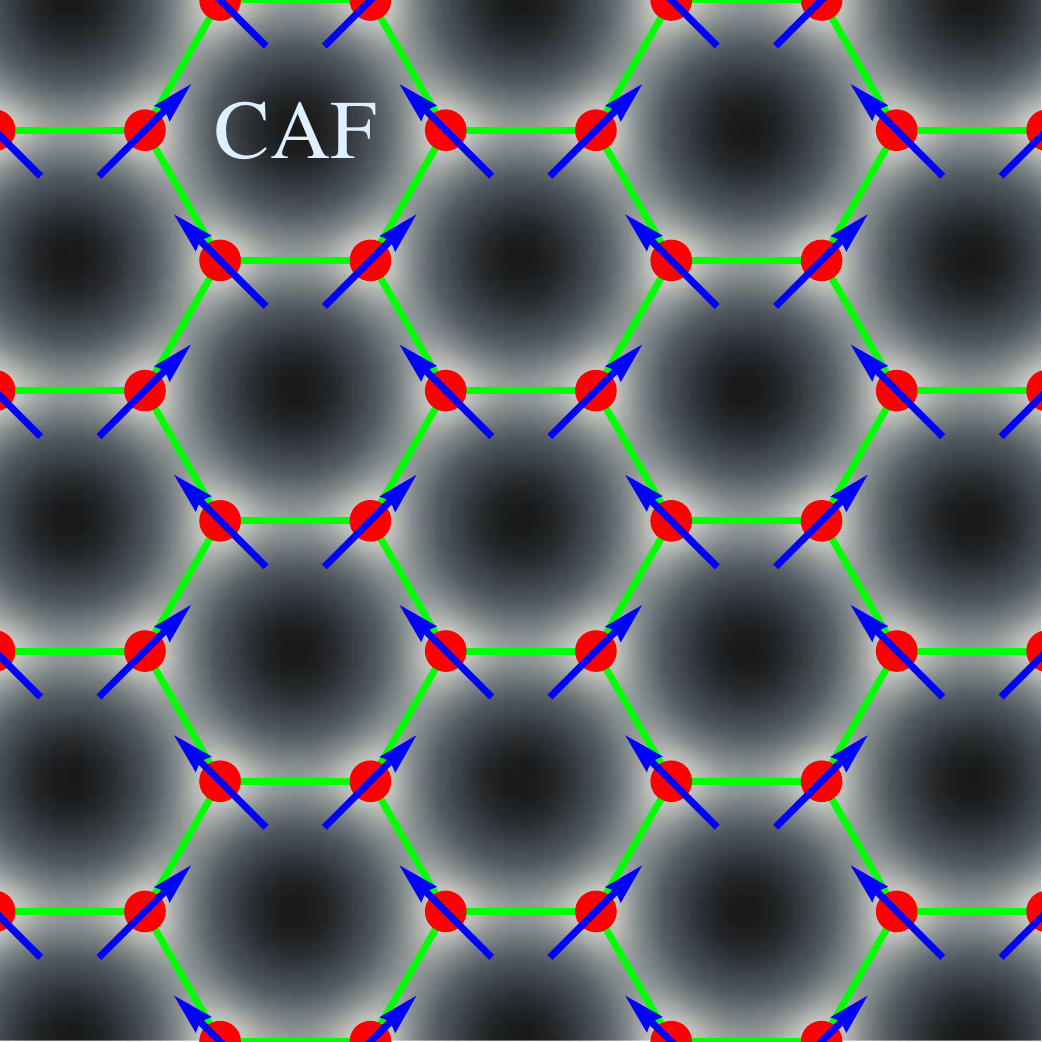}
\hspace{.2cm}
\includegraphics[width=.22\textwidth]{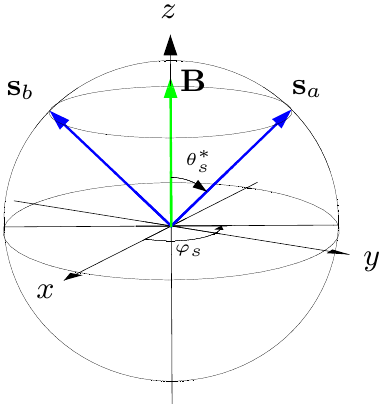}
}
\caption{Canted antiferromagnetic (CAF) phase of the $\nu=0$ QHFM.}
\label{fig:CAF}
\end{figure}

\begin{figure}
\includegraphics[width=.30\textwidth]{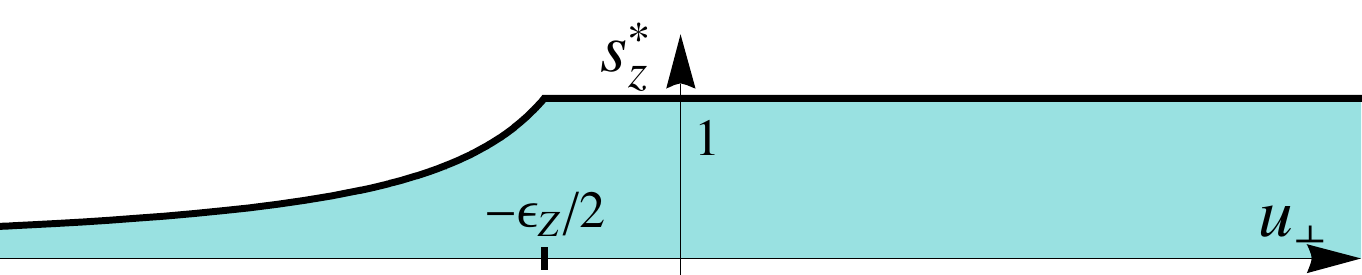}\\
\vspace{3mm}
\includegraphics[width=.30\textwidth]{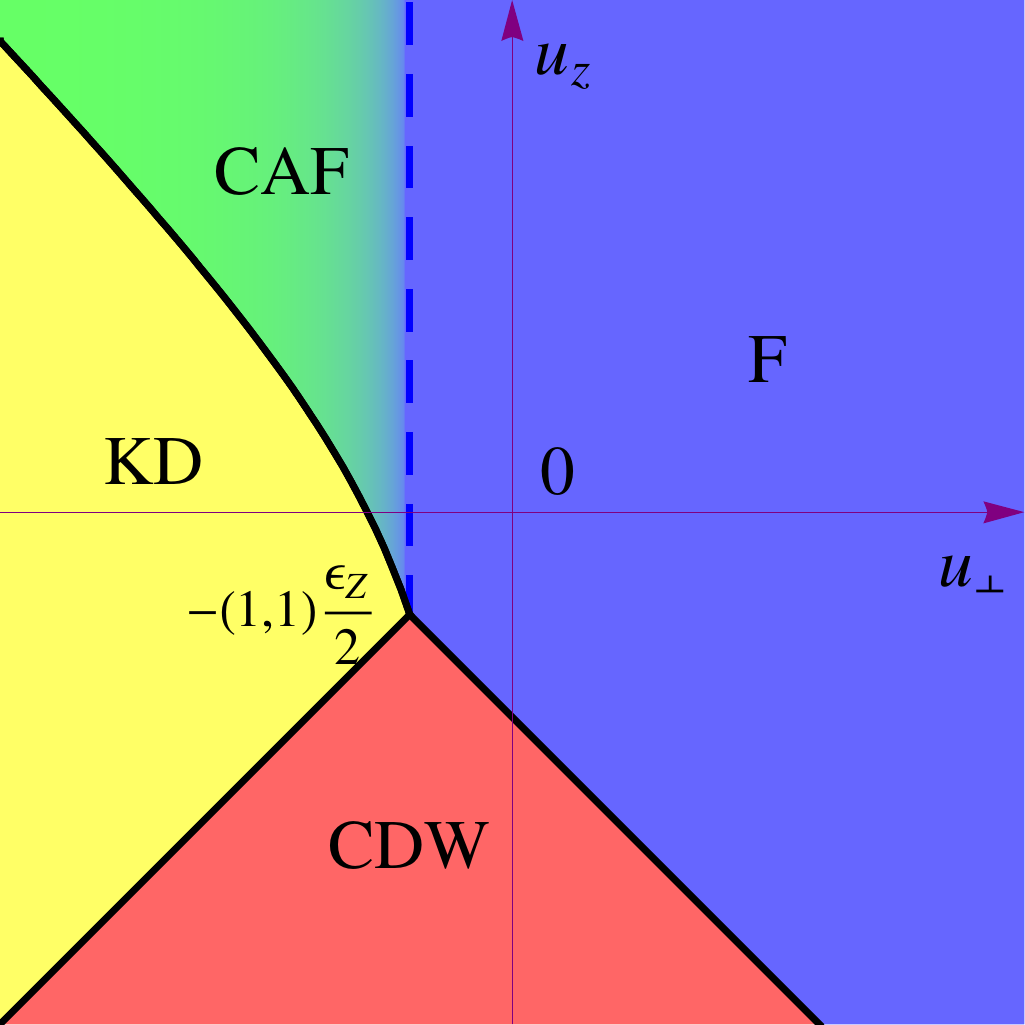}
\caption{Phase diagram of the $\nu=0$ quantum Hall ferromagnet in monolayer graphene in the presence of the isospin anisotropy and Zeeman effect,
in the space of the anisotropy energies $(u_\perp,u_z)$ and with the Zeeman energy $\e_Z$ as parameter.
Physical orders of the phases are shown in Figs.~\ref{fig:CDW}, \ref{fig:KD}, \ref{fig:F}, and \ref{fig:CAF}.
Top graph shows the optimal value $s_z^*$ [Eq.~(\ref{eq:szopt})] of the spin projection
on the direction of the magnetic field in the CAF and F phases.
}
\label{fig:phases}
\end{figure}

We first note that the spin-singlet CDW and KD phases are unaffected
by the Zeeman field and their total energy $\Ec(P)$ is equal to the anisotropy energy,
\beq
    \Ec^\text{CDW} = u_z, \mbox{ } \Ec^\text{KD}= u_\perp.
\label{eq:ECDWKD}
\eeq
One the other hand, the ``spin-active'' F and AF phases
are affected by the Zeeman field and, therefore, their whole sector has to be reconsidered.
Since both of these states belong to the family~(\ref{eq:Ps}), it is sufficient to minimize the energy
\beq
    \Ec(P^s)=
    - u_\perp-u_z - u_\perp \s_a \s_b -\e_Z (s_{a z}+s_{b z})
\label{eq:Es2}
\eeq
of $P^s$ state with $\nb = \nb_z $ but generally noncollinear spins $\s_{a,b}$.
One can easily check that for a given angle $2\theta_s$ between the spins, $\s_a \s_b =\cos (2\theta_s)$, $0 \leq \theta_s \leq \pi/2$,
which fixes the anisotropy energy $\Ec_\dm(P^s)$,
the Zeeman energy $\Ec_Z(P^s)$ is minimized by the spin orientations
\beq
    \s_{a,b} = (\pm \sqrt{1-s_z^2} \cos \vphi_s, \pm \sqrt{1-s_z^2} \sin \vphi_s, s_z),
\label{eq:sCAF}
\eeq
$s_z = \cos \theta_s$, that have equal projections on the direction of the magnetic field ($z$)
and are antiparallel in the perpendicular ($xy$) plane, as shown in Fig.~\ref{fig:CAF}.
The total energy (\ref{eq:Es2}) then equals
\beq
    \Ec(s_z) = -u_\perp-u_z-u_\perp (2 s_z^2 -1) - 2 \e_Z s_z.
\eeq
Minimizing it with with respect to $s_z$, we obtain that, for $u_\perp < - \e_Z /2 $,
a canted antiferromagnetic (CAF) state with the optimal angle $\theta^*_s$ between the spins (Fig.~\ref{fig:phases}),
\beq
    s_z^* = \cos \theta^*_s = \frac{\e_Z}{ 2|u_\perp|},
\label{eq:szopt}
\eeq
and energy
\beq
    \Ec^\text{CAF} = - u_z - \frac{\e_Z^2}{2 |u_\perp|}
\label{eq:ECAF}
\eeq
is realized, whereas for $u_\perp \geq - \e_Z/2$,
the fully spin-polarized F state ($\theta^*_s= 0$, $s_z^* =1$) with the energy
\beq
    \Ec^\text{F} = - 2 u_\perp -u_z - 2 \e_Z
\eeq
is realized.

We see that in the presence of the Zeeman effect, the total energy (\ref{eq:ECAF}) of the CAF state is indeed smaller than that
$\Ec^\text{CAF} = - u_z$ of the AF state with antiparallel spins:
by forming a noncollinear orientation (Fig.~\ref{fig:CAF}), electrons lose some of the anisotropy energy, but gain more in the Zeeman energy.
Therefore, the AF phase in Fig.~\ref{fig:phasesnoZ} is completely substituted by the CAF phase, in which
the optimal angle $\theta^*_s$ depends on the ratio $\e_Z/|u_\perp|$. The antiparallel orientation ($\theta^*_s=\pi/2$, $s_z^* =0$)
is reached asymptotically for $-u_\perp \gg \e_Z$.

We can now determine the boundaries between different phases.
As obtained above, the F and CAF phases are separated by the line
\beq
    u_\perp=-\frac{\e_Z}{2}.
\label{eq:CAFF}
\eeq
The separation line between the CDW and KD phases [Eq.~(\ref{eq:ECDWKD})] remains at
\beq
    u_\perp=u_z.
\label{eq:CDWKD}
\eeq
Next, comparing the energy (\ref{eq:ECAF}) of the CAF state with that (\ref{eq:ECDWKD}) of the KD state, we obtain
the separation line
\beq
    u_\perp+u_z=\frac{\e_Z^2}{2 u_\perp}.
\label{eq:CAFKD}
\eeq
Analogously, the phase boundary between the F and CDW phases is now given by
\beq
    u_\perp+u_z=-\e_Z.
\label{eq:FCDW}
\eeq
All four lines (\ref{eq:CAFF})-(\ref{eq:FCDW}) come together and terminate at point $(u_\perp, u_z) = - (1,1) \e_Z/2$.

\begin{table}
\begin{tabular}{|c|c|c|c|}
    \hline state    &   order parameter $P$ & $\Ec_\dm(P)+\Ec_Z(P)$  & symm.\\
    \hline F & $\hat{1} \otimes P_{\s_z}$  & $-2 u_\perp \!\!- u_z\! -\! 2 \e_Z $ & none \\
   \hline  KD  &    $P_{\nb_\perp} \otimes \hat{1}^s$ &  $  u_\perp $ & $\Utxt(1)_{KK'}$\\
   \hline  CDW  &   $ P_{\pm \nb_z} \otimes \hat{1}^s$ &  $  u_z $ & $Z_{2\,KK'}$\\
   \hline  CAF  &  $P_{\nb_z} \!\otimes\! P_{\s_a} + P_{-\nb_z}\! \otimes\! P_{\s_b}$
    & $\displaystyle -  u_z -\frac{\e_Z^2}{2|u_\perp|} $ & $\Utxt(1)_s$ \\
    \hline
\end{tabular}
\caption{
States minimizing the sum $\Ec_\dm(P) +\Ec_Z(P)$ of the isospin anisotropy and Zeeman energies
and forming the phase diagram in Fig.~\ref{fig:phases}. The last column (symm.) denotes the symmetries
of the ground states in the isospin ($KK'$) and spin ($s$) spaces.
}
\label{tab:phases}
\end{table}

This forms the phase diagram  of the $\nu=0$ quantum Hall ferromagnet in  monolayer graphene
in the presence of the generic isospin anisotropy and Zeeman effect, plotted in Fig.~\ref{fig:phases}.
This diagram constitutes the key result of the present work.
The total energy (\ref{eq:Etotal}) is minimized by the states with one of the following orders:
spin ferromagnetic (F),  charge-density-wave (CDW), Kekul\'{e} distortion (KD), or canted antiferromagnetic (CAF).
As compared to the situation without it (Fig.~\ref{fig:phasesnoZ}),
the Zeeman effect
(i) substitutes the AF phase with antiparallel spins by the CAF phase with noncollinear spins and
(ii)  naturally widens up the region of the F phase in the $(u_\perp, u_z)$ plane.
The discussion of the microscopic origins of the isospin anisotropies done in Sec.~\ref{sec:phasesnoZ}
can be directly carried over here. The key properties of the phases are summarized in Tab.~\ref{tab:phases}.

As far as the symmetries are concerned,
the fully isospin-polarized CDW and KD phases, unaffected by the Zeeman field,
retain their $\Utxt(1)$ and $Z_2$ degeneracies of the isospin orientation.
At the same time, the F phase becomes nondegenerate, with the spin $\s=\s_z = (0,0,1)$ directed along the Zeeman field
(i.e., the total magnetic field),
and the CAF phase is $\Utxt(1)$-degenerate with respect to simultaneous rotations of the spins $\s_{a,b}$
about the direction of the Zeeman field [angle $\vphi_s$ in Eq.~(\ref{eq:sCAF})].
The continuous $\Utxt(1)$ degeneracies of the CAF and KD phases could be subject to thermal fluctuations, which we do not addressed here.

Note that, while all phase transitions
in Fig.~\ref{fig:phasesnoZ}
and the rest of the transitions in Fig.~\ref{fig:phases} are first-order (black lines),
the CAF-F transition [Eq.~(\ref{eq:CAFF}), dashed blue line in Fig.~\ref{fig:phases}] is second-order:
upon increasing $\e_Z/|u_\perp|$, the CAF phase continuously crosses over to the F phase,
as the AF component ($\sim \!\sqrt{1-s_z^{*2}}$) of the CAF order parameter gradually decreases, while its F component ($\sim \!s_z^*$) grows;
eventually, at the CAF-F transition line ($s_z^*=1$), the AF component turns to zero,
while the F component saturates and experiences a jump in derivative (see top graph in Fig.~\ref{fig:phases}).

\subsection{Relation to earlier works}

Here, we discuss the connection of our results to earlier related studies.

In Refs.~\onlinecite{AF,JM}, the lattice effects of e-e interactions
on the $\nu=0$ QHFM were studied using the
tight-binding extended Hubbard model with adjustable interactions
at the lattice scale and asymptotically Coulomb interactions at large scales.
Within this model,
the overlap of the orbitals at different atomic sites is exactly zero and,
neglecting renormalizations, the anisotropy energy $u_\perp^{(\text{e-e})}=0$ vanishes exactly
and only $u_z^{(\text{e-e})}$ is present.
The cases $u_z^{(\text{e-e})}>0$ and $u_z^{(\text{e-e})}<0$ are realized
when, roughly, the sum of three nearest-neighbor repulsions is
smaller or greater than the on-site repulsion, respectively.
Accordingly, in Ref.~\onlinecite{AF}, in the presence of the Zeeman effect,
the competition between the F and CDW phases was predicted,
which agrees with the phase diagram in Fig.~\ref{fig:phases} at the line
$u_\perp=0$; the transition point is given by Eq.~(\ref{eq:FCDW}).
In Ref.~\onlinecite{JM}, the comparison between the CDW, F, and AF phases was done using numerical mean-field analysis.
The conclusion was reached that, ignoring the Zeeman effect,
depending on the details of interactions at lattice scale,
either CDW  or AF phase is favored, while the F phase has higher energy.
This is also consistent with the phase diagram in Fig.~\ref{fig:phasesnoZ} at $u_\perp=0$.
Note that, at  $u_\perp=0$ and $u_z>0$,
the system is right at the transition line between the AF and F phases
and therefore, even minor perturbations (numerical or other)
would drive the system into one of the phases.

One may also draw certain parallels between
the $\nu=0$ QHFM in graphene and
the semiconductor quantum Hall bilayers~\cite{CAFinQHB1,CAFinQHB2,CAFinQHB3,Ezawa_etal} (QHB)
at the total filling factor $\nu_{\text{QHB}} =2 $.
There, the role of the isospin is played by the layer degrees of freedom.
The leading anisotropy comes from the difference between the Coulomb
interactions within and between the layers.
The resulting ``capacitance'' effect is described by the anisotropy $u_z>0$,
while a minor $u_\perp>0$ due to the Coulomb interactions is usually neglected in theoretical studies.

The proximity of the layers results a finite overlap of the different-layer wave-functions and a possibility of tunneling.
In the QHFM theory, this is described by an extra isospin ``Zeeman'' term $\Ec_t(P) = - \e_t \tr [\Tc_x P ]$, $\e_t >0$,
in the energy $\Ec(P)$ [Eq.~(\ref{eq:Etotal})], i.e.,
tunneling by itself favors
the isospin-polarized (IP$_x$) state $P^{\text{IP}_x} = P_{\nb_x} \otimes \hat{1}^s$ with the isospin $\nb_x =(1,0,0)$ along the $x$ direction.
Remarkably, in addition to the F and IP$_x$ phases favored by the Zeeman effect and interlayer tunneling,
the CAF phase with the spin polarizations of the layers as in Fig.~\ref{fig:CAF}
was also predicted~\cite{CAFinQHB1,CAFinQHB2,CAFinQHB3} to exist in a finite-size region of the phase diagram
between the F and IP$_x$ phases.
According to Refs.~\onlinecite{CAFinQHB1,CAFinQHB2,CAFinQHB3},
the antiferromagnetic coupling between the spins in different layers favoring the CAF phase
has a super-exchange nature and arises from the correlated two-particle tunneling processes.

It should be emphasized that the physical origin
of KD and CAF phases in graphene and IP$_x$ and CAF phases in QHB is different.
On the one hand,  the possibility of tunneling between the layers is a necessary condition for the existence of both IP$_x$ and CAF phases in QHB.
An analogous ingredient is absent in the bulk of real graphene samples:
creating the  isospin ``Zeeman'' field directed in the $xy$ isospin plane
requires a {\em static} ``nano-engineered'' Kekul\'{e} distortion, Fig.~\ref{fig:phmodes}(right).
On the other hand, in graphene, the KD phase is favored by large enough negative anisotropy $u_\perp<0$, $|u_\perp|> |u_z|$,
while the AF phase (CAF, in the presence of the Zeeman effect) is favored by the  combined anisotropies $u_z>0$ and $u_\perp<0$ at $u_z>-u_\perp$.
The negative anisotropy $u_\perp<0$ arises from the attractive interactions in $\perp$-isospin channel,
provided by either the e-ph interactions or short-range e-e interactions that turned attractive upon renormalization.
Such mechanisms of $u_\perp <0$ are minor or absent in QHB.

\subsection{Inducing phase transitions in the $\nu=0$ QHFM\label{sec:transitions}}

Turning to potential practical applications of the present theory,
two important questions  can be addressed:
(i) which phase in Fig.~\ref{fig:phases} is realized in the experimentally observed insulating $\nu=0$ state;
(ii) whether one can induce transitions between different phases in a real system.
We address the latter question in this subsection and the former in Sec.~\ref{sec:conclusions}.

The ground state of the $\nu=0$ QHFM is determined by the relations between the parameters $(u_\perp,u_z, \e_Z)$,
the anisotropy $u_{\perp,z}$ and Zeeman $\e_Z$ energies,
and the question is whether one can change these relations in the experiment.
Changing the bare couplings $g_{\al\be}$ $\al,\be=\perp,z$,
in order to modify the relation between $u_\perp$ and $u_z$,
seems quite challenging as they are determined by the details of the interactions and band structure at atomic scale.
Varying the magnetic field, in magnitude or orientation,
is then virtually the only practical option.

Let us first neglect the Zeeman effect and inquire if the relation between $u_\perp(B_\perp)$ and $u_z(B_\perp)$
could be modified by varying the magnitude of the perpendicular magnetic field $B_\perp$.
As the RG analysis of Sec.~\ref{sec:renormu} shows,
the short-range e-e couplings $\gb_{\al\be}(l_B)$ [Eqs.~(\ref{eq:galperp}), (\ref{eq:galz}), and (\ref{eq:Fcee})]
do change their signs and relative values upon renormalization,
and, as a result, so do the anisotropy energies $u_{\perp,z}(B_\perp)$ [Eqs.~(\ref{eq:u}) and (\ref{eq:uee})]
upon varying the magnetic field $B_\perp$.
However, according to the estimates of Sec.~\ref{sec:implications},
these changes occur at magnetic length $l_B \sim a $ on the order of the lattice scale,
where the renormalizations are still moderate in magnitude,
$\Fc_\text{e-e}(l_B/a,w)\sim 1$, i.e.,
for unrealistically high  magnetic fields.
For all laboratory fields, $l_B / a \sim 10-100 $ and the renormalizations are strong, $\Fc_\text{e-e}(l_B/a,w) \gg 1$.
In this regime, the signs and relative value of the anisotropy energies  $u_{\perp,z}(B_\perp)$ [Eq.~(\ref{eq:ueestrong})]
cannot anymore be changed and, hence, the transitions cannot be induced, by varying the magnetic field.

A potentially  more  fruitful  way to induce phase transitions could be by tilting the magnetic field~\cite{AF,JM}.
The anisotropy energies $u_{\perp,z}(B_\perp)$ depend on the
magnetic field component $B_\perp$, perpendicular to the sample,
while the Zeeman energy $\e_Z(B)=\mu_B B$ depends on the total value $B = \sqrt{B_\perp^2+B_\parallel^2}$.
Therefore, tilting the magnetic field
increases the Zeeman energy $\e_Z$ relative to the anisotropy energies $u_{\perp,z}$, which makes the
F phase more favorable.
Provided at perpendicular field orientation the system is in a different phase,
eventually, upon increasing $B/B_\perp$, the system must end up in the F phase.

\begin{figure}
\includegraphics[width=.30\textwidth]{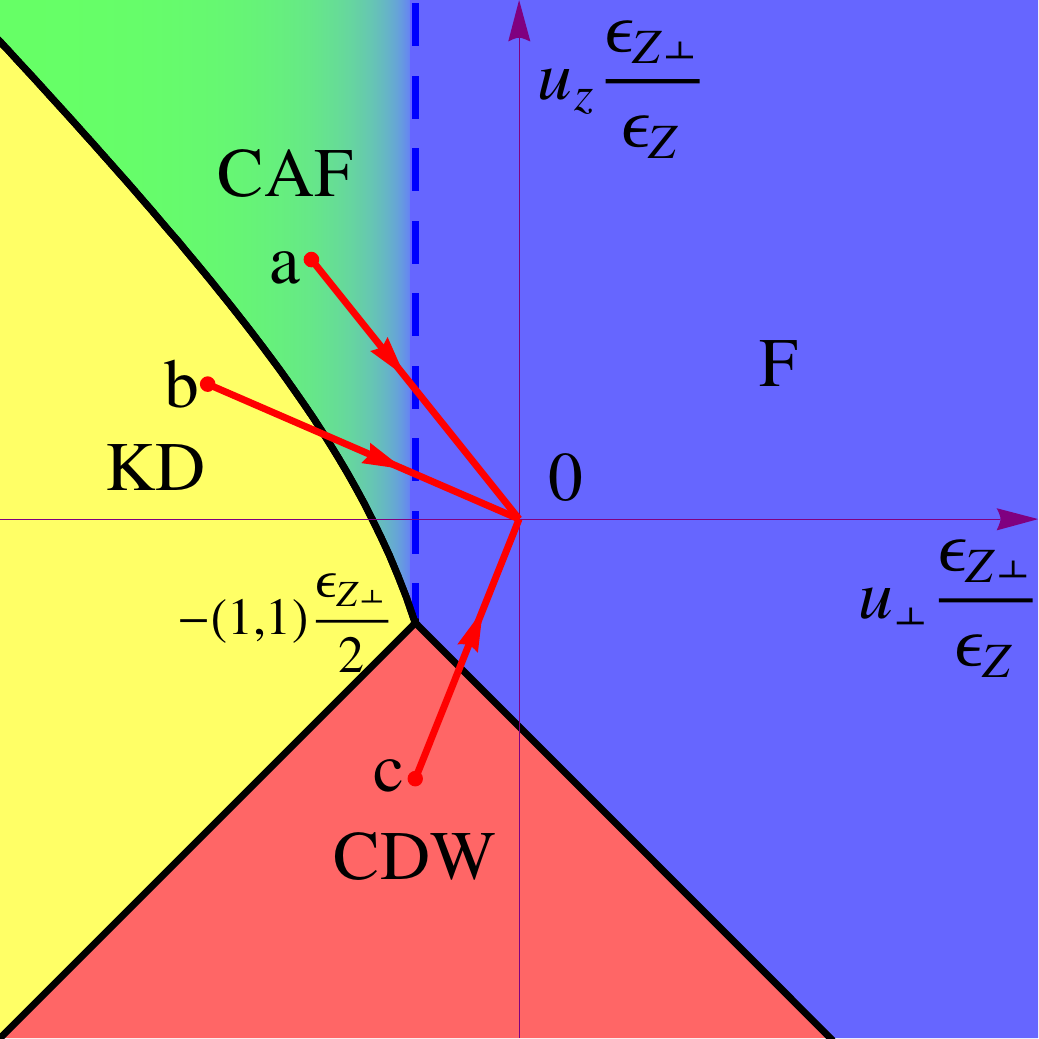}
\caption{Phase transitions induced by tilting the magnetic field, which increases the
Zeeman energy $\e_Z$ relative to the anisotropy energies $u_{\perp,z}$.
The transitions from  the CAF (point $a$) or CDW (point $c$) to the F phase occur directly,
while the transition from the KD (point $b$) to the F phase can occur only through the CAF phase.
}
\label{fig:transitions}
\end{figure}

Figure~\ref{fig:transitions} shows the evolution of phases as one applies the parallel component $B_\parallel$, while keeping $B_\perp$ fixed,
thus changing the Zeeman energy from $\e_{Z\perp} = \mu_B B_\perp$ to $\e_Z= \mu_B B > \e_{Z\perp}$ and keeping the anisotropy energies $(u_\perp,u_z)$ constant.
It is convenient to present the phase diagram in the units $\frac{\e_{Z\perp}}{\e_Z}(u_\perp,u_z)$:
this way, the phase boundaries remain fixed, while the phase point with constant  $(u_\perp,u_z)$ moves along
the straight line from its position at $B_\parallel=0 $  to the origin at $B/B_\perp \rtarr \infty$.
If the system is in either the CAF or CDW phases at $B_\parallel =0$, points $a$ or $c$,
respectively, the transition occurs directly into the F phase upon increasing $\e_Z/\e_{Z\perp}$,
via a continuous second-order transition in the former case and a first-order transition in the
latter case.
Remarkably, however, if the system is in the KD phase at $B_\parallel=0$, point $b$, the transition to the F phase cannot
occur directly: upon increasing $\e_Z$,  the system first makes a first-order phase transition into
the CAF phase, and then continuously transitions to the F phase.

We note, however, that, since the renormalizations of the isospin anisotropies are strong,
such transitions may be quite challenging to realize.
According to Sec.~\ref{sec:implications},
although the bare energies $u_{\perp,z}^{(0)} \sim B_\perp [\text{T}] \text{K}$ [Eq.~(\ref{eq:u0est})]
are comparable with the Zeeman energy $\e_{Z\perp} = \mu_B B_\perp$ for perpendicular
field orientation,  the renormalized energies $u_{\perp,z}$ [Eqs.~(\ref{eq:ueeest}) and (\ref{eq:uephest})]
can easily exceed $\e_{Z\perp}$ by one order of magnitude.
Therefore, achieving  the transition by applying the parallel component $B_\parallel$,
would generally require large ratios $B/B_\perp \gtrsim 10 $, i.e., large tilt angles.
Since one should, at the same, maintain a large enough perpendicular component $B_\perp$
to preserve the correlated quantum Hall state above the disorder level ($B_\perp \approx 3-5 \Ttxt$, according to Refs.~\onlinecite{nu0Andrei,nu0Kim}),
the practical maximum of the ratio $B/B_\perp$
is limited by the maximum achievable magnetic field and disorder in the system.
An exception concerns the special cases
when, for perpendicular field orientation, the system is ``anomalously'' close to one of the KD-CAF, CDW-F, or CAF-F transitions lines in Fig.~\ref{fig:phasesnoZ} [i.e., the point $(u_\perp,u_z)$ is $\sim \e_{Z\perp}$ away from the line] and is on the KD, CDW, or CAF side, respectively.
In this case, smaller ratios $B/B_\perp \sim 1$ would be sufficient for the phase transition to occur.

\section{Conclusions\label{sec:conclusions}}

In conclusion, in this paper, we studied the $\nu=0$ quantum Hall state in monolayer graphene in the framework of quantum Hall ferromagnetism,
with the key emphasis on the isospin anisotropies
that arise from the valley and sublattice asymmetric
short-range electron-electron and electron-phonon interactions.
The phase diagram in Fig.~\ref{fig:phases}, obtained in the presence of the generic isospin anisotropy and Zeeman effect
(neglecting thermal fluctuations),
consists of four phases characterized by the following orders: spin ferromagnetic (F), charge density wave (CDW),
Kekul\'{e} distortion (KD), and canted antiferromagnetic (CAF).
To the best of our knowledge, the CAF phase has not been predicted before in the context of the correlated quantum Hall states in graphene.
We took into account the Landau level mixing effects (Sec.~\ref{sec:renorm})
and found that they result in (i) the suppression of the stiffness $\rho_s$ due to screening and (ii)
critical renormalizations of the anisotropy energies $u_{\perp,z}$.
The latter has  crucial implications for the physics of the $\nu=0$ state.
First, the anisotropies are greatly enhanced and can significantly exceed the Zeeman energy.
Second, and most importantly, we conclude that the short-range electron-electron interactions could
favor any state on the generic phase diagram:
one cannot theoretically rule out any possibility
based just on the repulsive nature of the underlying Coulomb interactions.
The leading electron-phonon interactions, on the other hand, always favor the Kekul\'{e} distortion phase.

The main open practical question is then which of the phases
in Fig.~\ref{fig:phases} corresponds to the strongly insulating $\nu=0$ state
observed in transport experiments~\cite{nu0Ong,nu0Brookhaven,nu0Andrei,nu0Kim,ColumbiaBN}.
For high quality suspended graphene samples~\cite{nu0Andrei,nu0Kim},
the two-terminal resistance was instrument-limited at  $R \sim 10^9-10^{10} \text{Ohm}$
at $ B \sim 10\Ttxt$ and $T \lesssim 1 \text{K}$.
In the absence of an accurate estimate for the anisotropy energies,
one can try to infer about the nature of the
real $\nu=0$ state by addressing the transport properties of the phases in Fig.~\ref{fig:phases}.

While the charged excitations of the $\nu=0$ QHFM are gapped in the bulk for any order~\cite{Arovas_etal,AF,YDM,Goerbig},
the phases in Fig.~\ref{fig:phases} are expected to have markedly different edge transport behavior.
Existing studies~\cite{FB,Abanin} suggest that the F phase  has gapless edge excitations.
Therefore, an ideal sample with the F bulk order would have a two-terminal resistance $R = 1/(2 e^2/h) \sim 10^4 \text{Ohm}$,
with the factor 2 due to two edges.
On the other hand, the CDW~\cite{JM,edgestates2}, AF~\cite{JM,edgestates2}, and KD~\cite{edgestates3} phases were shown to have gapped single-particle mean-field edge excitations, which implies that these phase should exhibit insulating behavior.
The edge excitations of the CAF phase have not yet been addressed.
Note, however, that the spin-polarized F phase, being an isospin singlet, is the only phase  of the $\nu=0$ QHFM
that does not break the sublattice symmetry. This observation could be used as a  ``rule of thumb'' argument in speculating about the edge transport properties of the $\nu=0$ states. Based on this, one can expect all three phases, CDW, KD, and CAF, to have gapped edge excitations and exhibit
insulating behavior, which  makes them much stronger candidates than the F phase for the $\nu=0$ state realized experimentally.
Distinguishing further between the three insulating phases based on the transport properties
requires a more detailed analysis of their bulk and edge charge excitations, to be presented elsewhere~\cite{MKunpub}.

\section{Acknowledgements}
Author is thankful to Eva Andrei, Matt Foster, Piers Coleman, Andrea Young, and Peter Silvestrov for insightful discussions
and acknowledges the hospitality of the TPIII group at Ruhr-Universit\"{a}t Bochum, Germany, where part of the work was completed.
The work was supported by the US Department of Energy under Contracts DE-FG02-99ER45790 at Rutgers University 
and DE-AC02-06CH11357 at Argonne National Laboratory.


\begin{thebibliography}{99}
\bibitem{QHFMgroundstate}
It is worth noting that, for generic symmetric repulsive interactions and/or arbitrary integer filling factor,
a strict proof that the QHFM eigenstate is, in fact, a ground state does not exist in literature.
(This is definitely true for point interactions and singly occupied orbitals~\cite{YDM}, since in this case $E_0=0$
and the Hamiltonian of repulsive interactions is positive-semidefinite: $\lan \Psi | \Hh_{\text{e-e},0}^{(0)} |\Psi \ran \geq 0$
for any state $\Psi$.)
However, the physical argumentation in favor of this is provided by the very idea of the Hund's rule, which motivates the
form of the QHFM state.
\end{thebibliography}
\end{document}